\documentclass[10pt, conference, compsocconf]{IEEEtran}
\usepackage{latexsym}
\usepackage{times}
\usepackage{epsfig}
\usepackage{graphicx}
\usepackage{subfigure}
\usepackage{multirow}
\usepackage{amsfonts}
\usepackage{algorithm,algorithmic}
\usepackage{caption}
\usepackage{url}
\usepackage[usenames]{color}

\newcommand{\comment}[1]{}

\newsavebox{\boxone}
\newsavebox{\boxtwo}
\newsavebox{\boxthree}

\newlength{\narrow}
\newlength{\cnarrow}
\newcommand{\topline}{
  \hrule
  \vskip .5\baselineskip}
\newcommand{\bottomline}{
  \vskip 2pt
  \hrule}

\newcommand{\chbox}[2]{
  \hbox to #1{\hss\vtop{#2}\hss}}

\newcommand{\nchbox}[1]{
  \chbox{\narrow}{#1}}

\newcommand{\cnchbox}[1]{
  \chbox{\cnarrow}{#1}}

\newcommand{\fcode}[1]{
  
  \chbox{\textwidth}{\tgrind\input{#1}}}

\newcommand{\ncode}[1]{
  
  \chbox{\narrow}{\tgrind\input{#1}}}

\def\nfig#1#2#3{
  \vtop{\nchbox{#1}
  \hbox to\narrow{\parbox{\narrow}{\caption{#2}\label{#3}}}}}

\newcommand{\cncode}[1]{
  \chbox{\cnarrow}{\tgrind\input{#1}}}

\def\codefiggen[#1]#2#3#4#5#6{
  \begin{figure}[#1]
  #5
  \fcode{#2}
  \center\parbox{.9\textwidth}{\caption{#3}\label{#4}}
  #6
  \end{figure}}

\def\codefig[#1]#2#3#4{
  \codefiggen[#1]{#2}{#3}{#4}{}{}}

\def\codefigline[#1]#2#3#4{
  \codefiggen[#1]{#2}{#3}{#4}{\topline}{\bottomline}}

\def\doublefiggen[#1]#2#3#4#5#6#7#8#9{
  \begin{figure}[#1]
  #8
  \hbox to \textwidth{
  \nfig{#2}{#3}{#4}
  \hfil
  \nfig{#5}{#6}{#7}}
  #9
  \end{figure}}

\def\doublefig[#1]#2#3#4#5#6#7{
  \doublefiggen[#1]{#2}{#3}{#4}{#5}{#6}{#7}{}{}}

\def\doublefigline[#1]#2#3#4#5#6#7{
  \doublefiggen[#1]{#2}{#3}{#4}{#5}{#6}{#7}{\topline}{\bottomline}}

\def\doublecodefig[#1]#2#3#4#5#6#7{
  \doublefig[#1]{\ncode{#2}}{#3}{#4}{\ncode{#5}}{#6}{#7}}

\def\doublecodefigline[#1]#2#3#4#5#6#7{
  \doublefigline[#1]{\ncode{#2}}{#3}{#4}{\ncode{#5}}{#6}{#7}}

\newcommand{\codepair}[4]{\vbox{
  \hbox{\ncode{#1} \hfil \ncode{#3}}
  \vskip .3\baselineskip plus .3\baselineskip
  \hbox{\hbox to\narrow{#2\hfil} \hfil \hbox to\narrow{#4\hfil}}}}

\def\codepairfig[#1]#2#3#4#5#6#7{
  \begin{figure}[#1]
  \codepair{#2}{#3}{#4}{#5}
  \center\parbox{.9\textwidth}{\caption{#6}}
  \label{#7}
  \end{figure}}

\def\cncodepairfiggen[#1]#2#3#4#5#6#7{
  \begin{figure}[#1]
  #6
  \hbox{\cncode{#2}\hfil\cncode{#3}}
  \center\parbox{.9\columnwidth}{\caption{#4}\label{#5}}
  #7
  \end{figure}}

\def\cncodepairfig[#1]#2#3#4#5{
  \cncodepairfiggen[#1]{#2}{#3}{#4}{#5}{}{}}

\def\cncodepairfigline[#1]#2#3#4#5{
  \cncodepairfiggen[#1]{#2}{#3}{#4}{#5}{\topline}{\bottomline}}

\def\doublefigOnecap*[#1]#2#3#4#5{
  \begin{figure*}[#1]
  \hbox to \textwidth{
  \nchbox{#2}
  \hfil
  \nchbox{#3}}
  \caption{#4}
  \label{#5}
  \end{figure*}}

\def\doublefigOnecap[#1]#2#3#4#5{
  \begin{figure}[#1]
  \topline
  \hbox to \columnwidth{
  \cnchbox{#2}
  \hfil
  \cnchbox{#3}}
  \caption{#4}
  \label{#5}
  \bottomline
  \end{figure}}

\def\PSfig[#1]#2#3#4{
 \begin{figure}
 \centerline{\psfig{file=#2,width=\columnwidth}}
 \caption{{#3}}
 \label{#4}
 \end{figure}}

\def\PSfiglines[#1]#2#3#4{
 \begin{figure}[#1]
 \topline
 \centerline{\psfig{file=#2,width=\columnwidth}}
 \caption{{#3}}
 \label{#4}
 \bottomline
 \end{figure}}

\def\PSfiglinesht[#1]#2#3#4#5{
 \begin{figure}[#1]
 \topline
 \centerline{\psfig{file=#2,height=#3}}
 \caption{{#4}}
 \label{#5}
 \bottomline
 \end{figure}}

\def\doublePSfig[#1]#2#3#4#5#6{
  \doublefigOnecap[#1]
    {\cnchbox{\psfig{file=#2,height=#4}}}
    {\cnchbox{\psfig{file=#3,height=#4}}}
    {#5}
    {#6}}

\newlength{\boxwidth}
\newcommand{\bproof}{{\bf Proof Sketch: }}
\newcommand{\eproof}{\mbox{$\Box$}}

\def\tabdoublecode#1#2#3#4{
 \begin{figure*}[t]
 \topline\vs{-.4}
 \hbox to \columnwidth{
 \vtop{\small
 \begin{tabbing}
 #1
 \end{tabbing}}
 \hfil
 \hfil
 \hfil
 \vtop{\small
 \begin{tabbing}
 #2
 \end{tabbing}}
 }
 \caption{#3\label{#4}}
 \bottomline
 \end{figure*}
}
\def\tabtriplecode#1#2#3#4#5{
 \begin{figure}
 \topline\vs{-.4}
 \hbox to \columnwidth{
 \vtop{\small
 \begin{tabbing}
 #1
 \end{tabbing}}
 \hfil
 \hfil
 \hfil
 \vtop{\small
 \begin{tabbing}
 #2
 \end{tabbing}}
 \hfil
 \hfil
 \hfil
 \vtop{\small
 \begin{tabbing}
 #3
 \end{tabbing}}
 }
 \caption{#4\label{#5}}
 \bottomline
 \end{figure}
}

\newtheorem{lemma}{Lemma}
\newcommand{\blemma}{\begin{lemma}}
\newcommand{\elemma}{\end{lemma}}

\newtheorem{thm}{Theorem}
\newcommand{\bthm}{\begin{thm}}
\newcommand{\ethm}{\end{thm}}

\newtheorem{defin}{Definition}
\newcommand{\bdefin}{\begin{defin}}
\newcommand{\edefin}{\end{defin}}

\newtheorem{observation}{Observation}
\newcommand{\bobserv}{\begin{observation}}
\newcommand{\eobserv}{\end{observation}}

\newcommand{\vs}[1]{\vspace{#1cm}}
\newcommand{\be}{\begin{equation}}
\newcommand{\ee}{\end{equation}}

\newcommand{\bdesc}{\begin{description}}
\newcommand{\edesc}{\end{description}}
\newcommand{\benum}{\begin{enumerate}}
\newcommand{\eenum}{\end{enumerate}}
\newcommand{\bitem}{\begin{itemize}}
\newcommand{\eitem}{\end{itemize}}
\newcommand{\bcenter}{\begin{center}}
\newcommand{\ecenter}{\end{center}}
\newcommand{\btabular}{\begin{tabular}}
\newcommand{\etabular}{\end{tabular}}
\newcommand{\beqnarr}{
 \begin{eqnarray}}
\newcommand{\eeqnarr}{\end{eqnarray}}

\begin{document}

\date{}
\title{Network Backbone Discovery Using Edge Clustering}

\author{
\IEEEauthorblockN{Ning Ruan$^\dagger$ ~~~~~~~ Ruoming Jin$^\dagger$ ~~~~~~~ Guan Wang$^\dagger$}
\IEEEauthorblockA{$^\dagger$\mbox{ }Department of Computer Science\\
Kent State University\\
\{nruan,jin,gwang0\}@cs.kent.edu } \and \IEEEauthorblockN{Kun Huang$^\ddagger$}
\IEEEauthorblockA{$^\ddagger$\mbox{ }Department of Biomedical Informatics\\
Ohio State University\\
khuang@bmi.osu.edu} }


\maketitle

\begin{abstract}
In this paper, we investigate the problem of network backbone discovery.
In complex systems, a ``backbone'' takes a central role in carrying out the system functionality and carries the bulk of system traffic. 
It also both simplifies and highlight underlying networking structure. 
Here, we propose an itegrated graph theoretical and information theoretical network backbone model.
We develop an efficient mining algorithm based on {\em Kullback-Leibler} divergence optimization procedure and maximal weight connected subgraph discovery procedure.
A detailed experimental evaluation demonstrates both the effectiveness and efficiency of our approach. 
The case studies in the real world domain further illustrates the usefulness of the discovered network backbones. 
\end{abstract}

\section{Introduction}
\label{intro}

\comment{
As graph data is ubiquitous and is becoming larger and larger at a fast pace, their scale and complexity impose a great challenge on various graph mining and analysis tasks.
This problem directly leads to the emergence of research on reducing graph complexity or graph simplification, which has become an increasingly important topic~\cite{Zhou10,ToivonenMZ10,DBLP:conf/sigmod/SatuluriPR11,AggarwalWang10}.
Generally, graph simplification focuses on sparsifying graphs by reducing non-essential edges~\cite{Zhou10,ToivonenMZ10,DBLP:conf/sigmod/SatuluriPR11}, extracting key vertices~\cite{Tao11,Hutchinson:2003:EMD:639273.639278}, or collapsing substructures into supernodes~\cite{Navlakha:2008:GSB:1376616.1376661,AggarwalWang10}.
These simplified structures are able to facilitate many real-world applications, such as topology visualization~\cite{DBLP:conf/visualization/RafieiC05,Hennessey:2008} and computational speedup on various graph-centered tasks~\cite{Karande:2009:SUA,Tao11,DBLP:conf/sigmod/SatuluriPR11}.
}

In many man-made complex networking systems, a ``backbone'' takes a central role in carrying out the system functionality. The clearest examples are the highway framework in the transportation system and the backbone of the Internet.
When studying these systems, the backbone offers both a concise and highlighted view.
Furthermore, it also provides key insight on understanding how the entire system organizes and works.
Does ``backbone'' exist in natural or social network?  What should it look like? How we can discover them efficiently?
Interestingly, the backbone phenomena has been recently observed in several natural and social systems, including metabolic networks~\cite{CN:Almass04}, social networks~\cite{CN:Kossinets08}, and food webs~\cite{SgelesSerrano04212009}.
Unfortunately, there is no formal definition of the network backbone and no goodness function defined in all existing work.


\comment{
In this paper, we attempt to extract a backbone structure which takes a central role in the ``map'' of a complex network.
Intuitively, a ideal network backbone is able to survey a great deal of system-wide information traffic with very simple structure.
Like the highways in the transportation system and the backbone of the Internet, they offer both a concise and highlighted view of these man-made complex systems.
Interestingly, the backbone phenomenon has recently been observed in several natural and social systems, including metabolic networks~\cite{CN:Almass04}, social networks~\cite{CN:Kossinets08}, and food webs~\cite{SgelesSerrano04212009}.
Clearly, discovering and extracting network backbones is essential to understanding the organization principles of complex systems and a key step towards ``map'' functionality in complex network analysis.
Unfortunately, there is no formal definition of the network backbone.
All the existing work merely captures the backbone as whatever their particular heuristic procedures find.

Faloutsos {\em et al.} have studied the connection subgraph problem in trying to identify a concise subgraph to best describe the relationship between a pair of vertices~\cite{DBLP:conf/kdd/FaloutsosMT04} or a small number of vertices~\cite{DBLP:conf/kdd/TongF06}.
Their methods, however, cannot be utilized to discover the central structure (backbone) of the entire network.

In this work, we attempt to extract a backbone structure which takes a central role in such kind of structure.
Like highways in transportation system and backbone of Internet, they offer both a concise and highlighted view of these systems.
The backbone phenomenon also has been observed in several natural and social systems, including metabolic networks~\cite{CN:Almass04}, social networks~\cite{CN:Kossinets08}, and food webs~\cite{SgelesSerrano04212009}.
}

\comment{
Two goals for backbone discovery:
1, graph is too complicated, we want to simplify them and get a representative structure (core subgraph)
2, extract a structure to capture major traffic (highway in transportation network)
simply mention the non-local traffic or communication
In this paper, we generally consider the backbone discovery in non-local traffic or communication.

Complex network has been playing an increasingly important role in various real-world systems,
ranging from biocellular, ecological, and neurological systems, to the World Wide Web, social networks, and economical and financial markets.
It models these network-centered systems as a large number of objects (vertices) that are linked by non-trivial relationships (edges).
As a powerful abstraction, complex network not only explicitly describes the immediate interaction between two objects through edges,
but also implicitly characterize system-wide information traffic by the paths between objects.
Analyzing complex network offers a great opportunity to shape our understanding of system behaviors.
So far, many research efforts~\cite{Newman03,CN:GGLT04,DBLP:conf/kdd/LeskovecKGFVG07,DBLP:conf/kdd/SunFPY07,CN:APU07,DBLP:conf/kdd/LeskovecKF05}
have been devoted to explore complex network.
However, as networks are becoming larger and larger, with some even including hundreds of millions of vertices,
their scale and complexity impose a great challenge on complex network analysis.
A question naturally arises: {\em can we comprehend a network when we cannot really look at it?}
To address this fundamental call, we need a ``map'' to facilitate the complex network analysis and exploration.
A good map not only can both simplify and highlight the underlying structure, but also retain essential interactions among objects (i.e., system-wide information traffic) in the complex network.
Though existing clustering methods~\cite{CN:Handcock07,CN:Hastings06,CN:Karrer08} can decompose a complex network into densely connected components to simplify the original structure,
they cannot reveal the high-level system traffic flow from a map perspective as the inter-component interactions are omitted.

In this work, we attempt to extract a backbone structure which takes a central role in the ``map'' of a complex network.
Intuitively, a ideal network backbone is able to survey a great deal of system-wide information traffic with very simple structure.
Like the highways in the transportation system and the backbone of the Internet, they offer both a concise and highlighted view of these man-made complex systems.
Interestingly, the backbone phenomenon has recently been observed in several natural and social systems, including metabolic networks~\cite{CN:Almass04}, social networks~\cite{CN:Kossinets08}, and food webs~\cite{SgelesSerrano04212009}.
Clearly, discovering and extracting network backbones is essential to understanding the organization principles of complex systems and a key step towards ``map'' functionality in complex network analysis.
Unfortunately, there is no formal definition of the network backbone.
All the existing work merely captures the backbone as whatever their particular heuristic procedures find.

Faloutsos {\em et al.} have studied the connection subgraph problem in trying to identify a concise subgraph to best describe the relationship between a pair of vertices~\cite{DBLP:conf/kdd/FaloutsosMT04} or a small number of vertices~\cite{DBLP:conf/kdd/TongF06}.
Their methods, however, cannot be utilized to discover the central structure (backbone) of the entire network.
}

\comment{
As graph is becoming larger and larger, the scale and complex of graph data impose a great challenge on complex network analysis.
Considering the graph visualization task which is widely called on complex network analysis,
a visual representation of a graph even with hundred of vertices will be overwhelming and brings no real benefit to users.

Over the last decade, complex network analysis has emerged as a major interdisciplinary research field studying network-centered systems,
ranging from biocellular, ecological, and neurological systems, to the World Wide Web, social networks, and economical and financial markets.
A {\em complex network} models these systems as a large number of elements (vertices) that are joined by non-trivial relationships (edges).
An array of tools have been developed for complex network analysis~\cite{Newman03}.
Network modeling, modularity and network motifs are three representative tools: network modeling seeks simple generative rules to produce the emergent, scale-free, small-world, and even real spatial, weighted networks; network modularity reveals the underlying organizational principle by decomposing a complex network into densely connected (or functionally homogeneous) components; network motifs, which can be loosely described as over-represented subgraphs, are like small circuits, corresponding to the basic functional units in many different systems.
The data mining community has also contributed in a number of substantial ways to this emerging field, generally under the umbrella of graph mining, web mining, and social network analysis.
Some key contributions include the development of efficient frequent subgraph mining algorithms~\cite{WashioM03}, scalable graph clustering methods~\cite{CN:Xu07,CN:whites05}, and various methods to study the dynamics of the networks~\cite{CN:GGLT04,DBLP:conf/kdd/LeskovecKGFVG07,DBLP:conf/kdd/SunFPY07,CN:APU07,DBLP:conf/kdd/LeskovecKF05}, among many others.

However, it remains a far-reaching goal to fully decipher the relationship between
the network {\em structures} and the complex system {\em behaviors}.
These networks, even when abstracted to only vertices and edges, are still highly complex, with some as large as tens of millions of vertices.
How can we comprehend a network when we cannot really look at it?
How can we correlate the network structure with the system functionality and dynamics?
To answer these fundamental calls, we need a ``traffic-map'' for complex network exploration and analysis.
A good map can both simplify and highlight the underlying structures.
Integrating with traffic information, like the state-of-the-art Google and MapQuest maps, the dynamic behavior of the systems can also be characterized.
Though existing clustering methods can decompose a complex network into densely connected components, they cannot reveal the high-level system traffic flow from a map perspective.

Backbone structure takes a central role in the ``map'' of a complex network.
Like the highways in the transportation system and the backbone of the Internet, they offer both a concise and highlighted view of these man-made complex systems.
Interestingly, the backbone phenomenon has recently been observed in several natural and social systems, including metabolic networks~\cite{CN:Almass04}, social networks~\cite{CN:Kossinets08}, and food webs~\cite{SgelesSerrano04212009}.
Clearly, discovering and extracting network backbones is essential to understanding the organization principles of complex systems and a key step towards ``map'' functionality in complex network analysis.
Unfortunately, there is no formal definition of the network backbone.
All the existing work merely captures the backbone as whatever their particular heuristic procedures find.
}

\comment{
In this paper, we propose and develop a new tool for discovering {\em network backbones}, which form a core structure in linking the ``system traffic'' with the network topology.
A network backbone not only can simplify and highlight the underlying structures, but also carries the bulk of system traffic.
In many man-made complex systems, a ``backbone'' takes a central role in carrying out the system functionality. }



\begin{figure*}
    \centering
    \mbox{
        \subfigure[Network and its backbone]{\includegraphics[width=1.5in,height=1.2in]{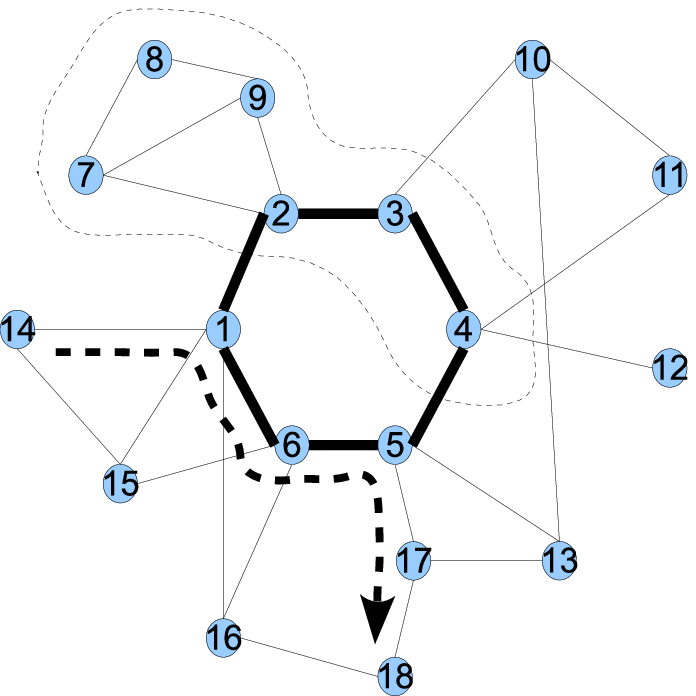}\label{f1}}
    }
    \mbox{
        \subfigure[Coding of subgraph]{\includegraphics[width=1.5in,height=1.2in]{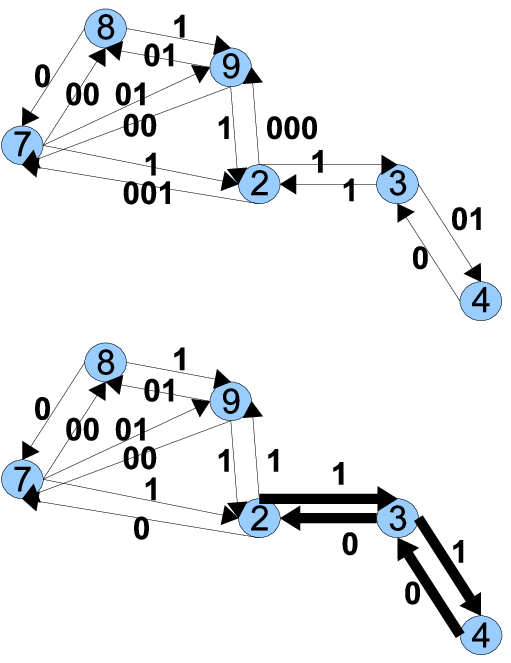}\label{f2}}
    }
    \mbox{
        \subfigure[Path coding example]{\includegraphics[width=1.5in,height=1.2in]{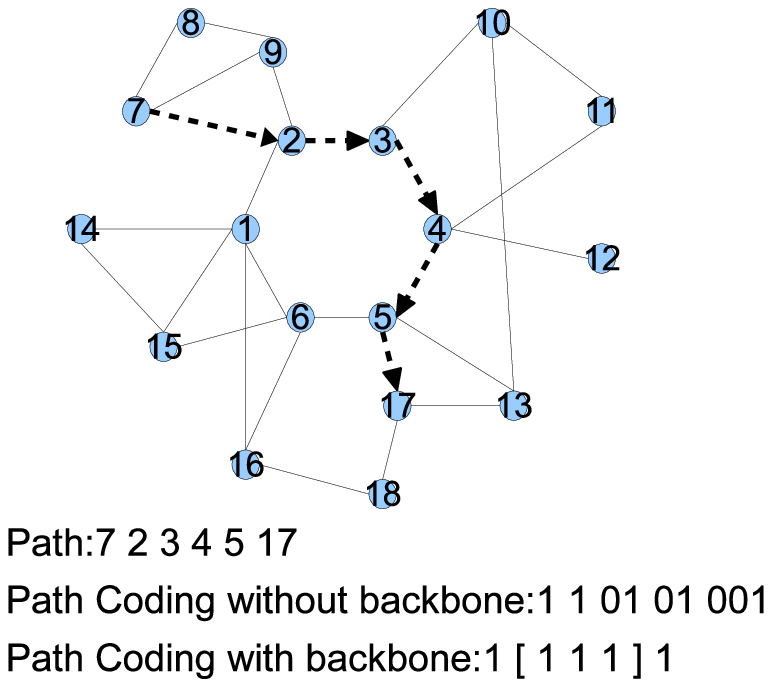}\label{f3}}
    }

    \caption{Path encoding with and without backbone}

\label{fig:graphcode}
\end{figure*}

In the paper, we propose the first theoretical network backbone model under an integrated graph theoretical and information theoretical framework.
Intuitively, network backbone is a connected subgraph which carries major network ``traffic''.
It can simplify and highlight underlying structures and traffic flows in complex systems.
A complex system's behavior relies on proper communication through the underlying network substrate,
which invokes a sequence of local interactions between adjacent pair of vertices.
These interactions thus form system-wide network traffic~\cite{CN:Rosvall08b}.
It is essential for a complex system to deliver such traffic in an energy efficient way~\cite{BaddiP99}.
Here, at least two energy costs should be considered:
1) communication cost and 2) organization cost in defining/recognizing communication path.
For the first cost, energy-efficient way for one vertex communicating with another vertex naturally points to shortest path,
i.e., the information flow over a network primarily follows the shortest ones~\cite{GirvanNewman02}.
The second cost can be described as a path-recognition complexity.
Based on information theory, we could depict it as the shortest coding length of a given path.
In general, the optimal code length for an edge $(i,j)$ can be bounded by $\log \frac{1}{p(i\rightarrow j)}$,
where $p(i\rightarrow j)$ is the probability of a shortest path taking edge $(i,j)$ when it goes through vertex $i$~\cite{CoverThomas91}.

Minimizing these two costs gives rise to the network backbone structure:
the shortest paths form a traffic flow which must efficiently travel from source to destination using the backbone.
Especially, if only the first cost is considered, only the edges with high betweenness ~\cite{GirvanNewman02} (roughly speaking, edge-betweenness defines the likelihood of any shortest path going through an edge) will be selected; however, those edges are not necessarily connected~\cite{GirvanNewman02}, how they should work together in delivering the system-wide traffic is unknown.
In this aspect, the second cost enables us to further constrain the backbone structure using the path-recognition complexity. A backbone that is too simple or has wrong topology is not an efficient route for vertex-vertex paths (first cost).
A backbone that is too complex is expensive to describe shortest paths (second cost).

Figure~\ref{fig:graphcode} illustrates a backbone and its usage in reducing the description length of a shortest path.
Since our computational framework implicitly evaluates path's information complexity by statistical likelihood,
it does not rely on any particular coding scheme (i.e., Huffman code is used here only for illustrative purpose).
Figure~\ref{f1} shows a network with its highlighted backbone.
Figure~\ref{f2} focuses on a subgraph, showing Huffman code for each edge, where the upper part is for the edge code without
utilizing backbone and the lower part is for using backbone.
For instance, edge $(9,2)$ with direction $9 \rightarrow 2$ is assigned with code $1$, and with direction $2 \rightarrow 9$ is assigned code $000$.
In Figure~\ref{f3}, we show path codings using and without using the backbone.
We basically list each edge code in the path consecutively.
Specifically, the subpath $(2,3,4,5)$ is in the backbone, and we use ``['' and ``]'' to denote the entering and exiting of the backbone, respectively.
It can also be observed that even with an extra coding cost for entering and exiting backbone, the lower coding cost for paths inside the backbone can still be beneficial.

We note that backbone discovery problem resonates with the recent efforts in the graph mining community on graph simplification.
To deal with the scale and complexity of graph data, reducing graph complexity or graph simplification is becoming an increasingly important research topic~\cite{Zhou10,ToivonenMZ10,DBLP:conf/sigmod/SatuluriPR11,AggarwalWang10}.
Generally, graph simplification focuses on sparsifying graphs by reducing non-essential edges~\cite{Zhou10,ToivonenMZ10,DBLP:conf/sigmod/SatuluriPR11}, extracting key vertices~\cite{Tao11,Hutchinson:2003:EMD:639273.639278}, or collapsing substructures into supernodes~\cite{Navlakha:2008:GSB:1376616.1376661,AggarwalWang10}.
These simplified structures are able to facilitate many real-world applications, such as topology visualization~\cite{DBLP:conf/visualization/RafieiC05,Hennessey:2008} and computational speedup on various graph-centered tasks~\cite{Karande:2009:SUA,Tao11,DBLP:conf/sigmod/SatuluriPR11}.
From this perspective, the backbone structure can potentially serve as a simplification approach, which
highlights a core set of vertices and edges in the original network.


\comment{

\subsection{Contributions}
\label{contribution}


\noindent 1) We propose the first theoretical backbone model based on a graph theoretical and information theoretical framework.
Optimizing this model leads to a ``good'' backbone simplifying underlying structure and capturing major information traffic. (Section \ref{problem})

\noindent 2) We develop two efficient discovering algorithms based on an interesting {\em Kullback-Leibler} divergence optimization procedure and maximum weight connected subgraph discovery procedure. (Section \ref{basics} and Section \ref{algorithm})

\noindent 3) We perform a detailed experimental evaluation and case study to demonstrate both the effectiveness and efficiency of our approach. (Section \ref{experiment})

\noindent
A detailed experimental evaluation demonstrates both the effectiveness and efficiency of our approach.
The case studies in the real world domain further illustrates the usefulness of the discovered network backbones.

}

\section{Statistical Learning  Framework for Backbones}
\label{problem}

\comment{
1) simplify notation, remove all stuff about long-distance traffic and traffic map;
2) before two simple model, just mention that $I(u)=O(u)$ in undirected graphs;
3) simplify the model description on bimodal markovian model and using the path from Figures 1 (also highlight it) as illustrative example in Figure 4);
4) after definition 2, mentioned that AIC or BIC is meaningless because the log value is significantly larger than parameter penalty.
And say that we will discuss the backbone size determination problem in empirical study part.
}

In this section, we introduce and refine the backbone description under a statistical learning framework.
Based on the well-known relationship between information complexity and statistical likelihood, the information complexity of each path (or a set of paths) can be equivalently represented as a likelihood function.
Further, this reformulation also generalizes the notion of backbone to optimize the information complexity for any given set of paths (information pathway) beyond the shortest paths.

\comment{
Without any prior knowledge, the shortest paths, which are particularly useful in characterizing system-wide traffic flow and
revealing the topological properties of a given network, become a nature choice.}

\subsection{Notations}
\label{notation}

To facilitate our discussion, we introduce the following notations.
Let $G=(V,E)$ be a undirected graph with edges $E \subseteq V \times V$.
Since each edge $(u,v)\in E$ in the graph may be assigned two codes, one for $(u \rightarrow v)$ and another for $(v \rightarrow u)$ (See Figure~\ref{fig:graphcode}),
it is more convenient to consider each undirected edge $(u,v)$ as two directed edges $(u \rightarrow v)$ and $(v \rightarrow u)$.
Therefore, we represent undirected graph $G$ as a bidirected graph, i.e., $G=(V,\mathcal{E})$ where $\mathcal{E}=\cup_{(u,v)\in E}\{ (u\rightarrow v), (v \rightarrow u) \}$.
Note that when we say an edge $(u,v)$ is a backbone edge in a undirected graph,
it suggests that both directed edges $(u \rightarrow v)$ and $(v \rightarrow u)$ from $\mathcal{E}$ are backbone edges in the bidirected graph.
The same holds for non-backbone edges.
This constraint does not hold for directed graphs, where each directed edge is evaluated independently.
Though this paper mainly focuses on undirected graphs, this backbone discovery framework can be easily generalized to directed graphs.

For a vertex $u \in V$, let $\mathcal{N}(u)$ be the immediate neighbors of $u$, i.e., $\mathcal{N}(u)=\{v | (u \rightarrow v) \in \mathcal{E} \}$.
Let $\mathcal{I}(u)$ be all the incoming edges of vertex $u$, i.e., $\mathcal{I}(u)=\{(w \rightarrow u) | w \in \mathcal{N}(u)\}$ and
let $\mathcal{O}(u)$ be all the outgoing edges of vertex $u$, i.e.,
$\mathcal{O}(u)=\{(u \rightarrow v) | v \in \mathcal{N}(u)\}$.
Note that, in bidirected graphs, $\mathcal{I}(u) = \mathcal{O}(u)$.
A path $P$ from vertex $u$ to vertex $v$ can be defined as a vertex-edge sequence, i.e., $(u_0,e_1,u_1,e_2, \cdots, u_{k-1},e_k,u_k)$, where $u=u_0$, $v=u_k$, and $(u_{i-1} \rightarrow u_i)=e_i$.
When no confusion arises, we use only the edge sequence to describe the path as $(e_1,e_2,\cdots,e_k)$.
In this paper, we only consider simple path such that no vertex appears more than once in a path.
The path length is defined as the number of edges in the path.
The shortest path from vertex $u$ to $v$ is the one with minimal path length which is denoted as $P_{uv}$.


Let $\mathcal{P}$ be a collection of paths in graph $G$ characterizing system-wide information flow.
Without prior knowledge, we consider $\mathcal{P}$ to contain shortest paths for each pair of vertices in graph $G$, i.e, $\mathcal{P}=\{P_{uv}\}$.
In case there is more than one shortest path between a pair of vertices, only one shortest path is added to $\mathcal{P}$.
This is consistent with the earlier assumption that any two pairs of vertices have equal communication frequency.
Alternatively, we may assign each path with an equal weight such that total weight of all shortest paths is one.

\subsection{Two Simple Models}
\label{simpmodels}

In this subsection, we consider two simple statistical models for generating a collection of paths in $\mathcal{P}$.
At a high level, both generative models try to assign each edge a probability (or multiple probabilities) such that the probability of each path can be derived by augmenting its edges' probabilities.

\noindent{\bf Edge Independent Model:}
In the first model, referred to as {\em edge independent model}, each outgoing edge $(u \rightarrow v$) of a vertex $u$ in the graph is assigned a probability $p(u\rightarrow v)$,
and the sum of the probabilities of all outgoing edges should be equal to one, i.e., $\sum_{(u \rightarrow v) \in \mathcal{O}(u)} p(u \rightarrow v)=1$.
Furthermore, we assume any two edges in a path are independent.
Given this, the probability of a path $P=(v_0,e_1,v_1,e_2,v_2, \cdots, v_{k-1},e_k,v_k)$ is $p(P)=$
{\small
\beqnarr
p(e_1) p(e_2) \cdots p(e_k) = p(v_0\!\rightarrow\! v_1) p(v_1\!\rightarrow\! v_2) \cdots p(v_{k-1}\!\rightarrow\! v_k)  \nonumber
\eeqnarr}
Given the collection of paths $\mathcal{P}$, the overall likelihood $L_I(\mathcal{P})$ of these paths being generated from this model is
{\small
\beqnarr
L_I(\mathcal{P})  =  \prod_{P \in \mathcal{P}} p(P)  =  \prod_{e \in \mathcal{E}} p(e)^{N_e}
\label{Likelihood}
\eeqnarr
}
where $N_e$ is the number of paths in $\mathcal{P}$ passing through edge $e$.
Note that if $\mathcal{P}$ includes all shortest paths in $G$, then $N_e$ is often referred to as the edge betweenness~\cite{GirvanNewman02}.
To maximize the overall likelihood $L_I(\mathcal{P})$, using the Lagrange multiplier method, it is easy to derive that the optimal parameters are
{\small
\beqnarr
p(e) = p(u\rightarrow v) = \frac{N_e}{M_u}, \nonumber
\eeqnarr
}
where $M_u$ is the number of paths going through $u$, i.e., reaching vertex $u$ and then continue to one of its neighbors.

Clearly, there are a total of $2|E|=|\mathcal{E}|$ parameters in the edge independence model.
Note that this model directly corresponds to the coding scheme where each edge $(u\rightarrow v)$ is assigned a unique code with length $\frac{1}{p(u \rightarrow v)}$.
Thus, under this scheme, the overall minimal coding length for all paths in $\mathcal{P}$ is simply the negative log-likelihood, $-\log L_I(\mathcal{P})$.

\noindent{\bf Edge Markovian Model:}
The edge independent model is one of the simplest models for describing the path probability.
However, it is also rather unrealistic and results in poor model performance.
A path itself is a correlation between edges, so the edge probability model should consider this.
In the second model, we replace the independent assumption with the Markovian property, i.e., the probability of an edge is determined by its immediately preceding edge in a path.
Given this, the probability of a path $P=(e_1,e_2,\cdots,e_k)$ is rewritten as
{\small
\beqnarr
p(P) = p(e_1) p(e_2|e_1) \cdots p(e_k|e_{k-1}) \nonumber
\eeqnarr
}
where $p(e_j | e_i)$ is the conditional probability of edge $e_j$ appearing after edge $e_i$ in the path.

Given the path collection $\mathcal{P}$, the likelihood function for generating all these paths can be written as
{\small
\beqnarr
\label{condprob}
L_M(\mathcal{P}) =  \prod_{P \in \mathcal{P}} p(P)
=\prod_{e \in \mathcal{E} } p(e)^{N_e^\prime} \prod_{e, e^\prime \in \mathcal{E}} p(e^\prime|e)^{N_{ee^\prime}} \nonumber
\eeqnarr}
where $N_e^\prime$ is the number of paths in $\mathcal{P}$ starting with edge $e$, and $N_{e e^\prime}$ is the number of paths with consecutive edges ($e,e^\prime)$.
We note that this model directly corresponds to a Markov chain where each edge represents a state and the conditional probability $p(e^\prime|e)$ represents a transition probability from state (edge) $e$ to state (edge) $e^\prime$.
However, though this model is more accurate at capturing the paths in graph $G$, it is also much more expensive in terms of number of parameters.
It requires $\sum_{v \in V} |\mathcal{I}(v)|\times |\mathcal{O}(v)|=\sum_{v \in V} |\mathcal{N} (v)|^2$ parameters.

\begin{figure}
    \centering
    \mbox{
        \subfigure[Backbone vertex $u$]{\includegraphics[width=1.4in,height=1.0in]{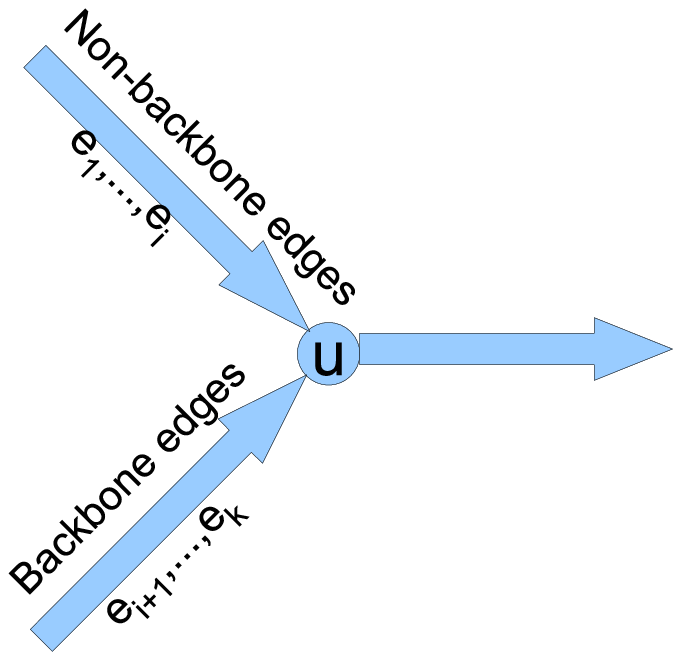}\label{localcode}}
    }
    \mbox{
        \subfigure[Probability table]{\includegraphics[width=1.6in,height=1.0in]{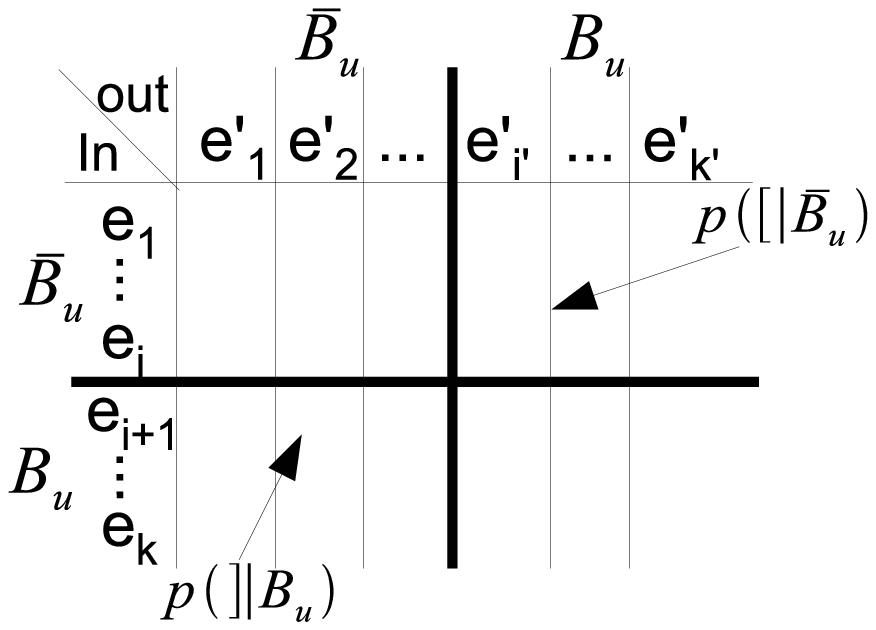}\label{probtab}}
    }

    \caption{Example of backbone vertex $u$}
    \label{fig:bicluster}

\end{figure}


\subsection{Bimodal Markovian Model}
\label{backbonemodel}

We propose a new model with reduced number of parameters to improve model performance.
This model is motivated by the observation on the usage of highway in the transportation system:
compared to complex local traffics, less information is needed to describe highway traffic.
Mapping back to modeling a path using backbone structure,
this suggests that its subpaths only consisting of edges from backbone can be described in a relatively coarse manner comparing to subpaths containing non-backbone edges.
Complying with this intuition's guidance,
we introduce Bimodal Markovian Model utilizing {\em backbone} structure to reduce the number of parameters in edge markovian model while minimizing the loss of its modeling accuracy.
It is termed {\em bimodal} because all vertices and edges would be categorized into two groups in the model.

Figure~\ref{fig:bicluster} illustrates the basic idea of this model.
In figure \ref{localcode}, for each vertex $u$, all its incoming edges are divided into two categories, the {\em backbone edges} and the {\em non-backbone edges}.
For each outgoing edge $e=(u \rightarrow v)$ from backbone vertex $u$,
it only has two probabilities, conditional on the categories of incoming edges,
denoted as $p(e|\mathcal{B} \cap \mathcal{I}(u) )$ and $p(e|\overline{\mathcal{B}} \cap \mathcal{I}(u) )$,
where $\mathcal{B}$ is the set of all backbone edges and $\overline{\mathcal{B}}$ is the set of all non-backbone edges ($\mathcal{B} \cup \overline{\mathcal{B}}=\mathcal{E}$).
In other words, multiple probabilities of edge $e$ conditional on different incoming edges are reduced to only two probabilities.
From clustering viewpoint, this can be considered as performing biclustering on each vertex's incoming edges.

It is worthwhile to note that only the conditional probability of directed edge starting from backbone vertex in edge markovian model would be affected by this model.
In other words, we still keep the conditional probability expressed in edge markovian model for edges starting with non-backbone vertex.
For instance, consider the shortest path $P=(14,1,6,5,17,18)$ highlighted in Figure~\ref{f1}.
The probability of this path can be expressed by bimodal markovian model as $p(P) =$
{\small
\beqnarr
p(14 \rightarrow 1) p(1\rightarrow 6|\overline{\mathcal{B}})  p(6\rightarrow 5|{\mathcal{B}}) p(5\rightarrow 17|{\mathcal{B}}) p(17 \rightarrow 18| 5 \rightarrow 17)
\nonumber
\eeqnarr
}
Note the conditioning on the first edge entering the backbone and the first edge after leaving the backbone:
even though $(1 \rightarrow 6)$ is a backbone edge, the probability we must use is $p(1\rightarrow 6|\overline{\mathcal{B}})$
and though $(5 \rightarrow 17)$ is a non-backbone edge, its probability is $p(5\rightarrow 17|{\mathcal{B}})$.
For consecutive edges $(5 \rightarrow 17 \rightarrow 18)$ without involving backbone vertices, the conditional probabilities of $(17 \rightarrow 18)$ follows what we used in edge markovian model.

Given this, the overall likelihood for a given collection of paths $\mathcal{P}$ with respect to the backbone subgraph $B=(V_B,E_B)$ is described as $L_B(\mathcal{P})=\prod_{P \in \mathcal{P}} p(P)$
{\small
\beqnarr
=\prod_{e \in \mathcal{E}} p(e)^{N^\prime_e} \prod_{u \notin V_B \wedge e^\prime \in \mathcal{O}(u) } p(e^\prime|e)^{N_{ee^\prime}^\prime} \nonumber \\
\prod_{u\in V_B \wedge e \in \mathcal{O}(u) } p(e| \mathcal{B})^{N_{\mathcal{B}e}}  \prod_{ u \in V_B \wedge e \in \mathcal{O}(u)} p(e | \overline{\mathcal{B}})^{N_{\overline{\mathcal{B}}e}}
\label{Likelihood3}
\eeqnarr
}
where $N_e^\prime$ is the number of paths in $\mathcal{P}$ starting from edge $e$,
$N_{e e^\prime}^\prime$ is the number of paths with consecutive edges $(e,e^\prime)$ while intermediate vertex connecting two edges are not backbone vertex,
$N_{\mathcal{B}e}$ and $N_{\overline{\mathcal{B}}e}$ denote the number of paths passing through $e$ when its starting vertex is backbone vertex or non-backbone vertex, respectively.
In connection with edge markovian model, we observe that for \textcolor{black}{backbone} vertex $u$ and its outgoing edge $e=(u \rightarrow v)$,
$N_{\mathcal{B}e} = \sum_{e^\prime \in \mathcal{B}\cap\mathcal{I}(u)}  N_{e^\prime e}$ and
$N_{\overline{\mathcal{B}}e} = \sum_{e^\prime \in \overline{\mathcal{B}}\cap\mathcal{I}(u)} N_{e^\prime e}$,
where $N_{e^\prime e}$ is the number of paths in $\mathcal{P}$ containing the consecutive edges $(e^\prime e)$.


In our framework, the overall number of parameters in bimodal markovian model is $\sum_{v \notin V_B} |\mathcal{I}(v) | \times |\mathcal{O}(v)| + \sum_{v \in V_B} 2|\mathcal{O}(v)| = \sum_{v \notin V_B} |\mathcal{N}(v)|^2 + \sum_{v \in V_B} 2|\mathcal{O}(v)|$.
Compared to edge markovian model, the saving regarding the number of parameters is \\
$\sum_{v \in V_B} (|\mathcal{I}(v)|-2) \times |\mathcal{O}(v)|$.



Given this, we formally define optimal backbone discovery problem based on bimodal markovian model:
\bdefin{\bf (Optimal $K$-Backbone Discovery Problem)}
\label{mdlprob}
Given a complex network $G=(V,E)$, the targeted path set $\mathcal{P}$ and the number of backbone vertices $K$,
the network backbone $B=(V_B,E_B)$ is a connected subgraph with $|V_B|=K$ such that $L_B(\mathcal{P})$ in Formula~\ref{Likelihood3} is maximized.
\edefin

Note that in this definition, we allow the user to define the number of vertices in the backbone structure.
Alternatively, as a model selection problem, we may use a parameter penalty to help determine the optimal backbone size.
For instance, if we use the Akaike information criterion (AIC), then, we simply want to optimize $-\log L(\mathcal{P}|B) + (|\mathcal{B}|+\sum_{v \in V_B} |N(v)|)$.
If we use the Bayesian information criterion (BIC), then, our goal is to optimize
$-2\log L(\mathcal{P}|B)+(|\mathcal{B}|+\sum_{v \in V_B} |N(v)|) \log |\mathcal{P}|$, where $|\mathcal{P}|$ is considered to be the sample size.
In this paper, our decision to study the optimal backbone model problem for any given number of vertices is based on several considerations.
First, though this model treats backbone discovery as a model selection problem, in many applications, the construction of backbones might involve other costs.
Thus, it is more convenient and flexible to set up an adjustable number of vertices.
Second, the solution of this problem forms the basis for solving the AIC or BIC criterion as we can utilize the solution with respect to different $K$ and then choose the overall optimal backbones.
Finally, it is important and useful to observe how the backbone grows (shrinks) when its size increases (decreases).
Therefore, in the reminder of the paper, we focus on studying the Optimal K-Backbone Discovery Problem.
We will empirically investigate the relationship between backbone size and the likelihood of bimodal markovian model in subsection~\ref{bmperform}.


\noindent{\bf Relationship to Path Coding Length:}
Before we work towards a solution for this problem, let us first confirm its relationship to the problem of optimizing path-recognition complexity (section \ref{intro}). 
Given a backbone structure $G_B$, it is not hard to see that $-\log L_B(\mathcal{P})$ serves as the corresponding coding length of $\mathcal{P}$ in network $G$.
When $L_B(\mathcal{P})$ is maximized, the optimal coding can be derived to describe network $G$.

In addition, recall that in the coding scheme, we have two special codes, representing entering the backbone (``['') and exiting the backbone (``]'').
To show how their code length is represented in the probabilistic model, let us take a look at the transition probability from a non-backbone edge to a backbone edge $p(u\rightarrow v|\overline{\mathcal{B}}\cap \mathcal{I}(u)), (u\rightarrow v) \in \mathcal{B}$, and the transition probability from a backbone edge to a non-backbone edge $p(w \rightarrow z|B \cap \mathcal{I}(u)), (w \rightarrow z) \in \overline{\mathcal{B}}$:
{\small
\beqnarr
p(u\rightarrow v| \overline{\mathcal{B}})= p^\prime(u\rightarrow v| \overline{\mathcal{B}})  \times p([|\overline{\mathcal{B}}_u), \mbox{ where }
p^\prime(u\rightarrow v| \overline{\mathcal{B}}) \nonumber \\
=\frac{p(u \rightarrow v|\overline{\mathcal{B}})}
{p([|\overline{\mathcal{B}}_u)},
p([|\overline{\mathcal{B}}_u)=\sum_{(u,v^\prime) \in \mathcal{B} \cap \mathcal{O}(u)} \ p(u \rightarrow v^\prime|\overline{\mathcal{B}}) \nonumber  \\
p(w\rightarrow z| \mathcal{B})=p^\prime(w\rightarrow z|\mathcal{B})  \times p(]|\mathcal{B}_w), \mbox{ where }
p^\prime(w\rightarrow z|\mathcal{B})\nonumber \\
=\frac{p(w \rightarrow z| \mathcal{B})}
{p(]|\mathcal{B}_w)},
p(]|\mathcal{B}_w)=\sum_{(w,z^\prime) \in \overline{B} \cap \mathcal{O}(w)} p(w \rightarrow z^\prime|\mathcal{B}) \nonumber
\eeqnarr
}
Here $\overline{\mathcal{B}}_u=\overline{\mathcal{B}}\cap \mathcal{I}(u)$ and $\mathcal{B}_w=\mathcal{B} \cap \mathcal{I}(w)$ correspond to the set of incoming non-backbone edges to $u$ and incoming backbone edges to $w$.
Table~\ref{fig:bicluster}(b) illustrates $p([|\overline{\mathcal{B}}_u)$ and $p(]|\mathcal{B}_u)$.
In other words, the code length ``$[ e_i $'' ($e_i =(u \rightarrow v) \in \mathcal{B}$) corresponds to the transition probability from a non-backbone edge to a backbone edge: $p([|\overline{\mathcal{B}}_u) p^\prime(e_i|\overline{\mathcal{B}})= p(e_i| \overline{\mathcal{B}})$. Similarly, the code length of ``$ e_j ]$'' ($e_j=(w \rightarrow z) \in \overline{\mathcal{B}}$) corresponds to the transition probability from a backbone edge to a non-backbone edge:
$p^\prime(e_j|\mathcal{B})  \times p(]|\mathcal{B}_w)=p(e_j| \mathcal{B})$.

\comment{
Alternatively, as a model selection problem, we may use a parameter penalty to help determine the optimal backbone size.
For instance, if we use the Akaike information criterion (AIC), then we simply want to optimize $-\log L_B(\mathcal{P}) + (|\mathcal{B}|+\sum_{v \in V_B} |\mathcal{N}(v)|)$.
However, through our experiments, we notice that $(|\mathcal{B}|+\sum_{v \in V_B} |\mathcal{N}(v)|)$ cannot serve as effective penalty in model selection
as the value of negative log-likelihood is always significantly greater than parameter penalty.
Since model selection formulation is not applicable to our scenario, we will discuss the backbone size determination problem through empirical studies.
}

\comment{
Since only vertices and edges appearing in $\mathcal{P}$ is used in Formula~\ref{Likelihood3},
the network backbone is essentially a subgraph of traffic network (Definition~\ref{trafficnetwork}).
As a model selection problem, we may use a parameter penalty to help determine the optimal backbone size.
For instance, if we use the Akaike information criterion (AIC), then our backbone discovery problem can be expressed as:
\bdefin{\bf (Optimal Backbone Discovery Problem)}
\label{mdlprobaic}
Given a complex network $G=(V,E)$ and the targeted path set $\mathcal{P}$,
the network backbone $B=(V_B,E_B)$ is a connected subgraph such that $-\log L_B(\mathcal{P}) + T$ is minimized, where $T$ is the number of parameters in bimodal markovian model.
\edefin
Of course, other measurements such as Bayesian information criterion (BIC) can be used to substitute AIC.
If we use the Bayesian information criterion (BIC), then, our goal is to optimize
$-2\log L_B(\mathcal{P})+T \log |\mathcal{P}|$, where $|\mathcal{P}|$ is considered to be the sample size.
In this paper, our decision to study the optimal backbone model problem for any given number of vertices is based on several considerations.
First, though this model treats backbone discovery as a model selection problem, in many applications, the construction of backbones might involve other costs.
Thus, it is more convenient and flexible to set up an adjustable number of vertices.
Second, the solution of this problem forms the basis for solving the AIC or BIC criterion as we can utilize the solution with respect to different $K$ and then choose the overall optimal backbones.
Finally, it is important and useful to observe how the backbone grows (shrinks) when its size increases (decreases).
Therefore, in the reminder of the paper, we focus on studying the Optimal K-Backbone Discovery Problem.
}

\comment{
\noindent{\bf Optimal Backbone Model:}
Note that in our framework, the number of parameters is
$ |\mathcal{E}|+\sum_{v \in V_B} |N(v)|$, where $V_B$ is the vertex set for backbone subgraph.
Since the backbone subgraph contains only a small fraction of the graph, the overall number of parameters does not increase much compared with the edge independent model and is much smaller than in the edge Markovian model.

Given this, we reformulate the backbone discovery problem as an optimal backbone model problem:

\bdefin{\bf (Optimal K-Backbone Model Problem)}
\label{mdlprob}
Given a complex network $G=(V,E)$, the targeted path set $\mathcal{P}$, and the number of backbone vertices $K$, the network backbone $B=(V_B,E_B)$ is a connected subgraph with $|V_B|=K$ such that $L(\mathcal{P}|B)$ in Eq. \ref{Likelihood3} is maximized.
\edefin

Note that in this definition, we allow the user to define the number of vertices in the backbone structure.
Alternatively, as a model selection problem, we may use a parameter penalty to help determine the optimal backbone size.
For instance, if we use the Akaike information criterion (AIC), then, we simply want to optimize $-\log L(\mathcal{P}|B) + (|\mathcal{B}|+\sum_{v \in V_B} |N(v)|)$.
If we use the Bayesian information criterion (BIC), then, our goal is to optimize
$-2\log L(\mathcal{P}|B)+(|\mathcal{B}|+\sum_{v \in V_B} |N(v)|) \log |\mathcal{P}|$, where $|\mathcal{P}|$ is considered to be the sample size.
In this paper, our decision to study the optimal backbone model problem for any given number of vertices is based on several considerations.
First, though this model treats backbone discovery as a model selection problem, in many applications, the construction of backbones might involve other costs.
Thus, it is more convenient and flexible to set up an adjustable number of vertices.
Second, the solution of this problem forms the basis for solving the AIC or BIC criterion as we can utilize the solution with respect to different $K$ and then choose the overall optimal backbones.
Finally, it is important and useful to observe how the backbone grows (shrinks) when its size increases (decreases).
Therefore, in the reminder of the paper, we focus on studying the Optimal K-Backbone Discovery Problem.
}

\section{Backbone Discovery based on
Vertex Betweenness}
\label{backbonevb}

In this section, we introduce a straightforward approach based on vertex betweenness and minimal steiner tree to discover backbone.
This approach also serves as the basic benchmark for backbone discovery.
Recall that, network backbone is a connected subgraph carrying the major traffic formed by shortest paths.
In the meanwhile, vertex betweenness has been widely used to evaluate the importance of a vertex by the number of shortest paths passing through it.
Given this, the straightforward solution for optimal $K$-backbone discovery problem is to consider $K$ vertices with highest betweenness as backbone vertices,
since they tend to captures more information flow following shortest paths.
Ideally, if these vertices are connected in the graph, its corresponding induced subgraph naturally forms the backbone,
where the edges included in the induced subgraph are considered as backbone edges.

However, these $K$ vertices are not necessarily connected to each other.
To obtain backbone structure, we want to build the connections among them while introducing minimal number of extra vertices.
Since we focus on unweighted graph in this paper, this problem is essentially an instance of minimal steiner tree problem.
The minimal steiner tree problem has been proved to be NP-hard, but an approximation algorithm has been introduced in~\cite{wu2004spanning}.
Applying this method, we are able to gain a set of connected vertices as the superset of backbone vertices.
The corresponding induced graph is treated as candidate backbone structure.
To discover backbone with exactly $K$ vertices, we can utilize a refinement strategy to remove extra backbone vertices iteratively.
Basically, in each iteration, we remove the vertex with smallest vertex betweenness from current graph.
If remaining graph is not connected, we consider the removal of vertex with second smallest betweenness.

\begin{algorithm}
\caption{BackboneDiscovery\_VB($G=(V,\mathcal{E})$,$K$)}
\label{alg:backboneben}
{\small
\begin{algorithmic}[1]
\REQUIRE{ $G$ is input network, $K$ is the backbone size}
\STATE Compute vertex betweenness for each vertex using method in~\cite{Brandes01afaster};
\STATE Select vertex set $V_s$ including $K$ vertices with largest vertex betweenness;
\STATE Construct minimal steiner tree $T=(V_T,E_T)$ on vertex set $V_s$ (i.e., $V_s \subseteq V_T$) by approximation algorithm~\cite{wu2004spanning};
\STATE $G_B$ is induced subgraph of $G$ on vertex set $V_T$;
\STATE $Q \leftarrow V_T$; \COMMENT{$Q$ is a queue which stores vertices in ascending order of their corresponding vertex betweenness}
\WHILE{$|V_T| > K$}
    \STATE $u \leftarrow Q.pop\_front()$;
    \STATE $G_B^\prime$ is the induced subgraph of $G_B$ on vertex set $V_T \setminus \{u\}$;
    \IF{$G_B^\prime$ is connected graph}
        \STATE $V_T \leftarrow V_T \setminus \{u\}$;
        \STATE $G_B \leftarrow G_B^\prime$;
    \ELSE
        \STATE $Q.push\_back(u)$;
    \ENDIF
\ENDWHILE
\RETURN $G_B$;
\end{algorithmic}
}
\end{algorithm}

We sketch this approach in Algorithm~\ref{alg:backboneben}.
The algorithm mainly consists of two steps: candidate backbone discovery step (Line 1 to Line 4) and refinement step (Line 5 to Line 15).
To begin with, the betweenness of each vertex is computed by fast approach in~\cite{Brandes01afaster} (Line 1).
We then select top $K$ vertices with largest betweenness and construct the minimal steiner tree $T=(V_T,E_T)$ over this vertex set (Line 2 to Line 3).
Now, we have a set of connected vertices $V_T$ and its corresponding induced subgraph $G_B$ serves as the candidate backbone (Line 4 to Line 5).
In the refinement step, we firstly store $V_T$ into a queue according to their corresponding betweenness in ascending order (Line 5).
During main loop, in each iteration, we consider the first vertex $u$ in the queue (Line 7).
If the removal of vertex $u$ and its incident edges will disconnect current graph, we push $u$ back into queue for further consideration (Line 13).
Otherwise, the remaining graph $G^\prime_B$ is used as current graph $G_B$ in next iteration.
The refinement procedure proceeds until only $K$ vertices remains in the graph.
This resulting graph is connected and returned as backbone.

\noindent{\bf Computational Complexity:}
In the candidate backbone discovery step, we take $O(|V||E|)$ time to compute vertex betweenness and $O(K^2 |V| )$ time to build steiner tree.
For refinement step, since we can remove at most one vertex from $V_T$ and connectivity checking on $G_B$ takes at most $O(|E_T|)$ in each iteration,
total running time is $O((|V_T|-K)|E_T|)$ in the worst case.
Putting both together, the overall computational complexity is $O(|V||E|+K^2 |V|)$.

Instead of optimizing likelihood function, this straightforward approach only employs vertex betweenness to approximate the contribution made by each vertex to $L_B(\mathcal{P})$.
In addition, this approach neglects the effect of edges on likelihood function by directly using the edge set of induced graph as backbone edges.
Therefore, the discovered backbone may not necessarily maximize likelihood function $L_B(\mathcal{P})$.
In the next section, we propose novel approaches to discover backbone by directly considering the optimality of likelihood function.

\section{Optimizing Bimodal Markovian Model}
\label{basics}

In order to solve the optimal $K$-backbone discovery problem, two key questions should be considered:
1) Can we identify a set of $K$ connected vertices as candidate backbone vertices based on their contribution to likelihood of bimodal markovian model?
2) For a given set of $K$ vertices, how to extract an optimal backbone such that $L_B(\mathcal{P})$ is maximized?
In other words, the edges in the induced subgraph of these $K$ vertices need to be classified into either backbone or non-backbone edges.
It turns out the second question is more fundamental as it can directly contribute to the solution of the first one.
Simply speaking, the solution of the second question offers an effective way to estimate individual vertex's contribution to the objective function $L_B(\mathcal{P})$ and thus is  very helpful in selecting the backbone vertices.

\noindent{\bf Backbone Edge Selection of $K$ Vertex Set:}
In the following, we introduce a log-likelihood representation of the objective function $L_B(\mathcal{P})$ to simplify our problem in discovering the optimal backbone given a set of $K$ vertices.
Formally, let $V_B$ be a subset of connected vertices in $G$ and $G[V_B]$ is corresponding induced graph of $G$.
The backbone graphs $G_B=(V_B,\mathcal{B}) \subseteq G[V_B]$, i.e., the edges in the backbone can only come from the edges in the induced subgraph by vertex set $V_s$.
Given path set $\mathcal{P}$, we want to extract edge set $E_B$ from $G[V_B]$ achieving maximum $L_B(\mathcal{P})$.
This problem turns out to be rather challenging on undirected graph due to large search space and
the edge consistent constraint that the categorization of both directed edges $(u \rightarrow v)$ and $(v \rightarrow u)$ of an undirected edge $(u,v)$ should be the same.
In fact, it is even non-trivial to categorize individual backbone vertex's incoming edges into backbone and non-backbone without consistent constraint.

To facilitate maximizing $L_B(\mathcal{P})$ (Formula~\ref{Likelihood3}),
we compare it with the likelihood of edge markovian model $L_M(\mathcal{P})$ (Formula~\ref{condprob}).
First of all, we rewrite likelihood of edge markovian model as $L_M(\mathcal{P})=$
{\small
\beqnarr
\prod_{e \in \mathcal{E}} p(e)^{N^\prime_e} \prod_{u \notin V_B \wedge e^\prime \in \mathcal{O}(u)} p(e^\prime|e)^{N_{ee^\prime}^\prime} 
\prod_{u \in V_B \wedge e^\prime \in \mathcal{O}(u)} p(e^\prime|e)^{N_{ee^\prime}^\prime} \nonumber
\eeqnarr
}
Now, we introduce the likelihood ratio (first two terms are canceled out for simplification) $LR(V_B)=\frac{L_B(\mathcal{P})}{L_M(\mathcal{P})}=$
{\small
\beqnarr
\label{lrvb}
\frac{\prod_{ u \in V_B \wedge  e \in \mathcal{O}(u)} p(e| \mathcal{B})^{N_{\mathcal{B}e}}  \prod_{ u \in V_B \wedge e \in \mathcal{O}(u)} p(e | \overline{\mathcal{B}})^{N_{\overline{\mathcal{B}}e}}}
{\prod_{u \in V_B \wedge e^\prime \in \mathcal{O}(u)} p(e^\prime|e)^{N_{ee^\prime}^\prime}}  \nonumber \\
= \prod_{u \in V_B} \frac{ \prod_{e \in \mathcal{O}(u)} p(e| \mathcal{B})^{N_{\mathcal{B}e}}
\prod_{e \in \mathcal{O}(u)}
p(e | \overline{\mathcal{B}})^{N_{\overline{\mathcal{B}}e}}}
{\prod_{e^\prime \in \mathcal{I}(u), e \in \mathcal{O}(u)} p(e|e^\prime)^{N_{e^\prime e}}}
\eeqnarr
}

Given graph $G$ and path set $\mathcal{P}$, $L_M(\mathcal{P})$ is a constant, assuming each of its parameters $p( e|e^\prime)$ is optimized for the maximal likelihood.
Therefore, maximizing the likelihood ratio $L_B(\mathcal{P})/L_M(\mathcal{P})$ is equivalent to maximize $L_B(\mathcal{P})$.
The following definition formalizes our problem:
\bdefin({\bf Optimizing Vertex Set Likelihood Ratio Problem})
\label{ovslrp}
Given graph $G=(V,E)$ and path set $\mathcal{P}$, for any connected vertex set $V_B$,
we would like to construct backbone subgraph $G_B=(V_B,\mathcal{B})$, where $\mathcal{B} \subseteq V_B \times V_B \cap E$ and $LR(V_B)$ (Formula~\ref{lrvb}) is maximized.
\edefin

\subsection{Clustering Incoming Edges of Individual Vertex}
To approach the problem(Definition ~\ref{ovslrp}), we start with relaxing the consistent constraint.
In other words, for directed edges $(u \rightarrow v)$ and $(v \rightarrow u)$ with opposite direction, we assume that each of them can be determined independently to be a backbone or non-backbone edge.
Then, the following rule is applied to enforce consistent constraint: we say $(u,v)\in E$ is backbone edge iff both $(u\rightarrow v)$ and $(v \rightarrow u)$ are backbone edges.

First of all, we rewrite $LR(V_B) = \sum_{u \in V_B} LR(u)$ where

\beqnarr
{\small
LR(u)=
\frac{ \prod_{e \in \mathcal{O}(u)} p(e| \mathcal{B})^{N_{\mathcal{B}e}}
\prod_{e \in \mathcal{O}(u)}
p(e | \overline{\mathcal{B}})^{N_{\overline{\mathcal{B}}e}}}
{\prod_{e^\prime \in \mathcal{I}(u), e \in \mathcal{O}(u)} p(e|e^\prime)^{N_{e^\prime e}}} \nonumber
}
\eeqnarr

Based on aforementioned relaxation, we can see that $LR(u)$ is independent of $LR(v)$ if $u \neq v$, i.e., the optimality of $LR(u)$ will not affect the optimality of $LR(v)$.
Therefore, optimizing each $LR(u)$ corresponding to vertex $u \in V_B$ individually is able to result in global maximization of $LR(V_B)$.
Given this, our problem is converted to categorizing each vertex $u$'s incoming edges as backbone edges or non-backbone edges in order to optimize $LR(u)$.
From the viewpoint of clustering, this essentially group each vertex's incoming edges into only two clusters (backbone or non-backbone).

Before proceeding to our solution, we first study how to compute optimal $L_B(u)$ assuming backbone edges are given.
Essentially, we want to figure out the optimal $p(e|\mathcal{B})$ and $p(e|\overline{\mathcal{B}})$ leading to maximal $L_B(u)$.
It is not hard to derive following result:
\blemma
For a backbone vertex $u$, assuming each incoming edge has been categorized as backbone or non-backbone, i.e., $\mathcal{B}$ and $\overline{\mathcal{B}}$, then the minimum of $-\log LR(u)$ is achieved when
{\small
\beqnarr
\label{optcenters}
& p(e|\mathcal{B}) = \frac{N_{\mathcal{B}e}}{\sum_{e^\prime \in \mathcal{O}(u)}N_{\mathcal{B}e^\prime}} \mbox{ and }
p(e | \overline{\mathcal{B}}) = \frac{N_{\overline{\mathcal{B}}e}}{\sum_{e^\prime \in \mathcal{O}(u)} N_{\overline{\mathcal{B}}e^\prime}}
\eeqnarr
}
\elemma

\bproof
The minimum of $- \log LR(u)$ is essentially finding the maximum value of likelihood function subject to certain probabilistic constraints.
Using Lagrange multiplier method with two probabilistic constraints $\sum_{e \in \mathcal{O}(u)} p(e| \mathcal{B}) = 1$ and \\
$\sum_{e \in \mathcal{O}(u)} p(e| \overline{\mathcal{B}} ) = 1$,
we are able to obtain the optimal values of $p(e| \mathcal{B})$ and $p(e| \overline{\mathcal{B}})$.
\eproof

\noindent{\bf Algorithm for Edge Clustering:}
We propose an iterative refinement algorithm to resolve this edge clustering problem on each vertex $u$.
Initially, each incoming edge $e$ is randomly assigned to be backbone edge or non-backbone edge.
Given such assignment, optimal value of $LR(u)$ can be computed based on corresponding optimal $p(e|\mathcal{B})$ and $p(e|\overline{\mathcal{B}})$ (Formula~\ref{optcenters}).
In the subsequent iterations, we iteratively refine each edge's cluster membership in order to achieve a better value of $LR(u)$.
The iterations terminate until no further improvement can be obtained.
Interestingly, we will show that this method is essentially a K-Means under {\em Kullback-Leibler} divergence measure (K=2).

To further explain the algorithm, we express $LR(u)$ in negative log-likelihood format as follows (for simplicity, $\mathcal{I}$ and $\mathcal{O}$ are used to replace $\mathcal{I}(u)$ and $\mathcal{O}(u)$):
$-\log LR(u) = $
{\small
\beqnarr
\label{funcF}
    \sum_{e \in \mathcal{O}}
( \sum_{e^\prime \in \mathcal{B} \cap \mathcal{I}} N_{e^\prime e} \log \frac{p(e|e^\prime)}{p(e|\mathcal{B})} + \sum_{e^\prime \in \overline{\mathcal{B}} \cap \mathcal{I}} N_{e^\prime e} \log \frac{p(e|e^\prime)}{p(e|\overline{\mathcal{B}})} ) \nonumber \nonumber \\
  = \sum_{e^\prime \in \mathcal{B} \cap \mathcal{I}} M_{e^\prime} \sum_{e \in \mathcal{O}} p(e|e^\prime) \log \frac{p(e|e^\prime)}{p(e|\mathcal{B})} + \nonumber \\
    \sum_{e^\prime \in \overline{\mathcal{B}} \cap \mathcal{I}} M_{e^\prime}
\sum_{e \in \mathcal{O}} p(e |e^\prime) \log \frac{p(e|e^\prime)}{p(e|\overline{\mathcal{B}})} \nonumber
\eeqnarr
}
where $M_{e^\prime}=\sum_{e \in \mathcal{O}(u)} N_{e^\prime e}$ is the total number of paths passing through edge $e^\prime$ and then continue to one of its neighbors.

Indeed, $\sum_{e \in \mathcal{O}} p(e|e^\prime) \log \frac{p(e|e^\prime)}{p(e|\mathcal{B})}$ simply corresponds to the well-known {\em Kullback-Leibler} divergence between two distributions  \\
$(p(e_1|e^\prime), \cdots, p(e_k|e^\prime))$ and
$(p(e_1|\mathcal{B}), \cdots, p(e_k|\mathcal{B}))$, where \\
$e_1, \cdots, e_k \in \mathcal{O}(u)$ and $k=|\mathcal{O}(u)|$.
Here, each incoming edge $e^\prime \in \mathcal{I}(u)$ corresponds to a point with $k$ features ($p(e|e^\prime), e \in \mathcal{O}(u)$) to be clustered.
In addition, $p(e|\mathcal{B})$ and $p(e|\overline{\mathcal{B}})$ are interpreted
as ``centers'' for the two clusters, the backbone clusters $\mathcal{B} \cap \mathcal{I}(u)$ and non-backbone clusters $\overline{\mathcal{B}} \cap \mathcal{I}(u)$.
In this sense, the objective function $-\log LR(u)$ actually serves as the within-cluster distance.
Now, we can utilize the K-Means type clustering to categorize incoming edges into backbone edges or non-backbone edges.
In each refinement iteration, each incoming edges is assigned to the cluster who results in smallest KL-divergence.
The procedure is outlined in Algorithm~\ref{alg:kmeans}.
Clearly, this algorithm will converge to a local minimum similar to classical K-Means algorithm.

\comment{
Based on Formula~\ref{lrvb}, to maximize $L_B(\mathcal{P})$, we can maximize each vertex $u \in V_B$ independently.
This relaxation produces an upper bound on the maximal contribution of each vertex to the overall likelihood.
In the following, we will show that the problem of categorizing each vertex's incoming edges corresponds to an interesting biclustering
under {\em Kullback-Leibler} divergence measure.

\bdefin({\bf Optimal Vertex Likelihood Ratio Problem})
Given graph $G$ and path set $\mathcal{P}$, for any backbone vertex $u$, categorize its incoming edges
as either backbone edges or non-backbone edges, such that
{\small
\beqnarr
LR(u)=
\frac{ \prod_{e \in \mathcal{O}(u)} p(e| \mathcal{B})^{N_{\mathcal{B}e}}
\prod_{e \in \mathcal{O}(u)}
p(e | \overline{\mathcal{B}})^{N_{\overline{\mathcal{B}}e}}}
{\prod_{e^\prime \in \mathcal{I}(u), e \in \mathcal{O}(u)} p(e|e^\prime)^{N_{e^\prime e}}}
\eeqnarr
}
is maximized.
\edefin
Here $LR(V_B)=\sum_{u \in V_B} LR(u)$. For edge $e=(u \rightarrow v)$, $p(e|e^\prime)=\frac{N_{e^\prime e}}{\sum_{e_1 \in \mathcal{O}(u)} N_{e^\prime e_1}}$.

To maximize $LR(u)$, it is more convenient to work on its negative logarithmic format.
When the meaning is clear, we may use the shorthand $\mathcal{I}$ and  $\mathcal{O}$ for \textcolor{black}{$\mathcal{I}(u)$ and  $\mathcal{O}(u)$}, respectively:
$-\log LR(u) = $
{ \small
\beqnarr
\label{funcF}
(\sum_{e^\prime \in \mathcal{B} \cap \mathcal{I}}
\sum_{e \in \mathcal{O}} N_{e^\prime e} \log p(e|e^\prime) -
\sum_{e \in \mathcal{O}} N_{\mathcal{B}e} \log p(e | \mathcal{B})) \nonumber \\
  + (\sum_{e^\prime \in \overline{\mathcal{B}} \cap \mathcal{I}}
\sum_{e \in \mathcal{O}} N_{e^\prime e} \log p(e|e^\prime) -
\sum_{e \in \mathcal{O}} N_{\overline{B}e} \log p(e| \overline{\mathcal{B}})) \nonumber \\
  =  \sum_{e \in \mathcal{O}}
( \sum_{e^\prime \in \mathcal{B} \cap \mathcal{I}} N_{e^\prime e} \log \frac{p(e|e^\prime)}{p(e|\mathcal{B})} + \sum_{e^\prime \in \overline{\mathcal{B}} \cap \mathcal{I}} N_{e^\prime e} \log \frac{p(e|e^\prime)}{p(e|\overline{\mathcal{B}})} ) \nonumber \nonumber \\
  = \sum_{e^\prime \in \mathcal{B} \cap \mathcal{I}} M_{e^\prime} \sum_{e \in \mathcal{O}} p(e|e^\prime) \log \frac{p(e|e^\prime)}{p(e|\mathcal{B})} + \nonumber \\
    \sum_{e^\prime \in \overline{\mathcal{B}} \cap \mathcal{I}} M_{e^\prime}
\sum_{e \in \mathcal{O}} p(e |e^\prime) \log \frac{p(e|e^\prime)}{p(e|\overline{\mathcal{B}})} \nonumber
\eeqnarr
}
where $M_{e^\prime}=\sum_{e \in \mathcal{O}(u)} N_{e^\prime e}$ is the total number of paths passing through edge $e^\prime$ and then continue to one of its neighbors.


It is not hard to derive the following results:
\blemma
For a \textcolor{black}{backbone} vertex $u$, supposing each incoming edge has been categorized as backbone or non-backbone, i.e., $\mathcal{B}$ and $\overline{\mathcal{B}}$, then the minimum of $-\log LR(u)$ is achieved when
{\small
\beqnarr
p(e|\mathcal{B}) = \frac{N_{\mathcal{B}e}}{\sum_{e^\prime \in \mathcal{O}(u)}N_{\mathcal{B}e^\prime}} \mbox{ and }
p(e | \overline{\mathcal{B}}) = \frac{N_{\overline{\mathcal{B}}e}}{\sum_{e^\prime \in \mathcal{O}(u)} N_{\overline{\mathcal{B}}e^\prime}} \nonumber
\eeqnarr
}
\elemma
Note that the same choice of $p(e|\mathcal{B})$ and $p(e|\overline{\mathcal{B}})$ also maximizes the overall likelihood $L_B(\mathcal{P})$.
Indeed, $\sum_{e \in \mathcal{O}} p(e|e^\prime) \log \frac{p(e|e^\prime)}{p(e|\mathcal{B})}$ simply corresponds to the well-known {\em Kullback-Leibler} divergence between two distributions $(p(e_1|e^\prime), \cdots, p(e_k|e^\prime))$ and
$(p(e_1|\mathcal{B}), \cdots, p(e_k|\mathcal{B}))$, \textcolor{black}{where $e_1, \cdots, e_k \in \mathcal{O}(u)$ and $k=|\mathcal{O}(u)|$.}
Thus, we can utilizes a k-means type iterative algorithm for minimizing $-\log LR(u)$.
Here, we can interpret $p(e|\mathcal{B})$ and $p(e|\overline{\mathcal{B}})$
as ``centers'' for the two clusters, the backbone clusters $\mathcal{B} \cap \mathcal{I}(u)$ and non-backbone clusters $\overline{\mathcal{B}} \cap \mathcal{I}(u)$.
Each incoming edge $e^\prime \in \mathcal{I}(u)$ corresponds to a point with $k$ features \textcolor{black}{($p(e|e^\prime), e \in \mathcal{O}(u)$)} to be clustered.
Further, the objective function $-\log LR(u)$ serves as the within-cluster distance to be minimized.
In addition, we note that each point (incoming edge $e^\prime$) is weighted by $M_{e^\prime}$, the number of paths coming from $e^\prime$.
The outline of this algorithm is illustrated in Algorithm~\ref{alg:kmeans}.
Clearly, this algorithm will converge to a local minimum similar to the classical k-means algorithm.
}

\begin{algorithm}[t!]
{\small
\caption{Bi-KL-Partition(Vertex $u$)}
\label{alg:kmeans}
\begin{algorithmic}[1]
\STATE randomly partition the incoming edges of $u$ into $\mathcal{B}$ and $\overline{\mathcal{B}}$;
\STATE compute $p(e|\mathcal{B})$ and $p(e|\overline{\mathcal{B}})$, $e \in \mathcal{O}(u)$;
\REPEAT
    \STATE assign each incoming edge $e^\prime$ to the cluster with the closest distance: $min(\sum_{e \in \mathcal{O}(u)} p(e|e^\prime) \log
\frac{p(e|e^\prime)}{p(e|\mathcal{B})}$, \\
 $\sum_{e \in \mathcal{O}(u)} p(e|e^\prime) \log
\frac{p(e|e^\prime)}{p(e|\overline{\mathcal{B}})})$;  \\
    \STATE calculate the two new centroids: \\
 $p(e|\mathcal{B}) = \frac{N_{\mathcal{B}e}}{\sum_{e^\prime \in \mathcal{O}(u)}N_{\mathcal{B}e^\prime}} \mbox{ and }
p(e | \overline{\mathcal{B}}) = \frac{N_{\overline{\mathcal{B}}e}}{\sum_{e^\prime \in \mathcal{O}(u)} N_{\overline{\mathcal{B}}e^\prime}}$ \\
\UNTIL{stop criteria is satisfied}
\end{algorithmic}
}
\end{algorithm}

\subsection{Clustering Undirected Edges}
Utilizing above clustering method, we are able to convert the global optimization problem into the local problem of optimizing each vertex independently.
Although the method is rather efficient, for each undirected edge $(u,v)$, both $(u \rightarrow v)$ and $(v \rightarrow u)$ cannot be guaranteed to be in the same category.
Thus, such clustering only generates an upper bound of each individual vertex's benefit brought to bimodal markovian model.
This upper bound will be utilized for selecting candidate backbone vertices in subsection~\ref{mcg}.

\begin{figure}
    \centering
    \mbox{
        \subfigure[Undirected graph] {\epsfig{figure=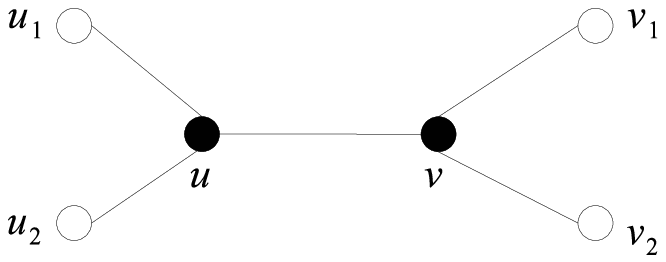,scale=0.6}
        \label{fig:ugvertexset}}
        \subfigure[Bidirected graph]{\epsfig{figure=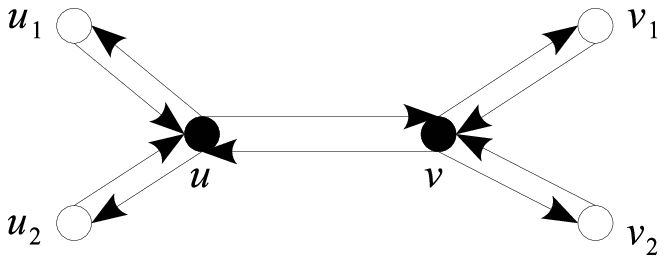,scale=0.6}
        \label{fig:bgvertexset}}
    }
    \label{fig:vertexset}

    \caption{Observations related to backbone vertices }

\end{figure}

To address the edge consistent constraint, we further explore the property of edges incident to $V_B$.
We observe that:
1) any undirected edge $e^\prime=(u, u_1) \in E$ with only one endpoint $u$ in $V_B$ (see Figure~\ref{fig:ugvertexset}) is not a candidate to be a backbone edge;
2) for each undirected edge $e^\prime=(u,v) \in E$ with both endpoints in $V_B$, we need $(u \rightarrow v)$ and $(v \rightarrow u)$ to be both in $\mathcal{B}$ or in $\overline{\mathcal{B}}$ (as shown in Figure~\ref{fig:bgvertexset}).
To distinguish those two types of edges, we decompose the likelihood ratio $LR(V_B)$ into three parts:
{\small
\beqnarr
\label{lr3part}
 -\log LR(V_B) =
 \sum_{e^\prime \in (E \backslash V_B \times V_B)} \sum_{e \in \mathcal{O}(u)} N_{e^\prime_v e} \log
\frac{p(e|e^\prime_v)}{p(e|\overline{\mathcal{B}})}  + \nonumber \\
 \sum_{e^\prime \in V_B \times V_B \cap E } F(e^\prime|\mathcal{B}) +\sum_{e^\prime \in V_B \times V_B \cap E } F(e^\prime|\overline{\mathcal{B}})
\eeqnarr
}
where
{\small
\beqnarr
F(e^\prime|\mathcal{B})=\sum_{e \in \mathcal{O}(u)}  N_{e^\prime_v e} \log \frac{p(e|e^\prime_v)}{p(e|\mathcal{B})} + \sum_{e \in \mathcal{O}(v)} N_{e^\prime_u e}  \log \frac{p(e|e^\prime_u)}{p(e|\mathcal{B})} \nonumber \\
F(e^\prime|\overline{\mathcal{B}})= \sum_{e \in \mathcal{O}(u)} N_{e^\prime_v e} \log
\frac{p(e|e^\prime_v)}{p(e|\overline{\mathcal{B}})} + \sum_{e \in \mathcal{O}(v)} N_{e^\prime_u e}  \log
\frac{p(e|e^\prime_u)}{p(e|\overline{\mathcal{B}})} \nonumber
\eeqnarr
}
Note that, first term of Formula~\ref{lr3part} relates to the probabilities of edges with one endpoint in $V_B$,
$F(e^\prime|\mathcal{B})$ and $F(e^\prime|\overline{\mathcal{B}})$ consider the cases of edge $e^\prime$ being backbone edge and non-backbone edge, respectively.
Given this, we derive the following result:
\blemma
For a connected vertex set $V_B$, supporting all edges of induced graph $G[V_B]$ have been categorized as backbone or non-backbone,
the minimum of $-\log LR(V_B)$ is achieved when
{\small
\beqnarr
p(e|\mathcal{B}) = \frac{N_{\mathcal{B}e}}{\sum_{e^\prime \in \mathcal{O}(u)}N_{\mathcal{B}e^\prime}} \mbox{ and } p(e
| \overline{\mathcal{B}}) = \frac{N_{\overline{\mathcal{B}}e}}{\sum_{e^\prime \in \mathcal{O}(u)}
N_{\overline{\mathcal{B}}e^\prime}} \nonumber
\eeqnarr
}
\elemma

We apply the same technique based on Lagrange multiplier for obtaining Lemma~\ref{optcenters}, to gain the value of $p(e | \mathcal{B})$ and $p(e|\overline{\mathcal{B}}$ for optimizing $-\log LR(V_B)$.
The probabilistic constraint here is the same, i.e., $\sum_{e \in \mathcal{O}(u)} p(e| \mathcal{B}) = 1$ and $\sum_{e \in \mathcal{O}(u)} p(e| \overline{\mathcal{B}} ) = 1$.

The first term of Formula~\ref{lr3part} is relatively stable on different backbone edge set as only $p(e^\prime|\overline{\mathcal{B}})$ is involved.
Therefore, the minimization of $-\log LR(V_B)$ can be approximately achieved by minimizing last two terms of Formula~\ref{lr3part}, serving as within-cluster distance.
Given this, we describe our generalized {\em Bi-KL-Partition} algorithm (Algorithm \ref{alg:gen2means}) to solve it, which again has an interesting convergence property.
The basic idea is to update the cluster membership of each candidate backbone edge $e^\prime=(u,v)\in V_B\times V_B \cap E$
based on $F(e^\prime|\mathcal{B})$ and $F(e^\prime|\overline{\mathcal{B}})$ with the optimal $p(e|\mathcal{B})$ and $p(e|\overline{\mathcal{B}})$
for the current clusters $\mathcal{B}$ and $\overline{\mathcal{B}}$.
In other words, $F(e^\prime|\mathcal{B})$ and $F(e^\prime|\mathcal{B})$ describe a generalized ``distance'' function from a point (edge) $e^\prime$ to corresponding centroids.

\begin{algorithm}[t!]
{\small
\caption{GBi-KL-Partition(Vertex Set $V_s$)}
\label{alg:gen2means}
\begin{algorithmic}[1]
\STATE Assign each edge with only one end in $V_s$ to cluster $\overline{\mathcal{B}}$;
randomly partition edges with both ends in $V_s$ into $\mathcal{B}$ and $\overline{\mathcal{B}}$; \\
\STATE for each vertex $u \in V_s$, compute optimal $p(e|\mathcal{B})$ and $p(e|\overline{\mathcal{B}})$, $e \in \mathcal{O}(u)$; \\
\REPEAT
    \STATE assign each edge $e^\prime$ to the cluster with the closest distance: $min(F(e^\prime|\mathcal{B}),F(e^\prime|\overline{\mathcal{B}}))$;  \\
    \STATE calculate new centroids of two clusters for each vertex $u \in V_s$: \\
 $p(e|\mathcal{B}) = \frac{N_{\mathcal{B}e}}{\sum_{e^\prime \in \mathcal{O}(u)}N_{\mathcal{B}e^\prime}} \mbox{ and }
p(e | \overline{\mathcal{B}}) = \frac{N_{\overline{\mathcal{B}}e}}{\sum_{e^\prime \in \mathcal{O}(u)} N_{\overline{\mathcal{B}}e^\prime}}$ \\
\UNTIL{stop criteria is satisfied}
\end{algorithmic}
}
\end{algorithm}

\blemma{\bf (Convergence Property)} \label{converge}
As we iteratively update membership of each edge $e^\prime \in V_B \times V_B \cap E$ in Algorithm \ref{alg:gen2means},
the function $-\log LR(V_B)$ converges to a local optimum in finite iterations.
\elemma


\bproof
Let us use $F_i^1$ and $F_i^2$ to denote the values of objective function obtained from step 1 (Line 4) and step 2 (Line 5) at $i$-th iteration in Algorithm~\ref{alg:gen2means}, respectively.
Clearly, $F_i^2$ records the value of objective function at the end of iteration $i$.
Assuming the algorithm just finishes iteration $i$, we will show that the value of $-\log LR(V_B)$ in iteration $i+1$ is no greater than the value obtained from iteration $i$.
Considering the step 1 in iteration $i+1$,  for each edge $e^\prime \in V_B \times V_B \cap E$, its within-cluster distance is reduced (i.e., $min (F(e^\prime|\mathcal{B}), F(e^\prime|\overline{\mathcal{B}}))$),
i.e., $ F_i^2 \ge F_{i+1}^1$.
Fixing backbone edge assignment, step 2 attempts to minimize the objective function by updating two clusters' centroids.
In other words, we guarantee that $F_{i+1}^1 \ge F_{i+1}^2$.
Considering both, we have $F_i^2 \ge F_i^1 \ge F_{i+1}^2$.
In this sense, the value of $-\log LR(V_B)$ cannot be increased when the number of iteration increases.
On the other hand, the number of possible edge assignment to be backbone or non-backbone is bounded by $2^{|E_s|}$ where $E_s$ is the edge set of induced graph based on vertex set $V_s$.
This implies that the number of iterations in Algorithm~\ref{alg:gen2means} is at most $2^{|E_s|}$.
Putting both together, the lemma holds.
\eproof

\comment{
\textcolor{blue}{
reverse the writing order:
1) propose the problem first (one paragraph to clearly and briefly describe the goal);
2) mention the major results (present them as lemmas);
3) describe the ideas and solve the major issues in the subsequential subsections
}

In this section, we study two basic questions with respect to \textcolor{black}{optimal bimodal markovian model}.
These two questions form the basis of the algorithm to discover optimal backbone, which will be discussed in Section~\ref{algorithm}.
The first question asks for any \textcolor{black}{backbone} vertex $u$ in a graph $G=(V,E)$,
how to classify its incoming edges (backbone or non-backbone) in order to optimize the parameters $p(e|\mathcal{B})$ and $p(e|\overline{\mathcal{B}})$ for maximizing overall
likelihood $L_B(\mathcal{P})$ (Formula~\ref{Likelihood3}).
The second question studies for a subset of connected vertices serving as backbone vertices $V_B$,
how to classify the edges of the induced subgraph $G^\ast[V_B]$ of traffic network $G^\ast$ as backbone or non-backbone, in order to maximize $L_B(\mathcal{P})$.
Recall that, we restrict our attention on traffic network as representative of original graph and assume that backbone is a subgraph of traffic network.

\subsection{Backbone Discovery on Incoming Edges}
\label{bediscovery}

Here, we study how to categorize its incoming edges into backbone $\mathcal{B}$ or non-backbones $\overline{\mathcal{B}}$ edges in order to maximize the overall likelihood $L_B(\mathcal{P})$.
The assumption is that each vertex and each directed edge in the graph \textcolor{black}{$G^\ast$} can be a backbone vertex or a backbone edge, respectively.
\textcolor{black}{In other words, for directed edges $(u \rightarrow v)$ and $(v \rightarrow u)$ with opposite direction, we assume that each of them can be determined independently to be a backbone or non-backbone edge.}
This relaxation produces an upper bound on the maximal contribution of each vertex to the overall likelihood.

In the following, we will first show how the overall likelihood can be expressed in terms of each \textcolor{black}{backbone} vertex in graph $G$.
Then, to optimize each \textcolor{black}{backbone} vertex, we will show that the problem of categorizing its incoming edges corresponds to an interesting clustering under the {\em Kullback-Leibler} divergence measure.

\comment{
Recall that our goal aims to reduce the number of parameters while preserving good modeling accuracy which is comparable with likelihood model in Eq.~\ref{condprob}.
Clearly, significant reduction of the number of parameters can be achieved utilizing backbone (as likelihood function $\mathcal{L}(p)$ in Eq.~\ref{bbmdl}).}

\noindent{\bf Likelihood Ratio and Its Decomposition:}
To facilitate maximizing $L_B(\mathcal{P})$ (Formula~\ref{Likelihood3}),
we compare it with the likelihood of the edge markovian model $L_M(\mathcal{P})$ (Formula~\ref{condprob}).
\textcolor{black}{
To be consistent with bimodal markovian model whose expression is based on backbone vertices,
we rewrite $L_M(\mathcal{P})$ of edge markovian model as follows:
{\small
\beqnarr
L_M(\mathcal{P}) = \prod_{e \in \mathcal{E}^\ast} p(e)^{N^\prime_e} \prod_{u \notin V_B \wedge e^\prime \in \mathcal{O}(u)} p(e^\prime|e)^{N_{ee^\prime}^\prime} \nonumber \\
\prod_{u \in V_B \wedge e^\prime \in \mathcal{O}(u)} p(e^\prime|e)^{N_{ee^\prime}^\prime} \nonumber
\eeqnarr
}
Clearly, the first two terms are the same with that of $L_B(\mathcal{P})$ since their probabilities are independent of backbone vertices.
Now, we formally introduce the following likelihood ratio (first two terms are canceled for simplification) $\frac{L_B(\mathcal{P})}{L_M(\mathcal{P})}=$
{\small
\beqnarr
\frac{\prod_{ u \in V_B \wedge  e \in \mathcal{O}(u)} p(e| \mathcal{B})^{N_{\mathcal{B}e}}  \prod_{ u \in V_B \wedge e \in \mathcal{O}(u)} p(e | \overline{\mathcal{B}})^{N_{\overline{\mathcal{B}}e}}}
{\prod_{u \in V_B \wedge e^\prime \in \mathcal{O}(u)} p(e^\prime|e)^{N_{ee^\prime}^\prime}}  \nonumber \\
=
\prod_{u \in V_B} \frac{ \prod_{e \in \mathcal{O}(u)} p(e| \mathcal{B})^{N_{\mathcal{B}e}}
\prod_{e \in \mathcal{O}(u)}
p(e | \overline{\mathcal{B}})^{N_{\overline{\mathcal{B}}e}}}
{\prod_{e^\prime \in \mathcal{I}(u), e \in \mathcal{O}(u)} p(e|e^\prime)^{N_{e^\prime e}}}
\eeqnarr
}
}


Given graph $G$ and path set $\mathcal{P}$, $L_M(\mathcal{P})$ is a constant, assuming each of its parameters $p( e|e^\prime)$ is optimized for the maximal likelihood.
Therefore, maximizing the likelihood ratio $L_B(\mathcal{P})/L_M(\mathcal{P})$ is equivalent to maximize $L_B(\mathcal{P})$.
Further, to maximize $L_B(\mathcal{P})$, we can maximize each \textcolor{black}{vertex $u \in V^\ast$} independently.
Specifically, we consider the following problem:

\bdefin({\bf Optimal Vertex Likelihood Ratio Problem})
Given graph $G$ and path set $\mathcal{P}$, for any \textcolor{black}{backbone} vertex $u$, categorize its incoming edges \textcolor{black}{appearing in $\mathcal{P}$}
as either backbone edges $\mathcal{B}$ or non-backbone edges $\overline{\mathcal{B}}$, such that
{\small
\beqnarr
LR(u)=
\textcolor{black}{
\frac{ \prod_{e \in \mathcal{O}(u)} p(e| \mathcal{B})^{N_{\mathcal{B}e}}
\prod_{e \in \mathcal{O}(u)}
p(e | \overline{\mathcal{B}})^{N_{\overline{\mathcal{B}}e}}}
{\prod_{e^\prime \in \mathcal{I}(u), e \in \mathcal{O}(u)} p(e|e^\prime)^{N_{e^\prime e}}}
}
\eeqnarr
}
is maximized.
\edefin
Here for edge $e=(u \rightarrow v)$,
$p(e|e^\prime)=\frac{N_{e^\prime e}}{\sum_{e_1 \in \mathcal{O}(u)} N_{e^\prime e_1}}$.

To maximize $LR(u)$, it is more convenient to work on its negative logarithmic format.
When the meaning is clear, we may use the shorthand $\mathcal{I}$ and  $\mathcal{O}$ for \textcolor{black}{$\mathcal{I}(u)$ and  $\mathcal{O}(u)$}, respectively:
{ \small
\beqnarr
\label{funcF}
-\log LR(u) = \sum_{e^\prime \in \mathcal{I}, e \in \mathcal{O}} N_{e^\prime e} \log p(e|e^\prime)  \nonumber \\
    - \sum_{e \in \mathcal{O}} ( N_{\mathcal{B}e} \log p(e |\mathcal{B}) + N_{\overline{\mathcal{B}}e} \log p(e | \overline{\mathcal{B}})) \nonumber \\
  = (\sum_{e^\prime \in \mathcal{B} \cap \mathcal{I}}
\sum_{e \in \mathcal{O}} N_{e^\prime e} \log p(e|e^\prime) -
\sum_{e \in \mathcal{O}} N_{\mathcal{B}e} \log p(e | \mathcal{B})) \nonumber \\
  + (\sum_{e^\prime \in \overline{\mathcal{B}} \cap \mathcal{I}}
\sum_{e \in \mathcal{O}} N_{e^\prime e} \log p(e|e^\prime) -
\sum_{e \in \mathcal{O}} N_{\overline{B}e} \log p(e| \overline{\mathcal{B}})) \nonumber \\
  =  \sum_{e \in \mathcal{O}}
( \sum_{e^\prime \in \mathcal{B} \cap \mathcal{I}} N_{e^\prime e} \log \frac{p(e|e^\prime)}{p(e|\mathcal{B})} + \sum_{e^\prime \in \overline{\mathcal{B}} \cap \mathcal{I}} N_{e^\prime e} \log \frac{p(e|e^\prime)}{p(e|\overline{\mathcal{B}})} ) \nonumber \nonumber \\
  = \sum_{e^\prime \in \mathcal{B} \cap \mathcal{I}} M_{e^\prime} \sum_{e \in \mathcal{O}} p(e|e^\prime) \log \frac{p(e|e^\prime)}{p(e|\mathcal{B})} + \nonumber \\
    \sum_{e^\prime \in \overline{\mathcal{B}} \cap \mathcal{I}} M_{e^\prime}
\sum_{e \in \mathcal{O}} p(e |e^\prime) \log \frac{p(e|e^\prime)}{p(e|\overline{\mathcal{B}})} \nonumber
\eeqnarr
}
where \textcolor{black}{$M_{e^\prime}=\sum_{e \in \mathcal{O}(u)} N_{e^\prime e}$} is the total number of paths passing through edge $e^\prime$ and then continue to one of its neighbors.

It is not hard to observe the following results:
\blemma
For a \textcolor{black}{backbone} vertex $u$, supposing each incoming edge has been categorized as backbone or non-backbone, i.e., $\mathcal{B}$ and $\overline{\mathcal{B}}$, then the minimum of $-\log LR(u)$ is achieved when
{\small
\beqnarr
\textcolor{black}{
p(e|\mathcal{B}) = \frac{N_{\mathcal{B}e}}{\sum_{e^\prime \in \mathcal{O}(u)}N_{\mathcal{B}e^\prime}} \mbox{ and }
p(e | \overline{\mathcal{B}}) = \frac{N_{\overline{\mathcal{B}}e}}{\sum_{e^\prime \in \mathcal{O}(u)} N_{\overline{\mathcal{B}}e^\prime}} \nonumber
}
\eeqnarr
}
\elemma
Note that the same choice of $p(e|\mathcal{B})$ and $p(e|\overline{\mathcal{B}})$ also maximizes the overall likelihood $L_B(\mathcal{P})$.
Indeed, $\sum_{e \in \mathcal{O}} p(e|e^\prime) \log \frac{p(e|e^\prime)}{p(e|\mathcal{B})}$ simply corresponds to the well-known {\em Kullback-Leibler} divergence between two distributions $(p(e_1|e^\prime), \cdots, p(e_k|e^\prime))$ and
$(p(e_1|\mathcal{B}), \cdots, p(e_k|\mathcal{B}))$, \textcolor{black}{where $e_1, \cdots, e_k \in \mathcal{O}(u)$ and $k=|\mathcal{O}(u)|$.}
Thus, we can utilizes a k-means type iterative algorithm for minimizing $-\log LR(u)$.
Here, we can interpret $p(e|\mathcal{B})$ and $p(e|\overline{\mathcal{B}})$
as ``centers'' for the two clusters, the backbone clusters $\mathcal{B} \cap \mathcal{I}(u)$ and non-backbone clusters $\overline{\mathcal{B}} \cap \mathcal{I}(u)$.
Each incoming edge \textcolor{black}{$e^\prime \in \mathcal{I}(u)$} corresponds to a point with $k$ features \textcolor{black}{($p(e|e^\prime), e \in \mathcal{O}(u)$)} to be clustered.
Further, the objective function $-\log LR(u)$ serves as the within-cluster distance to be minimized.
In addition, we note that each point (incoming edge $e^\prime$) is weighted by $M_{e^\prime}$, the number of paths coming from $e^\prime$.
The outline of this algorithm is illustrated in Algorithm~\ref{alg:kmeans}.
Clearly, this algorithm will converge to a local minimum similar to the classical k-means algorithm.

\begin{algorithm}[t!]
{\small
\caption{Bi-KL-Partition(Vertex $u$)}
\label{alg:kmeans}
\begin{algorithmic}[1]
\STATE randomly partition the incoming edges of $u$ into $\mathcal{B}$ and $\overline{\mathcal{B}}$;
\STATE compute $p(e|\mathcal{B})$ and $p(e|\overline{\mathcal{B}})$, $e \in \mathcal{O}(u)$;
\REPEAT
    \STATE assign each incoming edge $e^\prime$ to the cluster with the closest distance: $min(\sum_{e \in \mathcal{O}(u)} p(e|e^\prime) \log
\frac{p(e|e^\prime)}{p(e|\mathcal{B})}$, \\
 $\sum_{e \in \mathcal{O}(u)} p(e|e^\prime) \log
\frac{p(e|e^\prime)}{p(e|\overline{\mathcal{B}})})$;  \\
    \STATE calculate the two new centroids: \\
 $p(e|\mathcal{B}) = \frac{N_{\mathcal{B}e}}{\sum_{e^\prime \in \mathcal{O}(u)}N_{\mathcal{B}e^\prime}} \mbox{ and }
p(e | \overline{\mathcal{B}}) = \frac{N_{\overline{\mathcal{B}}e}}{\sum_{e^\prime \in \mathcal{O}(u)} N_{\overline{\mathcal{B}}e^\prime}}$ \\
\UNTIL{stop criteria is satisfied}
\end{algorithmic}
}
\end{algorithm}

The above partition process for the incoming edges of a vertex $u$ assumes that the directed edge in $\mathcal{E}^\ast$ can be categorized independently to be a backbone or not.
However, in an undirected graph, for each undirected edge $(u,v)$, both of its directed edges $(u \rightarrow v)$ and $(v \rightarrow u)$ should either be categorized as backbone or non-backbone edges consistently.
The above process cannot guarantee that.
Thus, such partition generates an upper bound of each individual vertex's benefit to the backbone model.
We will discuss how this partition result of individual vertices can be used to search for the optimal backbone with $K$ vertices in Section~\ref{algorithm}.

\subsection{Optimal Bimodal Markovian Model for Vertex Sets}
\label{vertexset}

In this subsection, we will develop an efficient algorithm to find the optimal backbone model when the backbone vertices are given.
Let $V_s$ be a subset of connected vertices in $G$.
For the backbone graphs $G_B=(V_B,E_B) \subseteq G[V_s]$.
In other words, the edges in the backbone can only come from the edges in the induced subgraph of vertex $V_s$.
In particular, we need to consider the consistent constraint that the categorization of both directed edges $(u \rightarrow v)$ and $(v \rightarrow u)$ of an undirected edge $(u,v)$ should be the same.

\bdefin({\bf Optimizing Vertex Set Likelihood Ratio Problem})
Given graph $G=(V,E)$ and path set $\mathcal{P}$, for any vertex set $V_s$,
we would like to construct backbone subgraph $G_B=(V_B,E_B)$,
where $V_B=V_s$ and $E_B=V_s \times V_s \cap E$, such that $LR(V_s)=\prod_{u \in V_s} LR(u) =$
{\small
\beqnarr
\prod_{u \in V_s}
\textcolor{black}{
\frac{ \prod_{e \in \mathcal{O}(u)} p(e| \mathcal{B})^{N_{\mathcal{B}e}}
\prod_{e \in \mathcal{O}(u)}
p(e | \overline{\mathcal{B}})^{N_{\overline{\mathcal{B}}e}}}
{\prod_{e^\prime \in \mathcal{I}(u), e \in \mathcal{O}(u)} p(e|e^\prime)^{N_{e^\prime e}}} \nonumber
}
\eeqnarr
}
is maximized.
\edefin


\begin{figure}
    \centering
    \mbox{
        \subfigure[Undirected graph] {\epsfig{figure=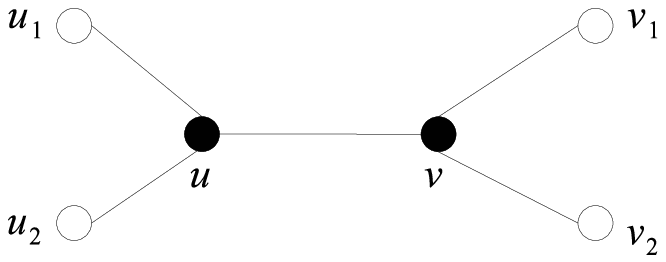,scale=0.6}
        \label{fig:ugvertexset}}
        \subfigure[Bidirected graph]{\epsfig{figure=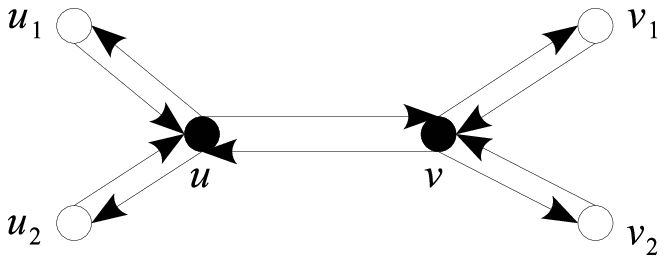,scale=0.6}
        \label{fig:bgvertexset}}
    }
    \label{fig:vertexset}
    \caption{\textcolor{black}{Observations related to backbone vertices} }
\end{figure}

\textcolor{black}{
Given a set of backbone vertices, we have following observations (see Figure~\ref{fig:vertexset}):
1) any undirected edge $e^\prime=(u, u_1) \in E$ with only one endpoint $u$ in $V_s$ (see Figure~\ref{fig:ugvertexset}) is not a candidate to be a backbone edge;
2) for each undirected edge $e^\prime=(u,v) \in E$ with both endpoints in $V_s$ ($(u,v) \in V_s \times V_s \cap E$), we need $(u \rightarrow v)$ and $(v \rightarrow u)$ from $\mathcal{E}^\ast$ to be both in $\mathcal{B}$ or in $\overline{\mathcal{B}}$ (as shown in Figure~\ref{fig:bgvertexset}).
To facilitate our discussion, for each undirected edge $e^\prime=(u,v) \in V_s \times V_s \cap E$,
we further define $F(e^\prime|\mathcal{B})$ and $F(e^\prime|\overline{\mathcal{B}})$
for the cases where both directed edges $e^\prime_u=(u\rightarrow v)$ and $e^\prime_v=(v \rightarrow u)$ are backbone or non-backbone edges, respectively:
{\small
\beqnarr
F(e^\prime|\mathcal{B})=\sum_{e \in \mathcal{O}(u)}  N_{e^\prime_v e} \log \frac{p(e|e^\prime_v)}{p(e|\mathcal{B})} + \sum_{e \in \mathcal{O}(v)} N_{e^\prime_u e}  \log \frac{p(e|e^\prime_u)}{p(e|\mathcal{B})} \nonumber \\
F(e^\prime|\overline{\mathcal{B}})=
\sum_{e \in \mathcal{O}(u)} N_{e^\prime_v e} \log \frac{p(e|e^\prime_v)}{p(e|\overline{\mathcal{B}})} + \sum_{e \in \mathcal{O}(v)} N_{e^\prime_u e}  \log \frac{p(e|e^\prime_u)}{p(e|\overline{\mathcal{B}})} \nonumber
\nonumber
\eeqnarr
}
}

Given this, we can represent $-\log LR(V_s)$ as follows:
\textcolor{black}{
{\small
\begin{align*}
& -\log LR(V_s) =  \nonumber \\
& \sum_{e^\prime \in (E \backslash V_s \times V_s)}
(\sum_{e \in \mathcal{O}(u)} N_{e^\prime_v e} \log \frac{p(e|e^\prime_v)}{p(e|\overline{\mathcal{B}})} +
\sum_{e \in \mathcal{O}(v)} N_{e^\prime_u e} \log \frac{p(e|e^\prime_u)}{p(e|\overline{\mathcal{B}})}) + \nonumber \\
& \sum_{e^\prime \in V_s \times V_s \cap E \cap \mathcal{B}} F(e^\prime|\mathcal{B})
+\sum_{e^\prime \in V_s \times V_s \cap E \cap \mathcal{B}} F(e^\prime|\overline{\mathcal{B}}) \nonumber
\end{align*}
}
}

We can easily derive that the minimum of $-\log LR(V_s)$ is achieved when
{\small
\textcolor{black}{
\beqnarr
p(e|\mathcal{B}) = \frac{N_{\mathcal{B}e}}{\sum_{e^\prime \in \mathcal{O}(u)}N_{\mathcal{B}e^\prime}} \mbox{ and }
p(e | \overline{\mathcal{B}}) = \frac{N_{\overline{\mathcal{B}}e}}{\sum_{e^\prime \in \mathcal{O}(u)} N_{\overline{\mathcal{B}}e^\prime}} \nonumber
\eeqnarr
}
}

We now describe our generalized {\em Bi-KL-Partition} algorithm (Algorithm \ref{alg:gen2means}), which again will have an interesting convergence property.
The basic idea is to update the cluster membership of each candidate backbone edge $e^\prime=(u,v)\in V_s\times V_s \cap E$ based on $F(e^\prime|\mathcal{B})$ and $F(e^\prime|\overline{\mathcal{B}})$ with the optimal $p(e|\mathcal{B})$ and $p(e|\overline{\mathcal{B}})$ for the current clusters $\mathcal{B}$ and $\overline{\mathcal{B}}$.
In other words, $F(e^\prime|\mathcal{B})$ and $F(e^\prime|\mathcal{B})$ describe a generalized ``distance'' function from a point (edge) $e^\prime$ to corresponding centroids.

\begin{algorithm}[t!]
{\small
\caption{GBi-KL-Partition(Vertex Set $V_s$)}
\label{alg:gen2means}
\begin{algorithmic}[1]
\STATE Assign each edge with only one end in $V_s$ to cluster $\overline{\mathcal{B}}$;
randomly partition edges with both ends in $V_s$ into $\mathcal{B}$ and $\overline{\mathcal{B}}$; \\
\STATE for each vertex $u \in V_s$, compute optimal $p(e|\mathcal{B})$ and $p(e|\overline{\mathcal{B}})$, $e \in \mathcal{O}(u)$; \\
\REPEAT
    \STATE assign each edge $e^\prime$ to the cluster with the closest distance: $min(F(e^\prime|\mathcal{B}),F(e^\prime|\overline{\mathcal{B}}))$;  \\
    \STATE calculate new centroids of two clusters for each vertex $u \in V_s$: \\
 $p(e|\mathcal{B}) = \frac{N_{\mathcal{B}e}}{\sum_{e^\prime \in \mathcal{O}(u)}N_{\mathcal{B}e^\prime}} \mbox{ and }
p(e | \overline{\mathcal{B}}) = \frac{N_{\overline{\mathcal{B}}e}}{\sum_{e^\prime \in \mathcal{O}(u)} N_{\overline{\mathcal{B}}e^\prime}}$ \\
\UNTIL{stop criteria is satisfied}
\end{algorithmic}
}
\end{algorithm}

\blemma{\bf (Convergence Property)}
\label{converge}
As we iteratively update membership of each edge $e^\prime \in V_s \times
V_s \cap E$ in Algorithm \ref{alg:gen2means}, the function $-\log LR(V_s)$
converges to a local optimum.
\elemma
\bproof
We can prove that for each edge $e^\prime \in V_s \times V_s \cap E$, its within-cluster distance is reduced (i.e., $min (F(e^\prime|\mathcal{B}), F(e^\prime|\overline{\mathcal{B}}))$), and then further when computing the new ``centroids''. So, the overall $-\log LR(V_s)$ cannot increase.
The detailed proof is omitted due to space limitation.
\eproof

Note that this algorithm will be used when we find a set of backbone vertices of size $K$ which may serve as the candidate backbone vertices.
In the next section, we will describe methods to utilize these two algorithms for discovering backbone with $K$ vertices to optimize bimodal markovian model.
}

\section{Algorithms for Backbone Discovery}
\label{algorithm}

\comment{
1) refine the algorithm description without long-distance path set assumption;
2) consider the connectivity checking in IterativeRefinement algorithm;
3) In the last paragraph for path set generation, slightly discuss the complexity of generating all pairs shortest paths in the worst case;
}

In this section, we will introduce two algorithms to discover the backbone with $K$ vertices for optimizing bimodal markovian model.
The first algorithm tries to choose a set of connected vertices as backbone vertices by certain criteria,
and then discover backbone edges among them in order to maximize $L_B(\mathcal{P})$.
Interestingly, the first step can be converted to an instance of {\em maximum weight connected subgraph} (MCG) problem~\cite{Lee96}.
However, the selected vertices in first algorithm cannot guarantee to produce ``good'' backbone.
We thus further propose second algorithm starting from above backbone and iteratively refine it to achieve better value of $L_B(\mathcal{P})$.

\comment{
In this section, we will describe an efficient algorithm to discover the optimal $K$-backbone models. The basic idea is
to iteratively find a list of candidate $K$-backbone models and then use the {\em GBi-KL-Partition} procedure to
discover the backbone edges for maximizing the likelihood function $L_B(\mathcal{P}|B)$. Especially, we will relate our
problem to the so-called {\em maximum weight connected subgraph} (MCG) problem~\cite{Lee96}.
}


\subsection{Backbone Discovery based on 
Maximal Weight Connected Subgraph}
\label{mcg}

The optimality of resulting backbone highly depends on the firstly selected backbone vertices.
How to choose ``good'' backbone vertices is a challenging problem,
as it is impossible to determine ``goodness'' of a set of backbone vertices in terms of $L_B(\mathcal{P})$ without backbone structure.
To tackle it, we utilize the upper bound of their contribution to $L_B(\mathcal{P})$ in order to approximate the true ``goodness''.
More specifically, we assign a score to each vertex which corresponds to maximal contribution by this vertex to the likelihood.
For a set of connected vertices, their overall weight (sum of vertex weight) serves as an upper bound of their true likelihood.
Larger upper bound potentially leads to better true value,
thus we attempt to find a set of connected vertices with maximal upper bound to effectively approximate their true contribution.
Then, {\em GBi-KL-Partition} procedure is used to discover backbone edges connecting them.

\noindent{\bf Upper Bound:}
Since most vertices will not be backbone vertices, we will rewrite our target maximal likelihood as
{\small
\beqnarr
\log L_B(\mathcal{P})= \log \frac{L_B(\mathcal{P})}{L_I(\mathcal{P})}+ \log L_I(\mathcal{P})  \nonumber
\eeqnarr
}
where $L_I(\mathcal{P})$ is the likelihood function for the edge independence model.
Given this, we introduce log-likelihood ratio $F(u)$ which represents the benefit for this vertex being a backbone vertex:
{\small
\beqnarr
\label{fu}
& F(u)=\sum_{e \in \mathcal{O}(u)}\left(N_{\mathcal{B}e} \log \frac{p(e|\mathcal{B})}{p(e)}+ N_{\overline{\mathcal{B}}e} \log \frac{p(e|\overline{\mathcal{B}})}{p(e)} \right)
\eeqnarr}
For simplicity, we omit the benefit for edge $e$ to be the first edge in any shortest path (this portion is very small).
It is easy to see that $\log \frac{L_B(\mathcal{P})}{L_I(\mathcal{P})}\approx \sum_{u \in V} F(u)$.
\textcolor{black}{In this case, we can invoke the {\em Bi-KL-Partition} procedure for each vertex $u$ and find its optimal bimodal markovian model.}
In the meanwhile, the values of $p(e|\mathcal{B})$, $p(e|\overline{\mathcal{B}})$ for each edge $e \in \mathcal{O}(u)$ are obtained.
Now, we can calculate $F(u)$ for each vertex $u$ and assign it as corresponding vertex weight in graph $G$.

Given this, the problem of choosing a set of connected vertices with maximal sum of vertex weight is converted to an instance of {\em maximum weight connected graph} (MCG) problem~\cite{Lee96}.
\bdefin({\bf Maximum Weight Connected Subgraph Problem} (MCG))
Given a graph $G=(V,E)$ where each vertex has a weight $w(v)$, and a positive integer $k$,
maximum weight connected subgraph problem tries to identify a connected subgraph $G^\prime=(V^\prime,E^\prime)$
where $|V^\prime|=k$ and $\sum_{v \in V^\prime} w(v)$ is maximized.
\edefin

The MCG problem has been proven to be NP-hard, but an efficient heuristic algorithm has been proposed to find a {\em maximal} weight connected subgraph~\cite{Lee98}.
We apply this heuristic method, termed MCG, on our vertex-weighted graph to extract a set of connected vertices as backbone vertices.
Following that, we utilize {\em GBi-KL-Partition} procedure to discover backbone edges for optimizing $L_B(\mathcal{P})$.

\begin{algorithm}
\caption{BackboneDiscovery($G=(V,\mathcal{E})$,$K$)}
\label{alg:backbonemcg}
{\small
\begin{algorithmic}[1]
\REQUIRE{ $G$ is input network, $K$ is the backbone size}
\FORALL {$u \in V$}
     \STATE invoke {\em Bi-KL-Partition}(u);
     \STATE compute $F(u)$;
\ENDFOR
\STATE $V_B \leftarrow MCG(V,K)$;
\STATE $\mathcal{B} \leftarrow$ {\em GBi-KL-Partition($V_B$)};
\STATE $G_B \leftarrow (V_B, \mathcal{B})$;
\RETURN $G_B$;
\end{algorithmic}
}
\end{algorithm}

The procedure to discover backbone based on maximal weight connected subgraph is outlined in Algorithm~\ref{alg:backbonemcg}.
To begin with, we invoke the {\em Bi-KL-Partition} procedure for each vertex $u$ to find its optimal bimodal markovian model and $p(e|\mathcal{B})$, $p(e|\overline{\mathcal{B}})$ for each $e \in \mathcal{O}(u)$ (Line 2).
Then, we calculate $F(u)$ for each vertex $u$ as its weight in graph $G$ (Line 3).
Following that, we use heuristic algorithm of MCG on vertex-weighted graph $G$ to identify backbone vertex set $V_B$ (Line 5).
Finally, {\em GBi-KL-Partition} procedure is applied to extract backbone edges $\mathcal{B}$ among $V_B$ (Line 6).

\noindent{\bf Computational Complexity: }
In the main loop, procedure {\em Bi-KL-Partition} dominates computational cost.
It takes at most \\
$\sum_{u \in V} O( c |\mathcal{I}(u)|\times |\mathcal{O}(u)|)=O(\sum_{u \in V} c |\mathcal{N}(u)|^2)=O( c |V| d^2)$ time,
where $c$ is the number of iterations repeated in {\em Bi-KL-Partition} and $d$ is the average degree of vertices in $G$.
Moreover, the heuristic algorithm of MCG takes $O(K^2 |V|^2+K^2 |V||{E}|)$ time.
Then, the procedure {\em GBi-KL-Partition} costs $\sum_{u \in V_B} O(c^\prime  |\mathcal{N}(u)|^2)=O(K d^2)$ time,
where $c^\prime$ is the number of iterations repeated in {\em GBi-KL-Partition}.
Putting together, overall time complexity of this backbone discovery procedure is $O(|V|d^2+K^2|V||{E}|)$.

However, discovered subset $V_B$ with maximal total weights 
$\sum_{u \in V_B} F(u)$ are not necessarily ``good'' backbone vertices for optimizing $L_B(\mathcal{P})$.
This is because we neglect the constraint that both $(u \rightarrow v)$ and $(v \rightarrow u)$ of undirected edge $(u,v)$ should be both as backbone edges or non-backbone edges in vertex weight computation.
In other words, if we apply procedure {\em GBi-KL-Partition}, then 
$\sum_{u \in V_B} F(u)$ may decrease by using the updated $p(e|\mathcal{B})$ and $p(e|\overline{\mathcal{B}})$.

\subsection{Backbone Discovery by Iterative Refinement}
\label{iterativerefine}

To address aforementioned issue, we propose a refinement strategy to improve the backbone in an iterative fashion.
The basic idea is to first discover a subgraph as search starting point by Algorithm~\ref{alg:backbonemcg}, then iteratively refine it by identifying a alternate backbone based on current one.
Specially, in each iteration, we randomly abandon one vertex from current candidate backbone and add a neighboring vertex with maximal value of $F(u)$ (Formula~\ref{fu}) to form a alternate backbone.
If new backbone leads to better value of $L_B(\mathcal{P})$, it would be used as current backbone for further refinement in the next iteration.

\begin{algorithm}
\caption{IterativeRefinement($G=(V,\mathcal{E})$,$K$)}
\label{alg:iterativealg}
{\small
\begin{algorithmic}[1]
\REQUIRE{ $G$ is input network, $K$ is the backbone size} \\
\COMMENT{Step 1: Preprocessing}
\STATE invoke {\em BackboneDiscovery}($G$,$K$) to obtain candidate backbone $G_B=(V_B, \mathcal{B})$;
\FORALL{$u \in V_B$}
    \STATE invoke {\em Bi-KL-Partition}(u);
    \STATE compute $F(u)$;
\ENDFOR
\STATE $W_H \leftarrow \sum_{u \in V_B} F(u)$;
\STATE $W_L \leftarrow \sum_{u \in V_B} F^\prime(u)$; \COMMENT{$F^\prime(u)$ is under the updated parameters of $p(e|\mathcal{B})$ and $p(e|\overline{\mathcal{B}})$}
\STATE $W \leftarrow W_L$; \\
\COMMENT{Step 2: Iterative Refinement}
\WHILE{$|V(G)| > K \wedge W_H > W$}
    \STATE $V_s \leftarrow V_B \backslash \{v\}$ and $V \leftarrow V \setminus \{v\}$; \COMMENT{randomly remove one vertex $v$ from $V_B$ and $G$}
    \STATE $u \leftarrow \arg\max_{u \in \mathcal{N}(V_B)} F(u)$;
    \STATE $V_B \leftarrow V_B \cup \{u\}$;
    \STATE $W_H \leftarrow  W_H-F(v)+ F(u)$;
    \STATE $\mathcal{B} \leftarrow$ {\em GBi-KL-Partition($V_B$)};
    \STATE $W_L=\sum_{u \in V_B} F^\prime(u)$;
    \IF{$W < W_L$}
        \STATE $W \leftarrow W_L$;
        \STATE $G_B \leftarrow (V_B,\mathcal{B})$; \COMMENT{keep the best result}
    \ENDIF
\ENDWHILE
\RETURN $G_B$;
\end{algorithmic}
}
\end{algorithm}


The overall procedure to discover backbone by iterative refinement scheme is outlined in Algorithm~\ref{alg:iterativealg}.
It consists of two key steps: preprocessing step to generate candidate backbone by procedure {\em BackboneDiscovery},
and refinement step for improving backbone by local search. 
In {\em Step 1}, we invoke procedure {\em BackboneDiscovery} to provide a subgraph $G_B$ serving as starting point of refinement step (Line 1).
For each vertex $u$ in $G_B$, we perform procedure {\em Bi-KL-Partition} to help compute $F(u)$ which is assigned as vertex weight (Line 2 to Line 5).
The sum $W_H$ of those vertex weights represents the upper bound of maximal likelihood achieved by $G_B$ (Line 6).
In the meanwhile, their true likelihoods are added together (i.e., $W_L$) to serve as lower bound of final backbone (Line 7).
Moreover, $W$ is used to denote the true maximal benefit achieved so far.
In {\em Step 2}, we try to seek alternate backbone vertices in an iterative manner for improving likelihood.
To speed up the search process, in each iteration, we randomly remove a vertex $v$ from the current candidate backbone and graph $G$ (eliminate its further consideration) (Line 11).
Then, we add neighboring vertex $u$ with maximal $F(u)$ to form new backbone vertex set (Line 12).
We update upper bound $W_H$ and lower bound $W_L$ accordingly (Line 13 to Line 15).
If a better backbone is found, current backbone will be kept as best result so far (Line 16 to Line 19).
The refinement loop terminates until either the remaining graph is too small or the upper bound is smaller than the true likelihood achieved so far (Line 9).
Note that since Step 2 involves a random search process, we can invoke it multiple times and choose the overall best backbone as our final result.

\noindent{\bf Computational Complexity: }
The cost of step 1 is dominated by procedure {\em BackboneDiscovery}, which takes $O(|V|d^2 + K^2 |V||{E}|)$ times, where $d$ is the average degree of vertices in $G$.
For step 2, in each iteration, it takes $O(dK)$ time to select best neighboring vertex to form a alternate backbone vertex set (Line 11),
and takes $O(Kd^2)$ times to perform {\em GBi-KL-Partition} supposing the number of iteration required in the procedure is a small constant.
Given this, assuming $c$ iterations are needed in the refinement loop, step 2 takes $O(c K d^2)$ time in total.
Overall, the time complexity of iterative refinement scheme to discover backbone is $O(|V|d^2 + K^2 |V||{E}|)$.

Finally, we note that in the above algorithm, we do not consider how to compute path set $\mathcal{P}$ and derive the basic probabilities,
such as $p(e)$ and $p(e|e^\prime)$ in edge independence model and edge markovian model, respectively.
In the worst case, assuming path set $\mathcal{P}$ includes the shortest path for each pair of vertices,
the most straightforward way is to invoke $|V|$ times BFS procedures in $O(|V|\times(|V|+|E|))$ time.
When we enumerate these paths, we can online compute $N_e$ (edge betweenness) and $N_{e^\prime e}$, so there is no need to materialize the entire path set $\mathcal{P}$.
The overall computational time is $O(|V|\times(|V|+|E|))$.

\comment{
\subsection{Overall Algorithm}
\label{overall}

Since the MCG of the entire graph does not necessarily correspond to the final backbone vertices,
we consider how to iteratively discover a list of those candidate vertex sets.
Specifically, we utilize MCG to discover a subgraph as a search starting point, and then we search its surrounding vertices to refine the discovered backbone.
The overall procedure to discover the backbone in a directed graph is sketched in Algorithm~\ref{alg:iterativealg}.
In {\em Step 1},  we invoke the {\em Bi-KL-Partition} procedure for each vertex $u$ to find its optimal backbone model and $p(e|\mathcal{B})$, $p(e|\overline{\mathcal{B}})$  for $e \in \mathcal{O}(u)$.
We then calculate $F(u)$ for each vertex $u$ in graph $G$.
In {\em Step 2}, we apply MCG on the weighted graph $G$ to identify the first candidate backbone vertex set $V_s$.
Then, we apply the {\em GBi-KL-Partition} procedure to compute the backbone edges $\mathcal{B}$ and updated parameters $p(e|\mathcal{B})$ and $p(e|\overline{\mathcal{B}})$.
We can then compute the true benefit from $V_s$, $W_L$.
In {\em Step 3}, we seek to iteratively identify alternative backbone vertices.
To speed up the search process, in each iteration, we randomly remove a vertex $v$ from the current candidate backbone and the graph $G$ (eliminate its further consideration).
Then, we try to add neighbor vertices with maximal $F_u$.
We compute $W_H=\sum_{v \in V_s} F(v) $ as the upper bound of the overall benefit and $W$ as the true maximal benefit discovered so far for any candidate backbones.
We stop the iteration when either the graph is too small or the upper bound is smaller than the true maximal benefit.
Note that since Step 3 involves a random search process, we can invoke it multiple times and choose the overall best backbone as our final result.

\begin{algorithm}
\caption{IterativeRefinement($G=(V,\mathcal{E})$,$K$)}
\label{alg:iterativealg}
{\small
\begin{algorithmic}[1]
\REQUIRE{ $G$ is input network, $K$ is the backbone size} \\
\COMMENT{Step 1: Preprocessing}
\STATE invoke {\em BackboneDiscovery}($G$,$K$) to obtain candidate backbone $G_B=(V_B, \mathcal{B})$;
\FORALL{$u \in V_B$}
    \STATE invoke {\em Bi-KL-Partition}(u);
    \STATE compute $F(u)$;
\ENDFOR
\STATE $W_H \leftarrow \sum_{u \in V_B} F(u)$;
\STATE $W_L \leftarrow \sum_{u \in V_B} F^\prime(u)$; \COMMENT{$F^\prime(u)$ is under the updated parameters of $p(e|\mathcal{B})$ and $p(e|\overline{\mathcal{B}})$}
\STATE $W \leftarrow W_L$; \\
\COMMENT{Step 2: Iterative Refinement}
\WHILE{$|V(G)| > K \wedge W_H > W$}
    \STATE $V_s \leftarrow V_B \backslash \{v\}$ and $V \leftarrow V \setminus \{v\}$; \COMMENT{randomly remove one vertex $v$ from $V_B$ and $G$}
    \STATE $u \leftarrow \arg\max_{u \in \mathcal{N}(V_B)} F(u)$;
    \STATE $V_B \leftarrow V_B \cup \{u\}$;
    \STATE $W_H \leftarrow  W_H-F(v)+ F(u)$;
    \STATE $\mathcal{B} \leftarrow$ {\em GBi-KL-Partition($V_B$)};
    \STATE $W_L=\sum_{u \in V_B} F^\prime(u)$;
    \IF{$W < W_L$}
        \STATE $W \leftarrow W_L$;
        \STATE $G_B \leftarrow (V_B,\mathcal{B})$; \COMMENT{keep the best result}
    \ENDIF
\ENDWHILE
\RETURN $G_B$;
\end{algorithmic}
}
\end{algorithm}


\noindent{\bf Computation Complexity:}
Step 1 needs $\sum_{u \in |V|} O( c |\mathcal{I}(u)|\times |\mathcal{O}(u)|)=O(\sum_{u \in |V|} c |\mathcal{N}(u)|^2)=O( c |V| d_v^2)$ time to perform the biclustering for all vertices in $G$, where $d_v$ is the average degree of each vertex in $G$.
Step 2 invokes a heuristic MCG algorithm, which takes $O(K^2 |V|^2+K^2 |V||E|)$ time.
Then, the {\em GBi-KL-Partition} costs $\sum_{u \in |V_s|} O(c^\prime  |\mathcal{N}(u)|^2)=O(K d_v^2)$ time.
For Step 3, let us assume  $c^\prime$ iterations are needed to terminate the procedure.
For each iteration, we need a total of $)(|V|+ c^\prime K  d_v^2)$ steps.
Overall, the time complexity of the backbone discovery scheme on a directed graph is $O(K^2 |V||E|+ K d_v^2)$.

Finally, we note that in the above algorithm, we do not consider  how to compute path set $\mathcal{P}$ and derive the basic probabilities, such as $p(e)$ and $p(e|e^\prime)$ in the edge independence and edge Markovian model, respectively.
Assuming the path set $\mathcal{P}$ includes the shortest path from each pair,
the most straightforward way is to invoke $|V|$ times BFS procedures in $O(|V|\times(|V|+|E|)$.
When we enumerate these paths, we can online compute $N_e$ (edge betweenness) and $N_{e^\prime e}$, so there is no need to materialize the entire path set $\mathcal{P}$.
The overall computational time is $(|V|\times(|V|+|E|))$.
Alternatively, we can simply compute all pairs of shortest distance and utilize the following observation to compute edge count $N_e$ or segment betweenness ($N_{e^\prime e}$)
\[dist(u,v) = dist(u,x) + dist(y,v) + distance(x,y)\]
That means if the shortest path from $u$ to $v$ passes through segment $x\rightarrow y$, then the above equation is satisfied, and vice versa.
In this case, the main computational cost is still $O(|V|\times(|V|+|E|)$.
However, the distance table needs to be materialized.
}

\section{Empirical Study}
\label{experiment}

In this section, we evaluate the performance of proposed backbone discovery algorithms:

\noindent 1) the basic backbone discovery based on vertex betweenness (referred to it as {\bf VB});

\noindent 2) the backbone discovery approach based on maximum weight connected graph (referred to it as {\bf MCG});

\noindent 3) the backbone discovery approach based on iterative refinement (referred to it as {\bf ITER}).

First, we study the performance of our methods from three aspects: modeling accuracy, parameter reduction and edge size in the backbone.
Note that, modeling accuracy is measured by ratio between edge markovian model and bimodal markovian model,
which is expressed as the logarithmic value of edge markovian model's ({\bf EM}) likelihood divides the one of bimodal markovian model extracted by VB, MCG and ITER (denoted by $EM/VB$, $EM/MCG$ and $EM/ITER$).
The closer to $1$ modeling accuracy is, the better results our methods achieve.
Then, we study the efficiency of our methods on large random graphs with power law degree distribution.
Finally, we perform case studies on co-author network and human protein-protein interaction (PPI) network to further demonstrate our approaches.

We implemented all algorithms using C++ and the Standard Template Library (STL).
All experiments were conducted on a 2.0GHz Dual Core AMD Opteron CPU with 4.0GB RAM running Linux.



\begin{figure*}[!htbp]
    \label{fig:realdata}
    \centering

    \mbox{
        \subfigure[Modeling accuracy on Yeast]{\epsfig{figure=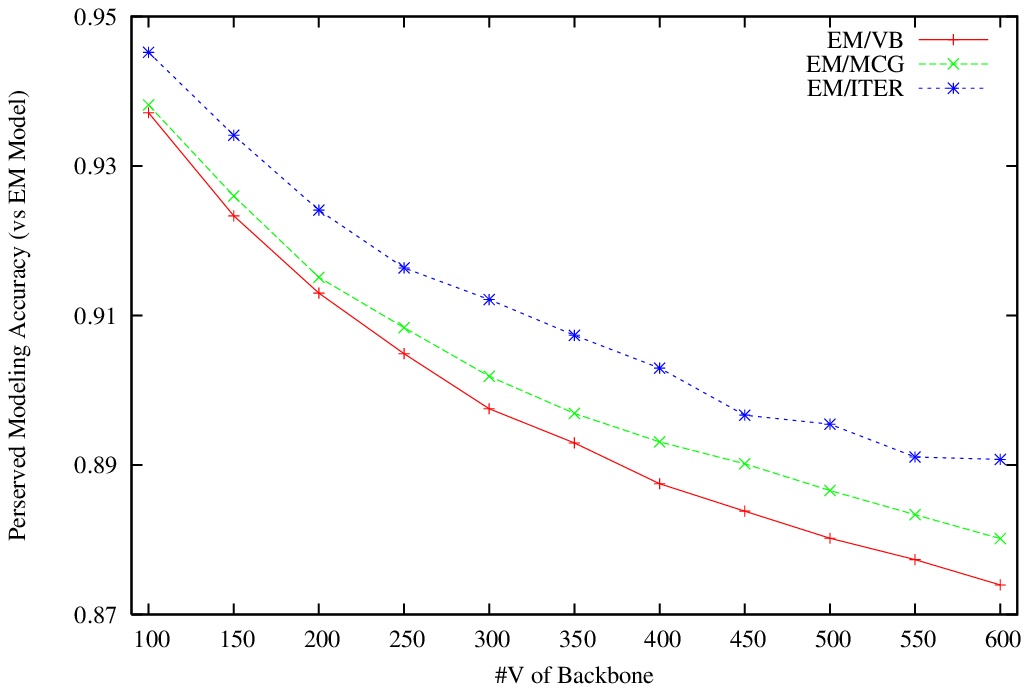,scale=0.57} \label{yeastratio}}
        \subfigure[Modeling accuracy on Net]{\epsfig{figure=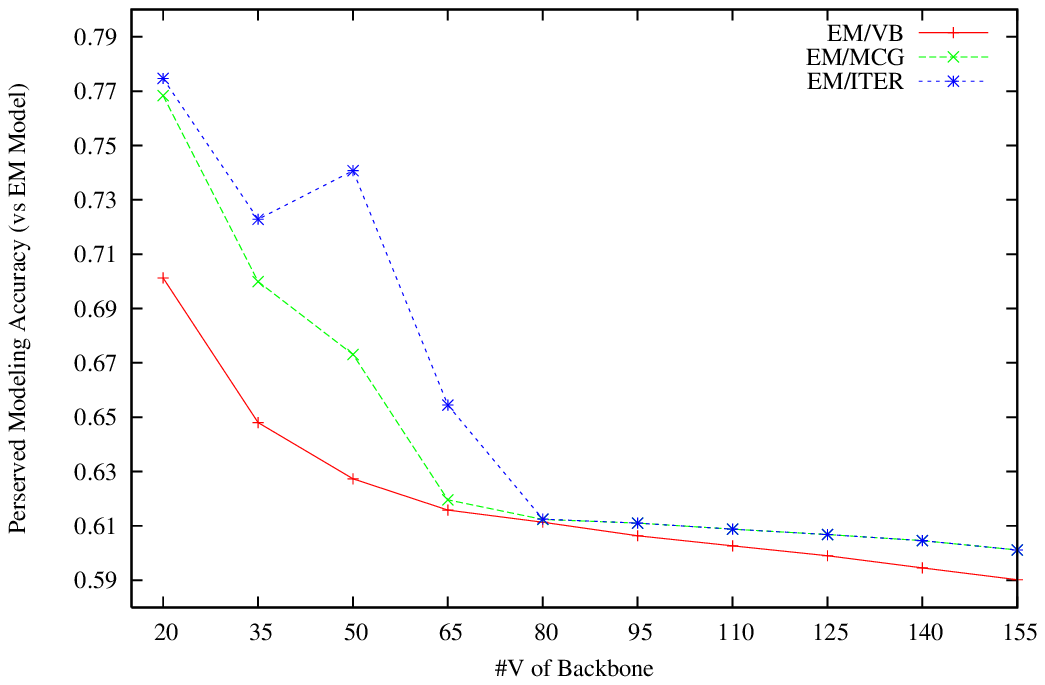,scale=0.57} \label{netratio}}
        \subfigure[Modeling accuracy on DM]{\epsfig{figure=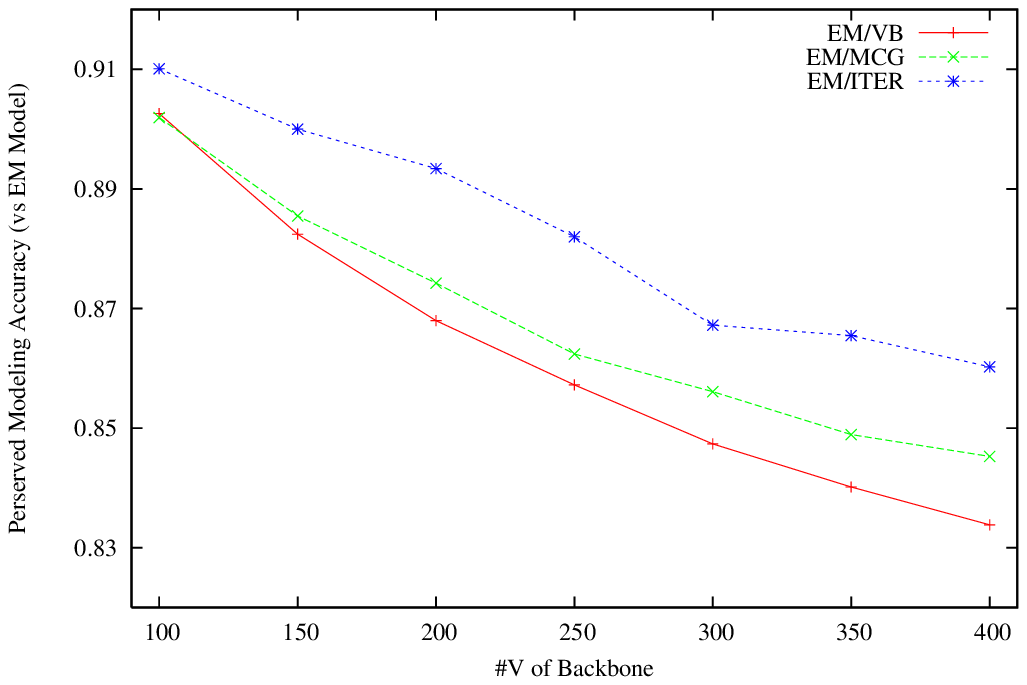,scale=0.57}\label{coauthorratio}}
    }
    \mbox{
        \subfigure[\#E of backbone (Yeast)]{\epsfig{figure=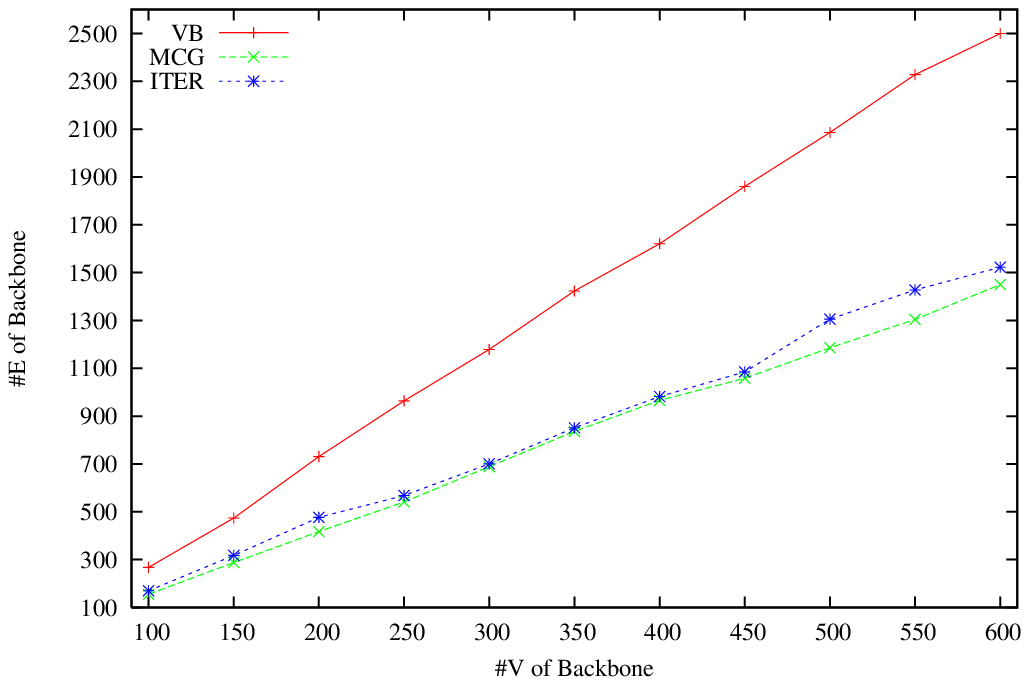,scale=0.57} \label{yeastedge}}
        \subfigure[\#E of backbone (Net)]{\epsfig{figure=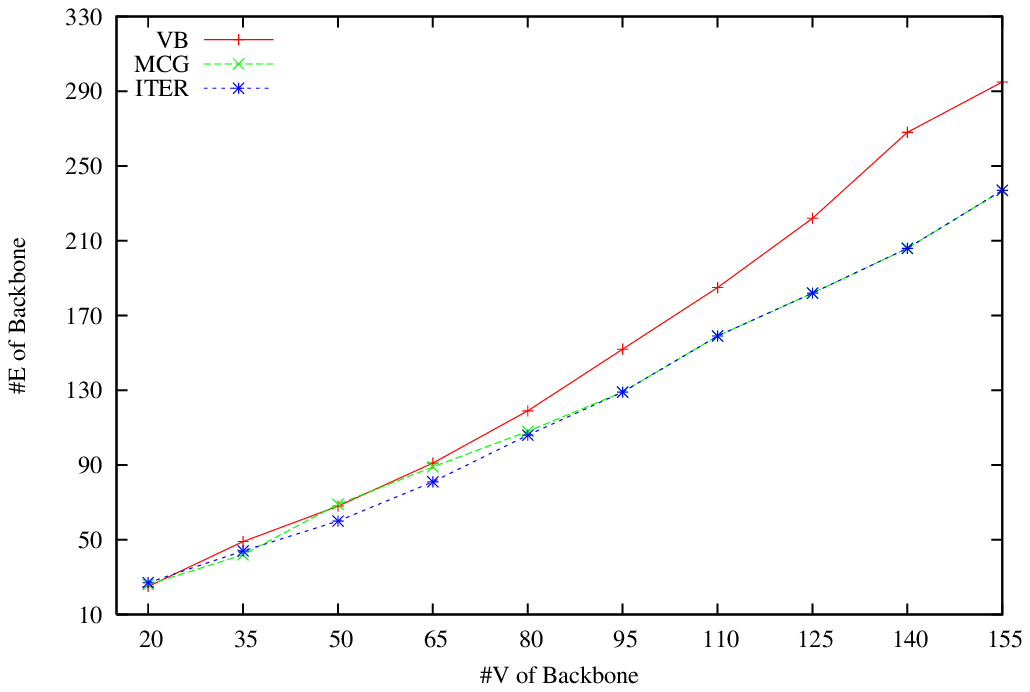,scale=0.57} \label{netedge}}
        \subfigure[\#E of backbone (DM)]{\epsfig{figure=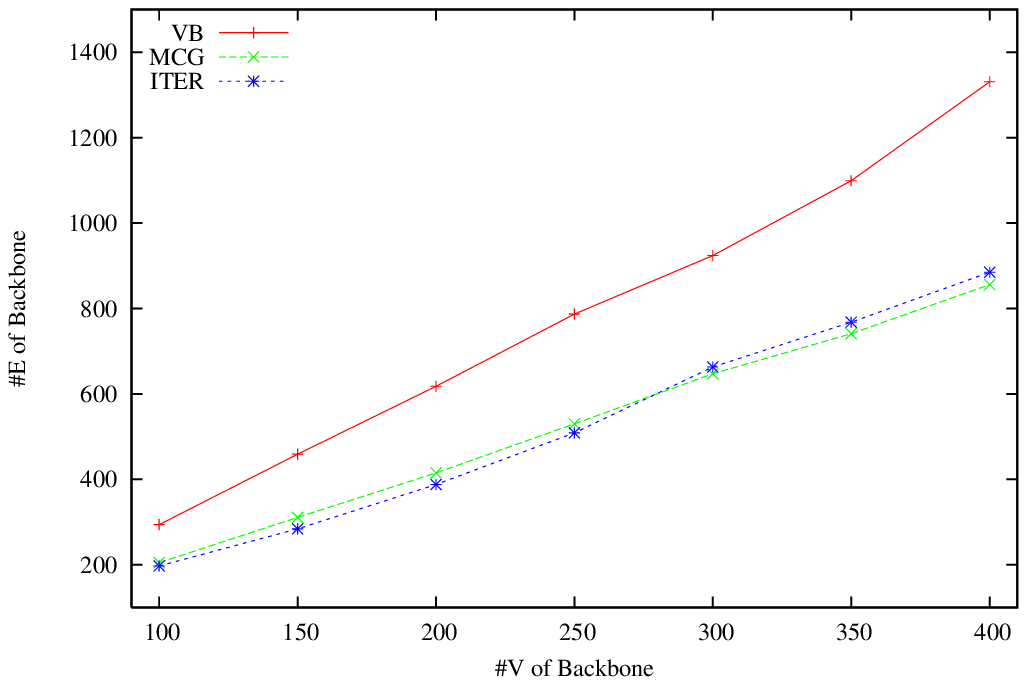,scale=0.57}\label{coauthoredge}}
    }    
    \mbox{
        \subfigure[\#Param. reduction on Yeast]{\epsfig{figure=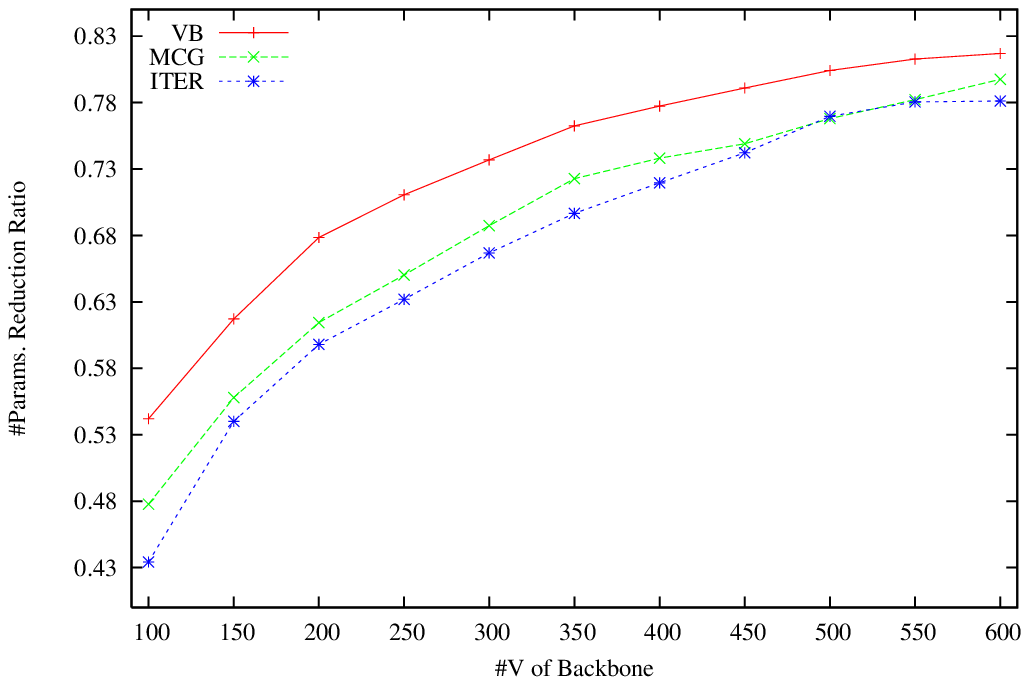,scale=0.57} \label{yeastparams}}
        \subfigure[\#Param. reduction on Net]{\epsfig{figure=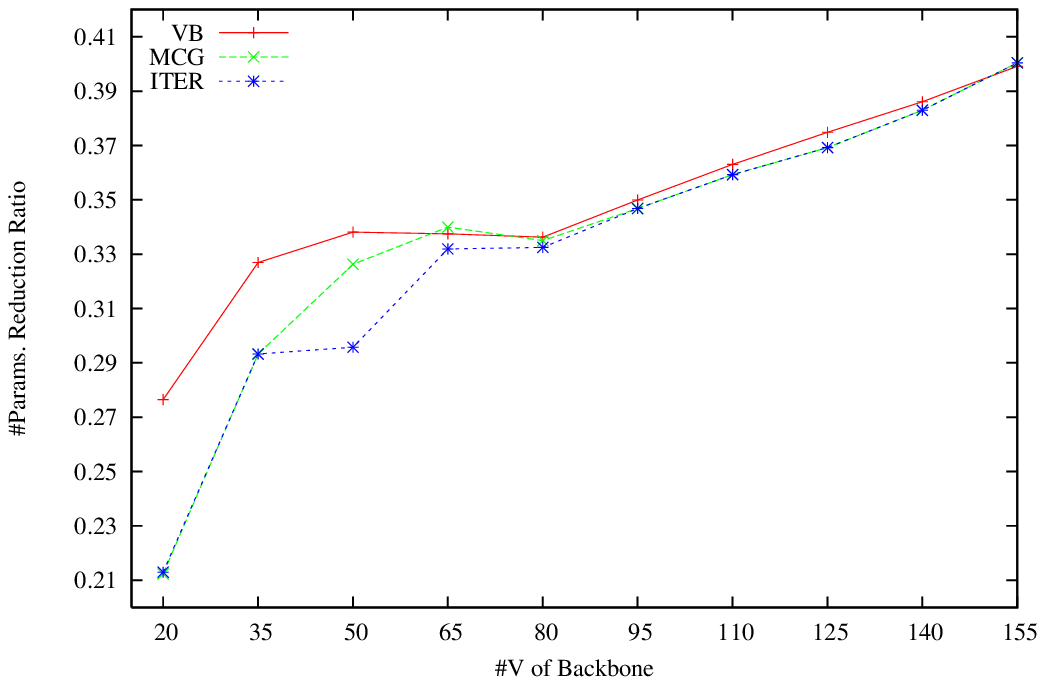,scale=0.57} \label{netparams}}
        \subfigure[\#Param. reduction on DM]{\epsfig{figure=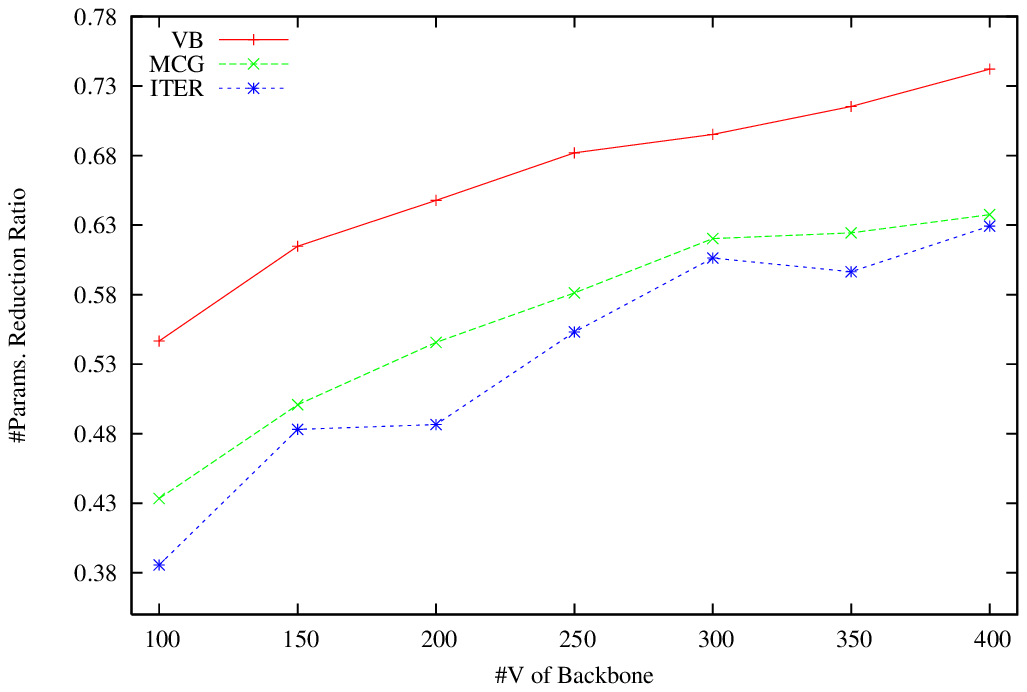,scale=0.57}\label{coauthorparams}}
    }

    \mbox{
        \subfigure[{\small 25-vertex backbone (Net)}]{\epsfig{figure=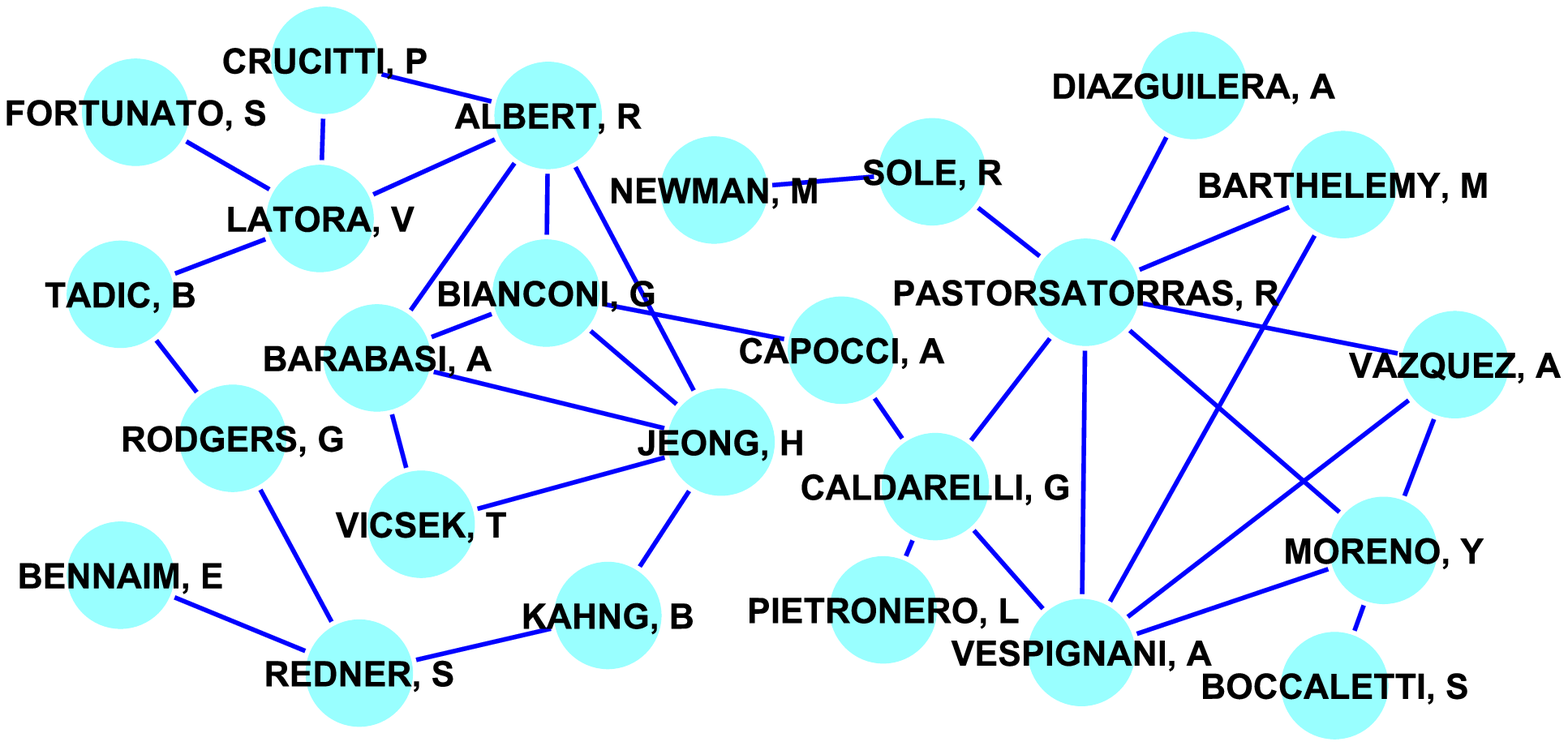,height=1.6in, width=3in} \label{fig:net25}}
        \subfigure[{\small 35-vertex backbone (Net)}]{\epsfig{figure=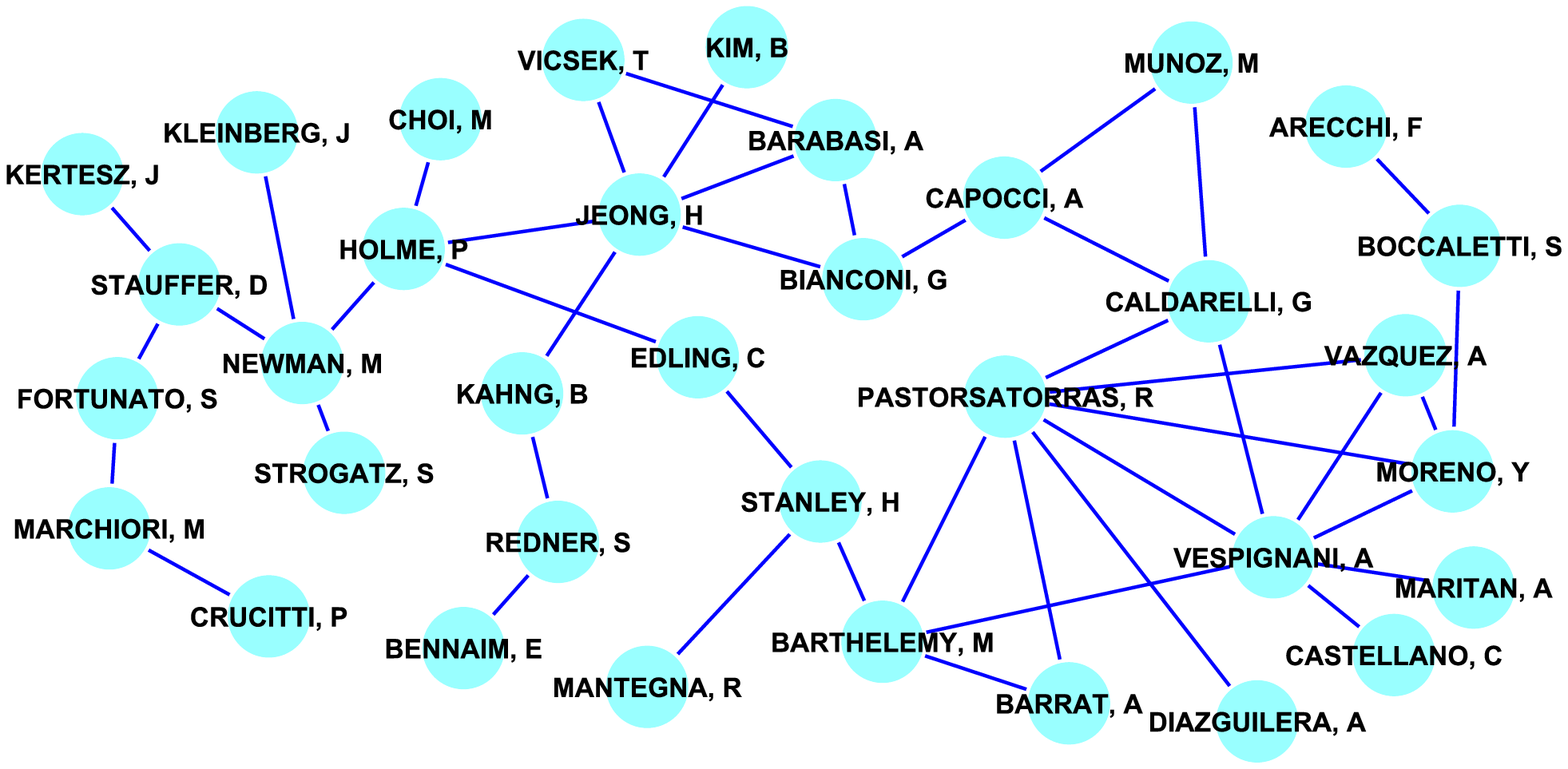,height=1.7in, width=3.8in} \label{fig:net35}}
    }
    \mbox{
        \subfigure[{\small 50-vertex backbone (Net)}]{\epsfig{figure=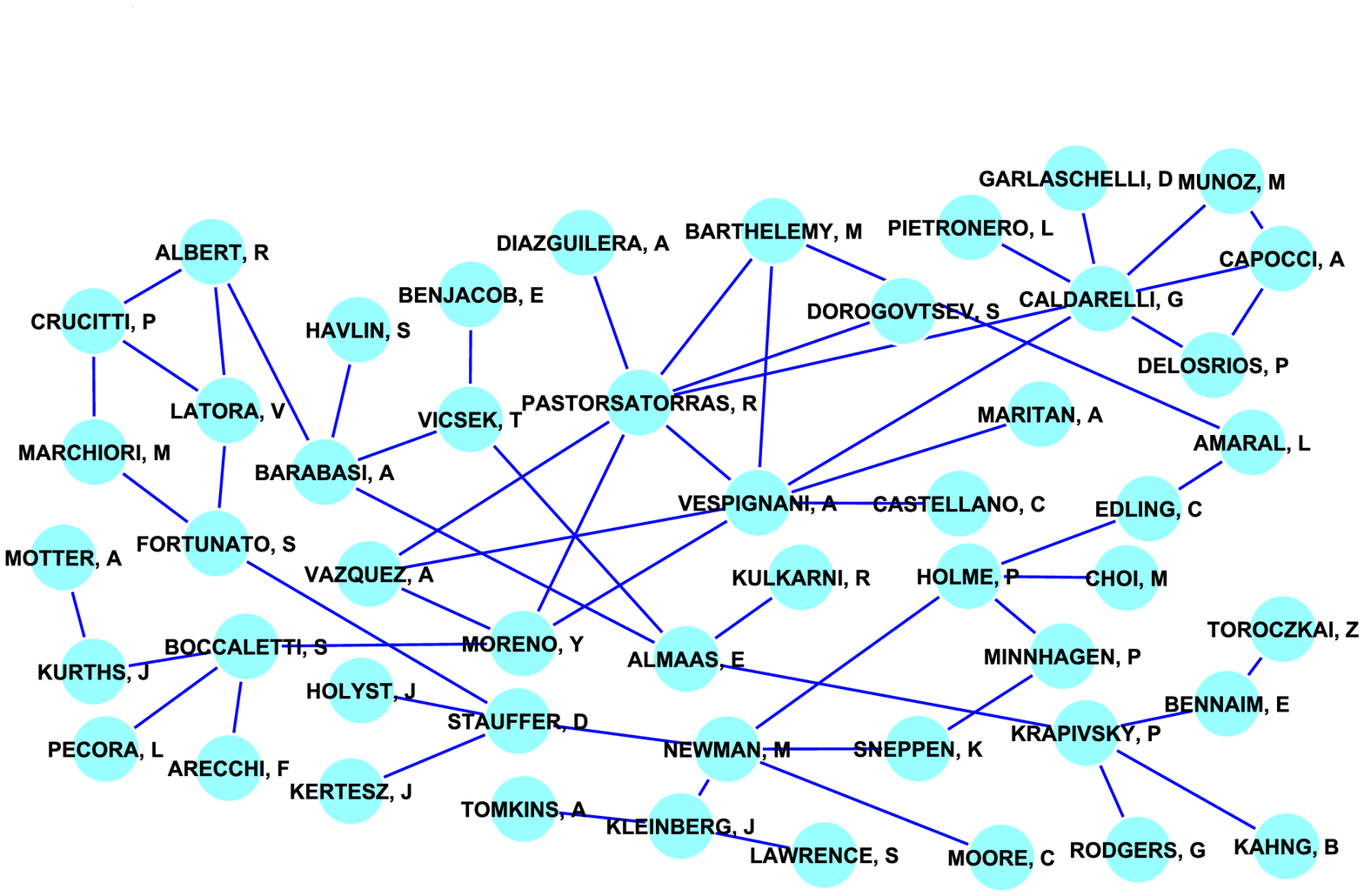,height=1.8in, width=3.4in} \label{fig:net50}}
        \subfigure[{\small 40-vertex backbone (DM)}]{\epsfig{figure=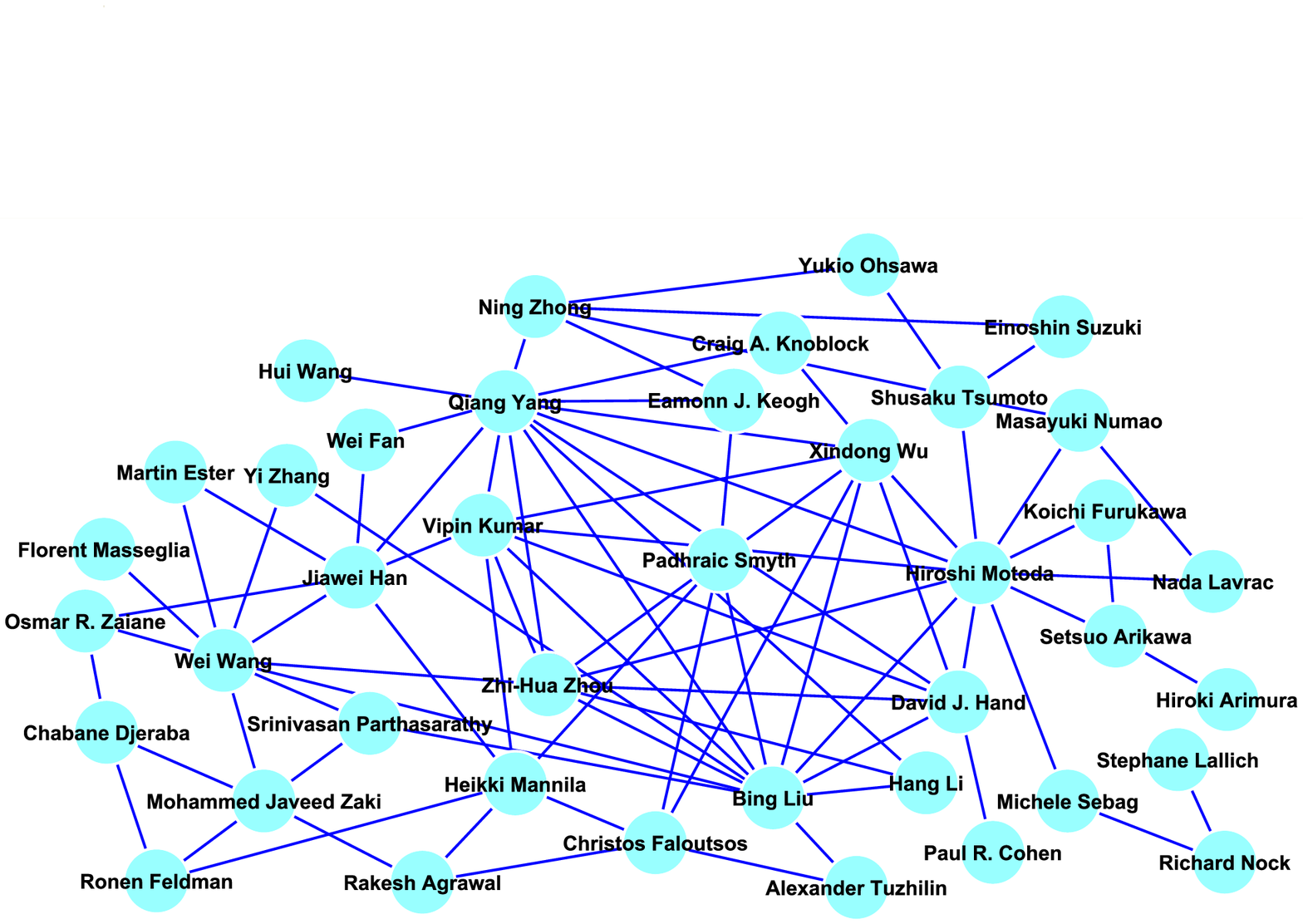,height=1.8in, width=3.6in} \label{fig:dm40}}
    }
    \caption{Backbone discovery on real-world datasets}
\end{figure*}

\subsection{Real Datasets}
\label{dataset}

We study the bimodal markovian model on three real-world datasets, including one biological network and two co-author networks on different research fields:

\noindent {\bf Yeast}: The yeast protein-protein interaction network~\cite{pajek} includes $2361$ vertices
and $6646$ edges. Each vertex indicates one protein and each edge denotes the interaction between two proteins. The
network's average pairwise shortest distance is $4.4$.

\noindent{\bf Net}: The coauthorship network~\cite{Newman06} of researchers who work in the field of network theory and experimentation, as collected by M. Newman.
An edge joins two authors if and only if these two have collaborated on at least one paper in this area.
Since the entire network consists of several disconnected components,
we extract the largest connected component with $379$ vertices and $1828$ edges for the experiment.
The network's average pairwise shortest distance is $6.1$.

\noindent{\bf DM}: The co-author network in the field of data mining~\cite{Tang08} which consists of $2000$ researchers.
        Each of the $10615$ edges indicates that the authors have co-authored at least one paper.
        The network's average pairwise shortest distance is $4.6$.

\subsection{Performance of Bimodal Markovian Model}
\label{bmperform}

In the following experiments, we investigate the performance of bimodal markovian model from several aspects.

\comment{
\noindent{\bf Modeling Performance on Various Path Sets}
We first study the performance of bimodal markovian model given different path sets under various path length thresholds (i.e., length of paths in the path set no less than threshold).
Note that this follows the intuition that the network backbone mainly serves the global system wide traffic, and thus in general only the ``non-local'' shortest paths (length no less than certain threshold) need to be considered.
We specify the size of backbone on Yeast, Net and DM to be $200$, $50$ and $200$, respectively.
Figure~\ref{yeastps}, Figure~\ref{netps} and Figure~\ref{coauthorps} show the preserved modeling accuracy on 3 datasets as the path length threshold varies from $0$ to $8$.
We can observe that the modeling accuracy is quite stable when the path length threshold is no more than average pairwise shortest distance (the average distances of Yeast, Net, DM are $4.4$, $6.1$, and $4.6$, respectively).
However, when the threshold is greater than the average pairwise shortest distance, the modeling accuracy has a significant drop.
This seems to indicate that the path set only consisting paths longer than the average pairwise shortest distance is not sufficient to capture the system-wide information traffics.
Since the larger the threshold, the smaller the input path set and the faster the algorithms run.
Thus, we choose the path length threshold to be the largest integer no more than the average pairwise shortest distance in the following experiments.
}

\noindent{{\bf Preserved Modeling Accuracy: }}
To verify the performance of bimodal markovian model on preserving modeling accuracy of edge markovian model while reducing its number of parameters,
we apply all approaches VB, MCG and ITER on above $3$ datasets.
The size of backbone is supposed to be small, thus we vary the number of vertices in backbone in the range from $5\%$ to around $20\%$ of the number of vertices in original networks.
Especially, for Yeast, the number of vertices in the backbone varies from $100$ to $600$.
The number of vertices in the backbone are set in the range from $20$ to $155$ and the range from $100$ to $400$, respectively, for datasets Net and DM.
We make the following observations:

Figure~\ref{yeastratio}, Figure~\ref{netratio} and Figure~\ref{coauthorratio} show that
bimodal markovian model based on backbones discovered by both ITER and MCG greatly preserve the modeling accuracy of edge markovian model (refer to it as {\bf EM}).
Their modeling accuracies are consistently better than the one of backbones discovered by VB.
In particular, for datasets Yeast and DM, the ratio between EM and bimodal markovian model based on backbones discovered by MCG and ITER are higher or very close to $90\%$ in most of cases.
Among two methods, ITER utilizing iterative refinement strategy consistently achieves better results than MCG on all datasets.
As the number of vertices in the backbone increases, the ratios of all methods are slowly decreased in Yeast and DM.
More vertices are considered as backbone vertices, the simpler the model becomes (this is confirmed by Figure~\ref{yeastparams} and Figure~\ref{coauthorparams}).
This directly leads to the coarser representation of paths and the decrease of bimodal markovian model's likelihood.
Interestingly, unlike the consistently decreasing trend observed from EM/VB and EM/MCG, the results of ITER on Net do not consistently decrease with the increasing number of vertices in the backbones.
This phenomena might be explained by two reasons: 1) our ITER method employing local search tries to achieve a local optimal solution while not global one;
2) larger backbone is possible to connect some important vertices which simplifies edge markovian model at the expense of less modeling accuracy.
Therefore, it is reasonable to see the climbing trend from the data point corresponding to $35$-vertex backbone to $50$-vertex backbone.


%

\noindent{\bf Backbone Complexity: }
We evaluate the backbone complexity based on the number of edges in the discovered backbone.
From Figure~\ref{yeastedge}, Figure~\ref{netedge} and Figure~\ref{coauthoredge}, we can see that the backbones generated by MCG and ITER are rather sparse.
Overall, the edge density (i.e., $|E|/|V|$) of backbones discovered by both MCG and ITER are very close to or small than $2.5$ on all three datasets.
The edge density of backbones in dataset Net is rather close to $1$, which suggests that discovered backbone is tree-like structure.
However, the edge density of backbone discovered by VB is much denser, which are around $4$ and $3.5$ in Yeast and DM.
In addition, though ITER achieves better results than MCG regarding the number of parameters (Figure~\ref{yeastparams}, Figure~\ref{netparams} and Figure~\ref{coauthorparams}),
the number of edges in the backbones generated by ITER is not guaranteed to be smaller than that of MCG (see Figure~\ref{netedge} and Figure~\ref{coauthoredge}).
This is reasonable because the parameter reduction relies on the number of edges incident to backbone vertices while is independent of the number of backbone edges.
In other words, for each backbone vertex $v$ with immediate neighbors $N(v)$, no matter how many incident edges are backbone edges,
the number of parameters in bimodal markovian model is fixed to be $2 \times |N(v)|$.

\noindent{\bf Parameter Reduction: }
For all three datasets, we compare the parameters reduction ratio between edge markovian model and bimodal markovian models based on backbones discovered by VB, ITER and MCG in Figure~\ref{yeastparams}, Figure~\ref{netparams} and Figure~\ref{coauthorparams}, respectively.
The parameter reduction ratio is computed by \\
$\frac{\#Param_{EM}-\#Param_{BM}}{\#Param_{EM}}$ where $\#Param_{EM}$ and $\#Param_{BM}$ denote the number of parameters used in edge markovian model and bimodal markovian model, respectively.
As we can see, all three approaches VB, ITER and MCG dramatically reduce the number of parameters in edge markovian model.
Among them, it is interesting to see that VB outperforms both ITER and MCG in all settings.
In VB, high-degree vertices tend to be selected as backbone vertices since they have high probability to lie in many shortest paths and have greater vertex betweenness.
Therefore, more conditional probabilities of edges incident to these vertices would be simplified compared to other two methods.
In datasets Yeast and DM, VB on average even reduces $73\%$ and $66\%$ parameters in EM model, respectively.
Both MCG and ITER also achieve good parameter reduction ratio.
In Yeast and DM, even though $5\%$ of vertices in original graphs are backbone vertices, around half of parameters in EM model are reduced while large portion of modeling accuracy is preserved.
Also, more parameters are reduced by MCG than that of ITER in most of cases.
As mentioned before, when the number of vertices in the backbones increases, more incident edges' conditional probabilities tend to be simplified.


Finally, we note that in general, the right backbone size is application-dependent.
Without any prior information, based on experimental results on those $3$ datasets, it seems that using around $10\%$ of vertices in original networks as backbone vertices is a reasonable choice.
There are significant losses of modeling accuracy for larger backbones and the number of parameter reduction is not high for smaller backbones.

\begin{figure*}[!htbp]
    \centering
    \mbox{
        \includegraphics[height=1.6in, width=5.0in]{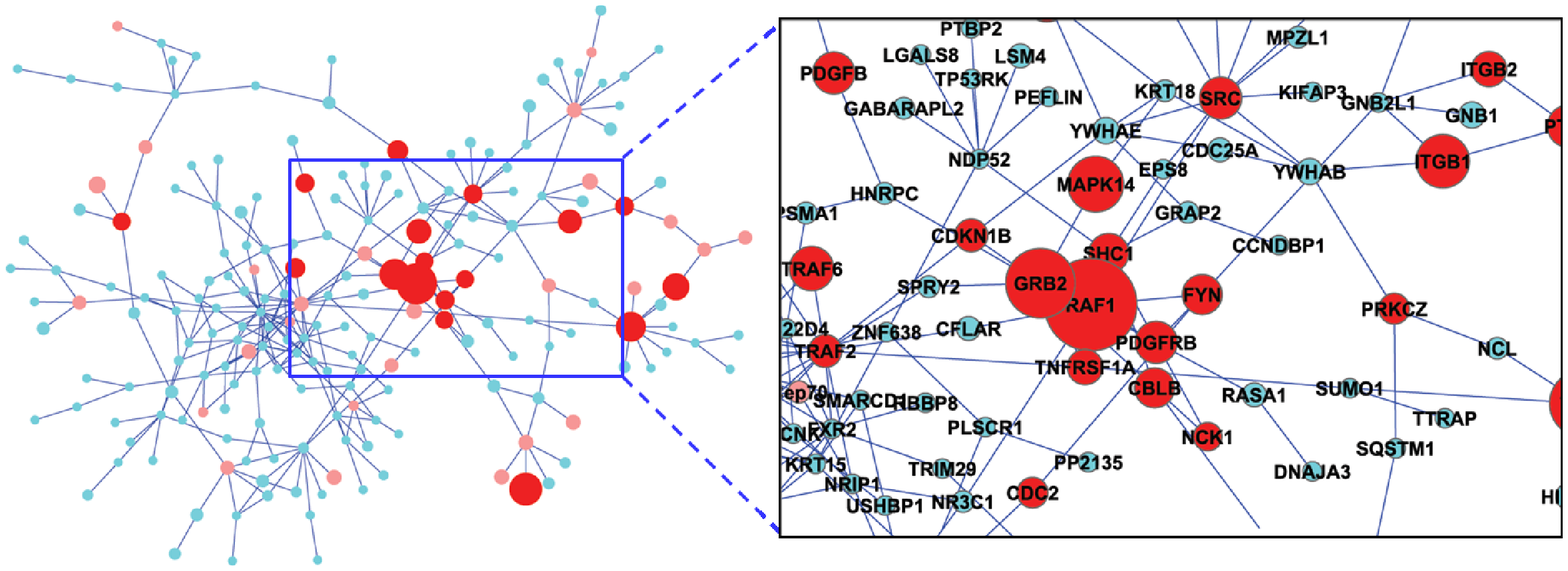}
    }
    \caption{Visualization of the 200 gene backbone for the PPI network.  Red color indicates genes involving in at least 4 KEGG pathways. The larger the nodes, the more KEGG pathways they are involved in.}
    \label{fig:NetVis}
\end{figure*}

\begin{figure*}
    \centering{
        \includegraphics[height=1.4in, width=2.2in]{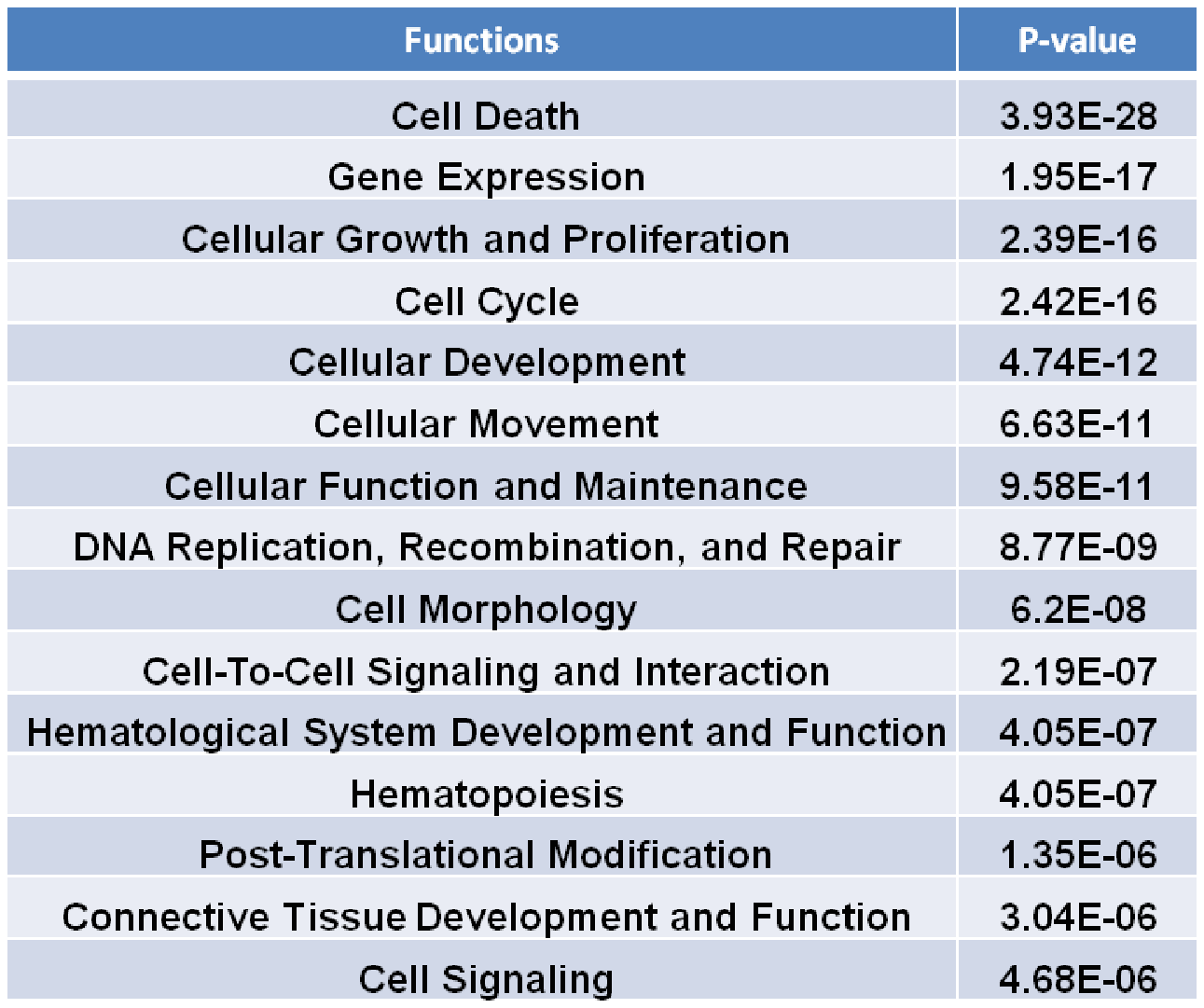}\label{IPA1}
        \includegraphics[height=1.4in, width=2.2in]{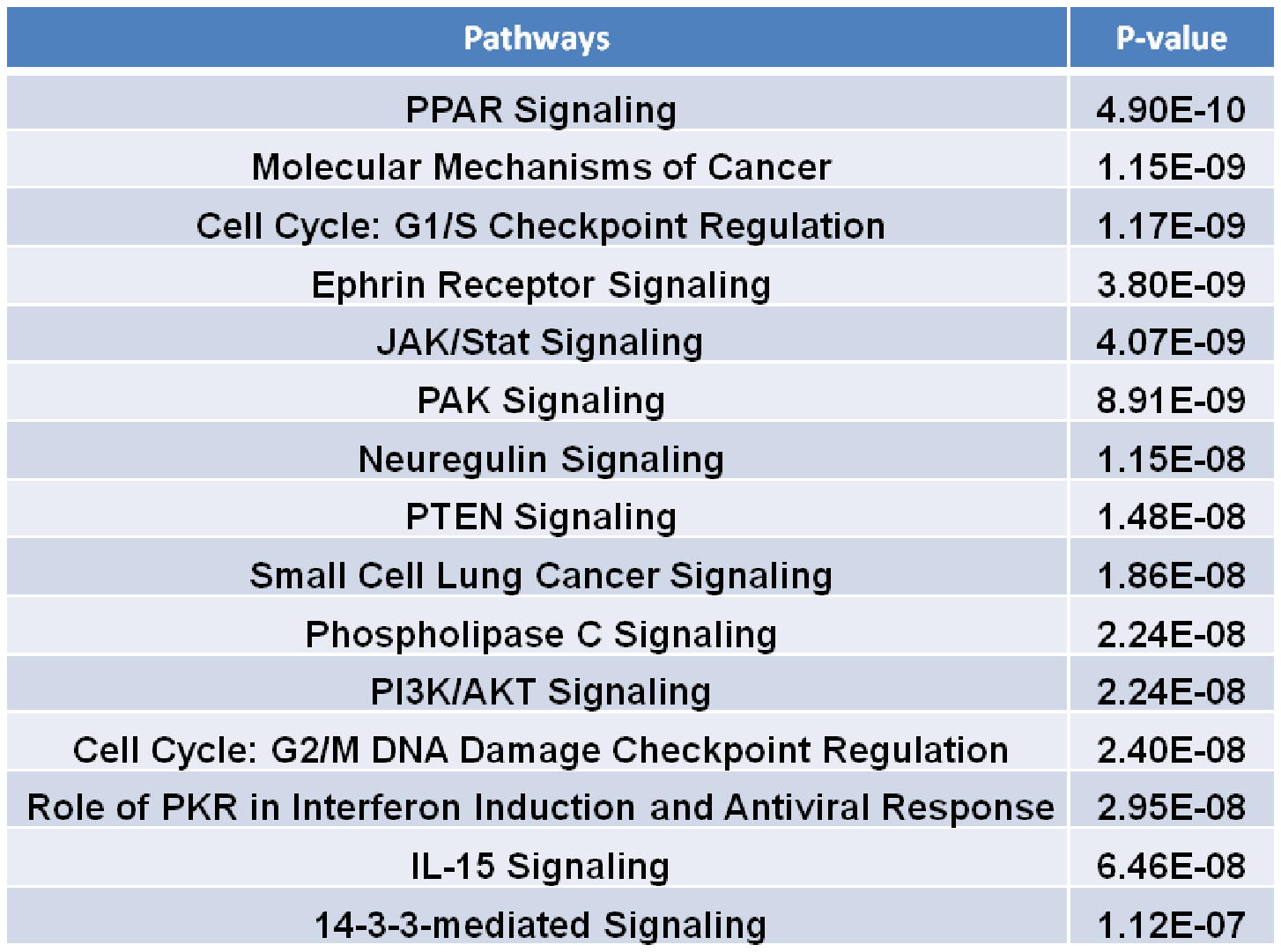}\label{IPA3}
        \includegraphics[height=1.4in, width=2.2in]{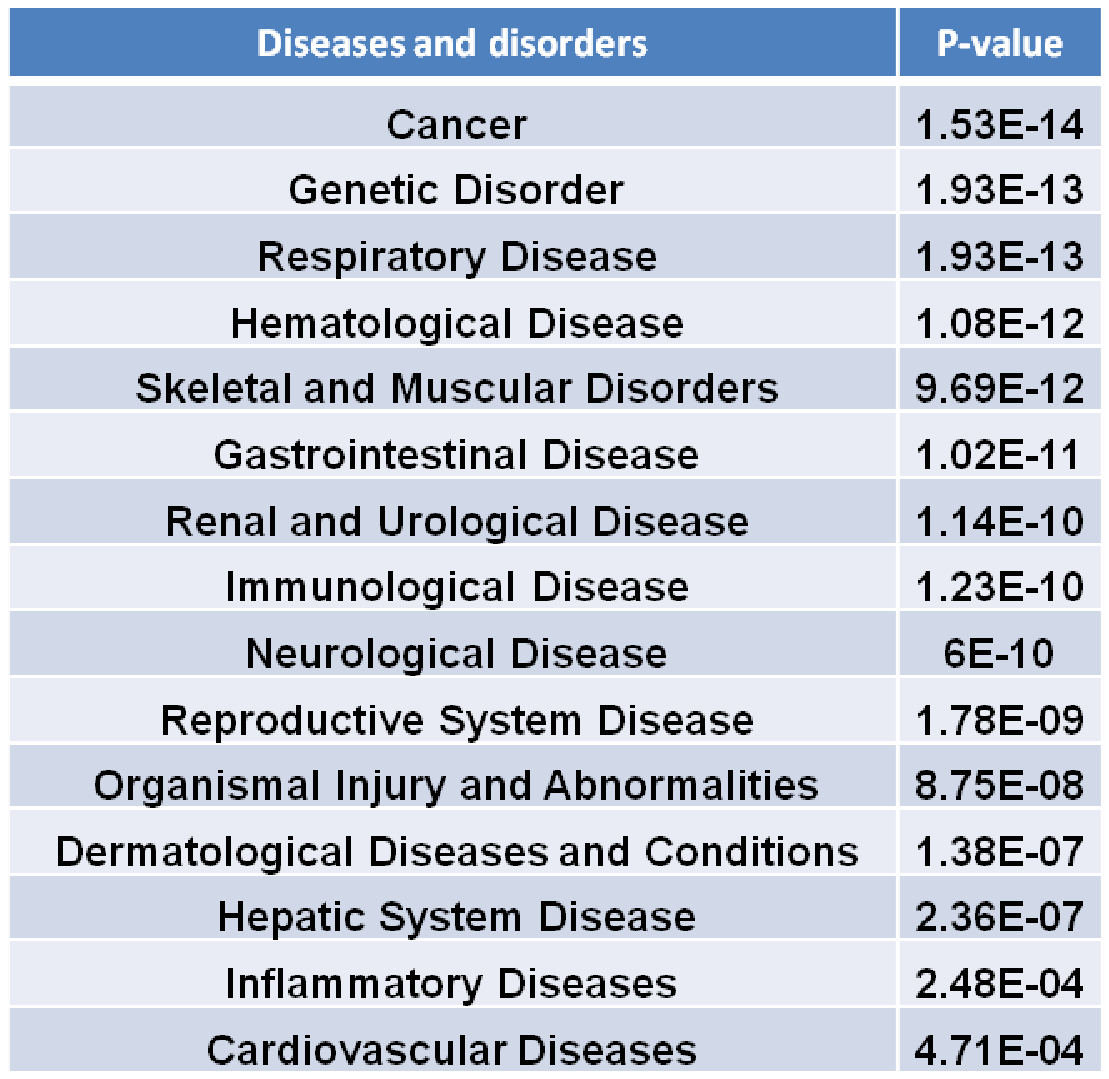}\label{IPA2}
    }
    \caption{Function and pathway enrichment analysis on the 200 backbone genes by IPA. Left: Top 15 enriched functional categories. Middle: Top 15 enriched canonical pathways. Right: Top 15 diseases and disorders.}
    \label{fig:IPA}
\end{figure*}

%
%

\begin{figure}[!htbp]
\centering
\begin{tabular}{c}
\psfig{figure=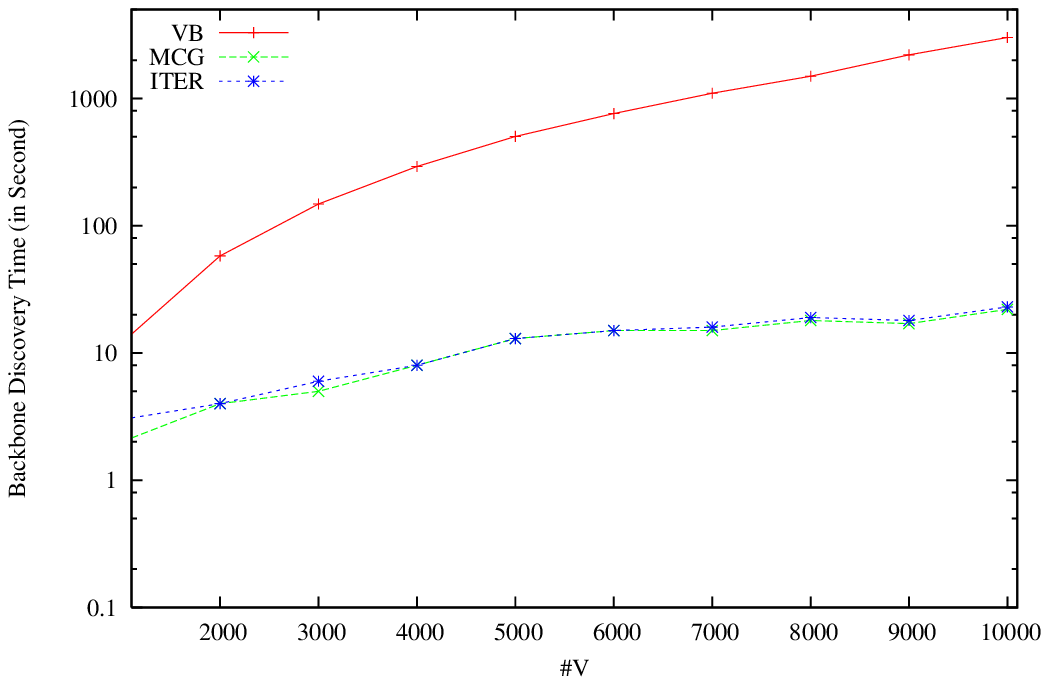,scale=0.6}
\end {tabular}

\caption {Running time for power-law graphs}

\label{fig:plbbtime}
\end{figure}

\subsection{Performance Study}
\label{performance}

To verify the scalability of our approach, we test a set of random undirected graphs with power-law degree distribution.
The graphs vary in size from $10K$ to $100K$ vertices and we set the edge density to be $4$.
We specified each backbone to have $100$ vertices.

We decompose the running time into two parts: preprocessing time (i.e., computing shortest paths and calculating edge or segment betweenness for basic probabilities) and backbone discovery time.
Figure~\ref{fig:plbbtime} shows the backbone discovery time of VB, ITER and MCG for random graphs with power-law degree distribution.
These results clearly demonstrate the scalability of approaches MCG and ITER.
In particular, the running time of ITER is very close to MCG,
because the extra computational cost of ITER (Algorithm~\ref{alg:iterativealg}) compared to MCG only depends on the size of backbone which is supposed to be small.
This also confirms our time complexity analysis on ITER and MCG.
Both are much faster than straightforward method VB by a factor of $59$, due to the high cost of building minimal steiner tree in VB.
The preprocessing step, especially computing the pairwise shortest distances, as expected is more expensive.
The preprocessing time of all methods varies from $30$ seconds to $121$ minutes.
We note that sampling seems to be an effective approach to avoid the full pairwise computation, thus speeding up the preprocessing time.
It is beyond the scope of this paper and will be investigated in future work.

\comment{
\begin{figure*}
    \centering
    {\small
    \mbox{
        \subfigure[10-vertex backbone]{\includegraphics[width=2.2in,height=1.4in]{Figures/netscience_b10.eps}\label{b1}}
        \subfigure[20-vertex backbone]{\includegraphics[width=2.2in,height=1.4in]{Figures/netscience_b20.eps}\label{b2}}
        \subfigure[25-vertex backbone]{\includegraphics[width=2.2in,height=1.4in]{Figures/netscience_b25.eps}\label{b3}}
    }
    }
    \vspace*{-2ex}
    \caption{Backbone discovery on network theory co-author network}
    \label{fig:netscience}
    \vspace*{-2.0ex}
\end{figure*}

\begin{figure*}
    \centering
    {\small
    \mbox{
        \subfigure[Time for Erd\"{o}s-R\'{e}nyi random graphs]{\includegraphics[width=2.2in,height=1.2in]{Figures/uniform_bbtime.eps}\label{fig:ubbtime}}
        \subfigure[Time for random graphs, power law degree distribution]{\includegraphics[width=2.2in,height=1.2in]{Figures/powerlaw_bbtime.eps}\label{fig:plbbtime}}
         \subfigure[Model Likelihood Ratio]{\includegraphics[width=2.2in,height=1.2in]{Figures/coauthor_ratio_results.eps}\label{fig:modellr}}
    }
    }
    \vspace*{-2ex}
    \caption{Experimental Results}
    \vspace*{-2.0ex}
\end{figure*}

\begin{figure*}
\centering
{\small
\begin{tabular}{c}
\psfig{figure=Figures/coauthor89_40.eps,width=5in,height=1.8in}
\end {tabular}
}
\vspace*{-2.0ex}
\caption {Backbone with 40 vertices from data mining co-author network}
\label{fig:coauthor40}
\vspace*{-3.0ex}
\end{figure*}

In this section, we empirically study the performance of our backbone discovery scheme.
First, we apply our method and discuss the results for 3 real-world datasets (biological network and two co-author networks).
Also, we show the optimality of our backbone model with respect to likelihood ratio between backbone model and Edge Markovian model.
Then, we study the efficiency of the method on large random undirected graphs: one with Erd\"{o}s-R\'{e}nyi random graph and one with a power law degree distribution\footnote{The power law graph generator can be downloaded from http://www.cs.ucr.edu/$\sim$ddreier/barabasi.html}.
We implemented all algorithms using C++ and the Standard Template Library (STL).
All experiments were conducted on a 2.0GHz Dual Core AMD Opteron CPU with 4.0GB RAM running Linux.

\vspace*{-1.0ex}
\subsection{Study Backbone in Biological Network}
\label{humanppi}
\vspace*{-1.0ex}

We applied the backbone discovery algorithm on the human protein-protein interaction (PPI) dataset obtained from ~\cite{HumanPPI05} to identify backbone of the human PPI.
This dataset consists of $3133$ genes and $12298$ edges indicating relationships among them.
Our algorithm returned the genes in the backbone with the user specified size (which is 200 in our test).
As shown in Figure~\ref{fig:NetVis}, the backbone genes contain many well known and important genes in cellular signaling transduction pathways including both kinases (e.g., {\em RAF1}, {\em MAPK14}, {\em SRC} and {\em FYN}) and receptors (e.g., {\em TRAF6}, {\em PDGFRB}) as well as signaling molecules such as {\em PDGFB}.
Unlike traditional gene set discovery studies for which we expect to obtain a group of genes with a small set of specifically enriched functions or pathways, we expect that the backbones genes of the PPI network would be engaged in many different functions and possibly pathways.
The functional and pathway analysis using tools such as the Ingenuity Pathway Analysis (IPA) indeed confirmed our expectation. As shown in Figure \ref{fig:IPA}, the 200 genes are highly enriched with a wide spectrum of important biological functions and are related to many different diseases with high statistical significance.
Moreover, they are involved in a large number of pathways, which is very rare for a gene list of this size. For the IPA canonical pathways, the 200 genes show enrichment with p-values (of the Fisher's exact test used by IPA) less than 0.0001 for more than 70 different pathways.
These observations suggest that many backbone genes may involve in more than one pathways.  Indeed we found that out of the 200 genes (of which 195 can be mapped to KEGG gene ids) 47 are involved in at least {\em four} KEGG pathways.
This is a highly significant enrichment comparing to the fact that a total 1,100 such genes can be found among the entire genome of 19,076 annotated human genes in the KEGG database ($p < 1.9\times 10^{-17}$ for hypergeometric test).
It can be conceived that perturbation on these genes can lead to serious disruption of important biological functions, which implies the involvement in diseases in human.
This is also confirmed as shown in Figure~\ref{fig:IPA}. Therefore, our experimental study on the PPI network backbone discovery demonstrated the effectiveness of our approach and its great potential as a new gene ranking tool.

\vspace*{-1.0ex}
\subsection{Study Backbone in Co-author Networks}
\label{real}
 \vspace*{-1.0ex}

{\bf Co-author Network (network theory and experimentation field): }
The first dataset is a coauthorship network \cite{Newman06} of researchers who work in the field of network theory and experimentation, as collected by M. Newman.
An edge joins two authors if and only if these two have collaborated on at least one paper in this area.
Since the entire network consists of several disconnected components, we extract the largest connected component with $379$ vertices and $1828$ edges for this experiment.
Figure \ref{fig:netscience} shows the discovered backbones from this component with backbone sizes varying from $10$ to $25$.
These figures precisely depict the backbone generation and growth process.
In Figure \ref{b1}, the backbone is a sparse chain-like subgraph with only $10$ vertices.
Comparing this backbone to the full connected component\footnote{A figure of the largest connected component can be found at http://www-personal.umich.edu/$\sim$mejn/centrality/},
we see that these vertices serve as the essential connectors among several ``small world'' components.
As the backbone expands to 20 vertices in Figure \ref{b2}, all 10 of the earlier vertices are retained, while several important researchers, such as J. Kleinberg and G. Caldarelli are added.
Then, the backbone is slightly expanded from Figure \ref{b2} to Figure \ref{b3}.
Another interesting observation is that some of the researchers in the backbone are not necessarily the best-known scientists nor do they have a high number of collaborators.
For instance,{\em G. Bianconi} and {\em C. Edling} in the backbone only have 4 and 5 collaborators in this network, respectively.
In contrast to the traditional research which aims to discover highly correlated components, our backbone model studies complex networks from a new angle.
The discovered backbone essentially captures the communication path among different highly correlated communities.

{\bf Optimality of Backbone Model on Co-author Network (network theory and experimentation field): }
In order to demonstrate the optimality of our backbone model, we show the likelihood ratio between our backbone model (referred to as {\bf EB}) and two benchmarks, Edge Independent Model (referred to as {\bf EI}) and Edge Markovian Model (referred to as {\bf EM}) in Figure \ref{fig:modellr}.
The likelihood ratio between backbone model and Edge Markovian model is expressed by the value of likelihood value of backbone model divides the one of Edge Markovian model (denoted by $EB/EM$).
Similar calculation is applied for {\bf EI} model and backbone model (denoted by $EB/EI$).
As discussed early, a smaller value of $EB/EM$ suggests that our backbone model can be used to effectively compress Edge Markovian model while slightly sacrificing the accuracy.
Overall, the likelihood ratio of $EB/EM$ is on average approximately $1.1$ which is rather close to the ideal value $1$, while the likelihood ratio of $EB/EI$ is averagely around $0.7$.
This demonstrates the effectiveness of our backbone model which can achieve ``good'' modeling accuracy while significantly reducing the number of parameters (subsection \ref{backbonemodel}) comparing to Edge Markovian model.
Finally, we can see that the likelihood ratio between backbone model and Edge Markovian model is very consistent with respect to different $K$ (i.e, the number of vertices in the backbone).

{\bf Co-author Network (data mining field): }
The second dataset is a co-author network in the field of data mining \cite{Tang08} which consists of $2000$ researchers.
Each of the $10615$ edges indicates that the authors have co-authored at least one paper.
From Figure \ref{fig:coauthor40}, we can see that many of the discovered researchers, like {\em Jiawei Han}, {\em Rakesh Agrawal} and {\em Christos Faloutsos} are prominent scientists in data mining.
Compared to the sparse backbone in Figure \ref{fig:netscience}, the backbone from the data mining co-author network is much denser.
This indicates the different collaboration styles in different research fields.
In the field of network theory and experiment, researchers tend to collaborate within small groups while a few of them have connections among different groups.
However, in the data mining area, many scientists work in several different directions which results in more wide-ranging collaborations among them.

\vspace*{-1.0ex}
\subsection{Performance Study}
\label{performance}
 \vspace*{-1.0ex}

To verify the scalability of our approach, we tested all three approaches on .
The graphs vary in size from $10K$ to $100K$ vertices.
The first set of graphs follows the Erd\"{o}s-R\'{e}nyi random graph model with edge density of $2$.
The second set of graphs changes the density to $4$ and applies the power-law degree distribution.
We specified each backbone to have $100$ vertices.

We decompose the running time into two parts: preprocessing time (i.e., computing shortest paths and calculating edge or segment betweenness for basic probabilities) and backbone discovery time.
Figure \ref{fig:ubbtime} and Figure \ref{fig:plbbtime} show the backbone discovery time for our random graphs with uniform degree distribution and power-law degree distribution, respectively.
These results clearly demonstrate the scalability of our developed approach.
The preprocessing step, especially computing the pairwise shortest distances, as expected is more expensive.
Preprocessing varies from $2$ minutes to $411$ minutes.
We note that sampling seems to be an effective approach to avoid the full pairwise computation, thus speeding up the preprocessing time.
It is beyond the scope of this paper and will be investigated in future work.
}

\section{Case Studies}
\label{humanppi}

In this section, we report network backbones in co-author networks and the PPI network discovered by ITER method.

\noindent{\bf Co-author Networks (Net and DM): }
Figure~\ref{fig:net25}, Figure~\ref{fig:net35} and Figure~\ref{fig:net50} show the discovered backbones with the number of backbone vertices varying from $25$ to $50$.
These figures precisely depict the backbone generation and growth process.
In Figure~\ref{fig:net25}, the backbone is a sparse subgraph with only $25$ vertices.
Comparing this backbone to the full connected component\footnote{A figure of the largest connected component can be found at http://www-personal.umich.edu/$\sim$mejn/centrality/},
we see that these vertices serve as the essential connectors among several ``small world'' components.
As the backbone expands to $35$ vertices in Figure~\ref{fig:net35}, most of the earlier vertices are retained, while several important researchers, such as J. Kleinberg and P. Holme are added.
Then, the backbone is slightly expanded from Figure~\ref{fig:net35} to Figure~\ref{fig:net50}.
Another interesting observation is that some of the researchers in the backbone are not necessarily the best-known scientists nor do they have a high number of collaborators.
For instance, {\em C. Edling} in the backbone only has 5 collaborators in this network.
In contrast to the traditional research which aims to discover highly correlated components, our backbone model studies complex networks from a new angle.
The discovered backbone essentially captures the communication path among different highly correlated communities.

From Figure \ref{fig:dm40}, we can see that many of the discovered researchers, like {\em Jiawei Han}, {\em Rakesh Agrawal} and {\em Christos Faloutsos} are prominent scientists in data mining.
Compared to relatively sparse backbones on Net, the backbone from the data mining co-author network is denser.
This indicates the different collaboration styles in different research fields.
In the field of network theory and experiment, researchers tend to collaborate within small groups while a few of them have connections among different groups.
However, in the data mining area, many scientists work in several different directions which results in more wide-ranging collaborations among them.

\noindent{\bf Human PPI Network: }
We applied the backbone discovery algorithm ({\bf ITER}) on the human protein-protein interaction (PPI) dataset obtained from ~\cite{HumanPPI05} to identify backbone of the human PPI.
This dataset consists of $3133$ genes and $12298$ edges indicating relationships among them.
Our algorithm returned the genes in the backbone with the user specified size (which is 200 in our test).
As shown in Figure~\ref{fig:NetVis}, the backbone genes contain many well known and important genes in cellular signaling transduction pathways including both kinases (e.g., {\em RAF1}, {\em MAPK14}, {\em SRC} and {\em FYN}) and receptors (e.g., {\em TRAF6}, {\em PDGFRB}) as well as signaling molecules such as {\em PDGFB}.
Unlike traditional gene set discovery studies for which we expect to obtain a group of genes with a small set of specifically enriched functions or pathways, we expect that the backbones genes of the PPI network would be engaged in many different functions and possibly pathways.
The functional and pathway analysis using tools such as the Ingenuity Pathway Analysis (IPA) indeed confirmed our expectation. As shown in Figure \ref{fig:IPA}, the 200 genes are highly enriched with a wide spectrum of important biological functions and are related to many different diseases with high statistical significance.
Moreover, they are involved in a large number of pathways, which is very rare for a gene list of this size. For the IPA canonical pathways, the 200 genes show enrichment with p-values (of the Fisher's exact test used by IPA) less than 0.0001 for more than 70 different pathways.
These observations suggest that many backbone genes may involve in more than one pathways.  Indeed we found that out of the 200 genes (of which 195 can be mapped to KEGG gene ids) 47 are involved in at least {\em four} KEGG pathways.
This is a highly significant enrichment comparing to the fact that a total 1,100 such genes can be found among the entire genome of 19,076 annotated human genes in the KEGG database ($p < 1.9\times 10^{-17}$ for hypergeometric test).
It can be conceived that perturbation on these genes can lead to serious disruption of important biological functions, which implies the involvement in diseases in human.
This is also confirmed as shown in Figure~\ref{fig:IPA}. Therefore, our experimental study on the PPI network backbone discovery demonstrated the effectiveness of our approach and its great potential as a new gene ranking tool.

\section{Conclusion}
\label{conc}

In this paper, we introduce a new backbone discovery problem and propose novel discovering approaches based on vertex betweenness and $KL$-divergence.
We believe the backbone approach opened a new way to study complex networks and systems, and also presents many new research questions for both data mining and complex network research:
How do network backbone and modularity coexists and how they affect each other?   How robust is the backbone, and how will it change?  What information is carried in the backbone?
We plan to work on these fascinating questions in the future.


\bibliographystyle{plain}
\bibliography{bib/ComplexNetwork,bib/ComplexNetwork2,bib/ConnectedSubgraph,bib/simplification}

\end{document}